\documentclass[a4paper,11pt]{article}
\pdfoutput=1 

\usepackage{jheppub} 

\usepackage[T1]{fontenc} 

\usepackage{braket}
\usepackage{comment}
\usepackage{braket}
\usepackage{float}
\usepackage{nicematrix}
\usepackage{amsmath}
\usepackage{enumitem}
\title{\boldmath Spacing Statistics of Energy Spectra: Random Matrices, Black Hole Thermalization, and Echoes}

\author[a,b,c]{Krishan Saraswat}
\author[a,b,c]{and Niayesh Afshordi}

\affiliation[a]{Department of Physics and Astronomy, University of Waterloo, 200 University Ave W, Waterloo, Canada}
\affiliation[b]{Waterloo Centre for Astrophysics, University of Waterloo, Waterloo, ON, N2L 3G1, Canada}
\affiliation[c]{Perimeter Institute For Theoretical Physics, 31 Caroline St N, Waterloo, Canada}

\emailAdd{ksaraswat@pitp.ca}
\emailAdd{nafshordi@pitp.ca}

\abstract{Recent advances in AdS/CFT holography have suggested that the near-horizon dynamics of black holes can be described by random matrix systems. We study how the energy spectrum of a system with a generic random Hamiltonian matrix affects its early and late time thermalization behaviour using the spectral form factor (which captures the time-dependence of two-point correlation functions). We introduce a simple statistical framework for generating random spectra in terms of the nearest neighbor spacing statistics of energy eigenvalues, enabling us to compute the averaged spectral form factor in a closed form. This helps to easily illustrate how the spectral form factor changes with different choices of nearest neighbor statistics ranging from the Poisson to Wigner surmise statistics. We suggest that it is possible to have late time oscillations in random matrix models involving $\beta$-ensembles (generalizing classical Gaussian ensembles). We also study the form factor of randomly coupled oscillator systems and show that at weak coupling, such systems exhibit regular decaying oscillations in the spectral form factor making them interesting toy models for gravitational wave echoes. We speculate on the holographic interpretation of a system of coupled oscillators, and suggest that they describe the thermalization behaviour of a black hole geometry with a membrane that cuts off the geometry at the stretched horizon.}

\begin{document} 
\maketitle

\section{Introduction}
\label{introsec}
\subsection{Background and Context}
Black holes serve as an important testing ground and challenge for our understanding of the quantum nature of gravity. The application of quantum field theory on black hole backgrounds leads to the conclusion that pure states can evolve into mixed states through the mechanism of Hawking radiation \cite{Hawking:1975vcx,Hawking:1976ra}. On the other hand, theoretical developments in string theory and AdS/CFT suggest that quantum gravity should be a unitary theory. The conflict with the expectation that quantum gravity should be a unitary theory and the apparent violation of unitarity by Hawking's calculation has given rise to the black hole information problem \cite{Mathur:2009hf,Harlow:2014yka,Stoica:2018uli}. Explorations into the information problem have led to the idea of quantum gravity effects modifying the near horizon description of black holes, most notably in the form of some kind of microstructure near the horizon (e.g. fuzzballs, firewalls, non-locality etc.) \cite{Almheiri:2012rt,Polchinski:2016hrw,Stoica:2018uli,Mathur:2009hf,Mathur:2013bra,Hayden:2020vyo,Neuenfeld:2021bsb,Giddings:2021qas}.  

The idea of quantum effects modifying the near horizon physics of a black hole have recently also been considered from an experimental perspective. If quantum effects near the horizon give rise to the possibility of partial reflection rather than complete absorption of a perturbation then it could give rise to ``echoes'' in the ringdown behaviour of a black hole after being perturbed. In particular, such echoes may be detected by LIGO in gravitational waves emitted during black hole binary merger events \cite{Abedi:2016hgu,Cardoso:2016oxy,Cardoso:2016rao,Conklin:2017lwb,Westerweck:2017hus,Oshita:2018fqu,Abedi:2018pst,Abedi:2018npz,Cardoso:2019rvt,Abedi:2020ujo,Abedi:2020sgg}. In most studies, these modifications are modelled by cutting off the black hole geometry near the horizon. At the cutoff, semi-reflective boundary conditions are enforced and perturbations on the cutoff background are studied\footnote{In these studies the time scale over which the echoes manifest after the black hole is perturbed is called the ``echo time scale.'' It depends on how close the cutoff is placed relative to the horizon. For a cutoff placed a proper radial Planck length from a Schwarzschild black hole it can be shown the echo time takes the form of the scrambling time $\beta\ln(S_{BH})$ \cite{Saraswat:2019npa} (e.g. for a 10 solar mass Schwarzschild black hole the scrambling or echo time is of the order of 0.2 secs.). }  \cite{Mark:2017dnq,Wang:2018gin,Saraswat:2019npa,Dey:2020lhq,Dey:2020wzm,Rahman:2021kwb}. Studies that do this make an implicit assumption that the quantum effects that encode the unitary nature of black holes are ``localized'' near the horizon and that they effectively manifest themselves as a cutoff near the horizon with non-standard boundary conditions. Although such models are completely classical, they highlight the interesting possibility of finding observable imprints of horizon scale quantum gravity effects in the ringdown behaviour of black holes.  

In this paper, we will explore the idea of echoes in the context of the thermalization behaviour of quantum chaotic systems. The motivation behind this study is the idea that the near horizon dynamics of a black hole is generally captured by quantum chaotic degrees of freedom. When a perturbation begins to probe and excite the microstructure near the horizon, it could lead to observable deviations from the standard ringdown as predicted by quasi-normal mode decay. If the microstructure effectively behaves as a semi-reflective cutoff near the horizon, then one would get echoes in the unitary description of the black hole. The goal of this paper is to explore the properties a quantum chaotic system, if it were to exhibit echoes in its thermalization behaviour. To aid in this exploration, we will make use of recent advances and tools in AdS/CFT that have placed  interesting constraints on the energy spectrum of black holes. In the next subsection, we will give a brief overview of these developments and tools.
\subsection{Black Holes as Quantum Chaotic Systems}
The AdS/CFT correspondence, which was first formulated by Maldacena \cite{Maldacena:1997re} suggests that (quantum) gravitational systems in AdS have a non-perturbative formulation in terms of a strongly coupled CFT \cite{Polchinski:2010hw,Ramallo:2013bua,VanRaamsdonk:2016exw}. This in turn provides a way to explore various aspects of the quantum nature of black holes from the perspective of a thermal CFT \cite{Maldacena:2001kr,Horowitz:1999jd,Mathur:2009hf,Stoica:2018uli,Polchinski:2016hrw}. 

One of the earliest explorations into the quantum spectrum of a black hole from the perspective of AdS/CFT involved the thermofield double (TFD) state \cite{Maldacena:2001kr}. The TFD is a state which is conjectured to describe a two sided eternal black hole in AdS and is written as:
\begin{equation}
    \ket{TFD}=\frac{1}{\sqrt{\mathcal{Z}(\beta)}}\sum_{n}e^{-\beta E_n/2}\ket{n}_L\otimes\ket{n}_R,
\end{equation}
the state above describes two identical systems (one living on the left boundary and the other on the right) which are entangled. The states $\ket{n}_{L,R}$ are eigenstates of a Hamiltonian with eigenvalues $E_n$, and the quantity $\mathcal{Z}(\beta)$ is the partition function. Using this, one should expect the decay of quasi-normal modes of the dual black hole to be consistent with the decay of two-point thermal correlators. 

It was noted in \cite{Maldacena:2001kr}, that the spectrum of a CFT which is dual to a black hole (with a spherical horizon) lives on a compact space and as a result its spectrum will be discrete. The discreteness of the spectrum implies that correlation functions cannot decay to zero which is an indication of unitary evolution in the bulk \cite{Maldacena:2001kr,Harlow:2014yka,Barbon:2003aq, Solodukhin:2004rv}. The inability for correlators to decay to zero comes as a surprise from the bulk perspective because quasi-normal modes for an AdS black hole decay exponentially to zero \cite{Horowitz:1999jd}. It was argued in \cite{Maldacena:2001kr} that correlators will decay in a manner consistent with the semi-classical calculation of quasi-normal modes of the black hole until the correlator roughly becomes of the order $e^{-S}$, where $S$ is the entropy. Afterwards, the discreteness of the spectrum will become important and the semi-classical calculation will no longer agree with the boundary calculation of the two-point function. In a semi-classical context, it was shown that to prevent the continual decay in the bulk, one can include other saddle point geometries in the Euclidean path integral approach \cite{Maldacena:2001kr}. It was also demonstrated that one can avoid continual decay of the two-point function and make the spectrum discrete by cutting off the spacetime near the horizon \cite{tHooft:1996rdg,Barbon:2003aq, Solodukhin:2004rv}.

Quantum chaos is another important feature of thermal CFT systems that are dual to black holes. In particular, CFT systems that are dual to black holes are conjectured to be maximally chaotic\footnote{The SYK model is an example of a maximally chaotic system which has been of recent interest \cite{Polchinski:2016xgd,Garcia-Garcia:2016mno,Garcia-Garcia:2017bkg,Sarosi:2017ykf}.} \cite{Maldacena:2015waa}. Quantum chaos manifests itself in the spectrum of a quantum system through nearest neighbor eigenvalue ``repulsion.'' In particular, it is conjectured that many statistical properties of the spectrum of a quantum chaotic system follow the spectrum statistics of random matrices drawn from a suitably chosen ensemble \cite{Bohigas:1983er,Eynard:2015aea,livan2017introduction}. The connection between gravity and random matrix theory has been further explored in the context of 2D Jackiw-Teitelboim (JT) gravity by showing the partition function of JT gravity can be expressed as a random matrix integral \cite{Saad:2018bqo,Saad:2019lba,Stanford:2019vob,Blommaert:2019wfy}. These recent results corroborate the conjecture that random matrix theories are capable of describing the statistical properties of the spectrum of quantum chaotic systems such as black holes. 

A particularly useful observable that has been computed in the context of quantum chaos, is the normalized spectral form factor which is defined by:
\begin{equation}
 \frac{\mathcal{Z}(\beta+it)\mathcal{Z}(\beta-it)}{\mathcal{Z}(\beta)^2}=\frac{\sum_{n,m}e^{-\beta(E_m+E_n)}e^{i(E_n-E_m)t}}{\sum_{n,m}e^{-\beta(E_m+E_n)}},
\end{equation}
where $\mathcal{Z}(\beta)$ is the partition function in the canonical ensemble with inverse temperature $\beta$. It is a useful quantity since it can be used to diagnose the discreteness of a spectrum. In the context of the TFD state, if one considers the two point function of a Hermitian operator of the form $\mathcal{I}_R\otimes A$ (where $\mathcal{I}_R$ is the identity on the right boundary and $A$ is a Hermitian operator living on the left boundary) we have: \begin{equation}
\begin{split}
    &\braket{TFD|[\mathcal{I}_R\otimes A(t)][\mathcal{I}_R\otimes A(0)]|TFD}=\frac{1}{\mathcal{Z}(\beta)}\sum_{n,m}e^{-\beta E_n}e^{i(E_n-E_m)t}\left|\braket{n|A|m}\right|^2\\
    &=\frac{1}{\mathcal{Z}(\beta)}\left[\sum_{n}e^{-\beta E_n}\left|\braket{n|A|n}\right|^2 +\sum_{n,m;n\neq m} e^{-\beta E_n}e^{i(E_n-E_m)t}\left|\braket{n|A|m}\right|^2\right],\\
    \end{split}
\end{equation}
we can see the time dependence (up to matrix elements which will depend on the energy) of the form factor and the two point function are governed by the details in the energy differences in the spectrum. If the matrix elements are smooth functions of energy that vary slowly, we can use the form factor as a proxy for understanding how a perturbation thermalizes for a TFD state. This is interesting because by identifying a Hamiltonian and its spectrum of states we can construct the associated TFD state. If we conjecture that the TFD is dual to a two-sided black hole geometry, then we can roughly interpret the early time behaviour of the spectral form factor as a description of the ringdown of the black hole after being perturbed (i.e. the quasi-normal modes). From this perspective, different choices of Hamiltonian give different types of ringdowns of the conjectured dual black hole. Usually, at early times, the discrete spectrum can be replaced by a coarse grained smooth density - and one initially sees decay in the form factor (controlled by the decay of quasi-normal modes of the black hole). At very late times, the behaviour of the spectral form factor is governed primarily by the small energy differences between nearby eigenvalues -  and in general it produces very erratic oscillations which never decay to zero. 

\begin{figure}[h]
\centering
\includegraphics[width=120mm]{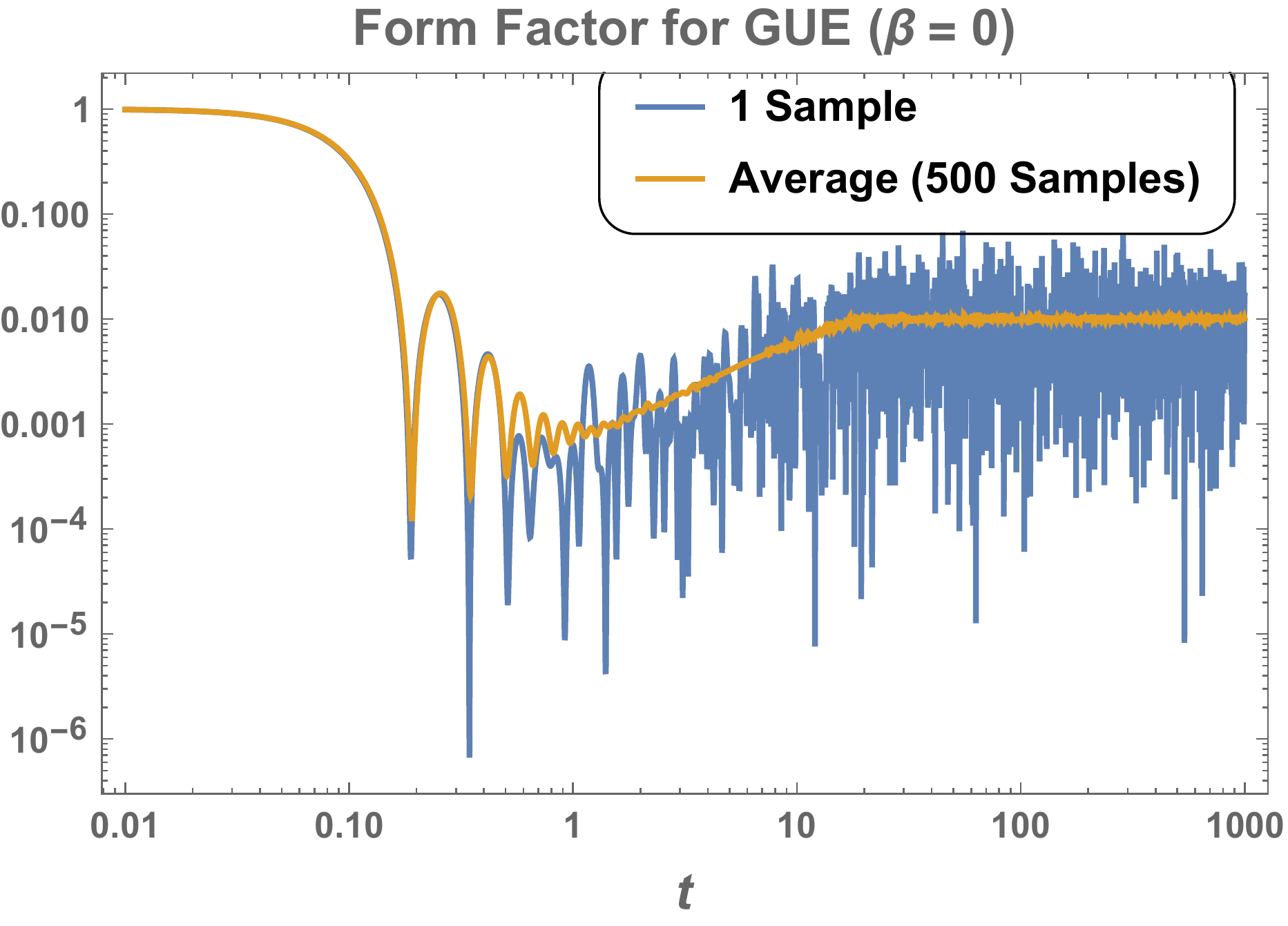}
\caption{We plot the infinite temperature form factor for a spectrum generated by $100 \times 100$ matrices pulled from the Gaussian unitary ensemble (GUE). The blue line is the form factor for a single sample matrix from the GUE and the yellow line is the 500 sample average. One can see that at late times, the averaged form factor exhibits a linear ramp followed by a plateau.  \label{ExmapleGUEIntro}}
\end{figure}

In the context of a random matrix ensemble (RME), one is interested in the normalized averaged spectral form factor (throughout this paper we shorten the name and call it the ``form factor'' unless otherwise stated) given by:    
\begin{equation}
    \frac{\braket{\mathcal{Z}(\beta+it)\mathcal{Z}(\beta-it)}_{RME}}{\braket{\mathcal{Z}(\beta)^2}_{RME}},
\end{equation}
where $\braket{\cdot}_{RME}$ is an average over a random matrix ensemble\footnote{Note that we are taking the average over the numerator and denominator separately. This particular way of averaging is called ``annelled'' disorder averaging and this is in contrast to ``quench'' disorder average which is obtained by doing an average of the entire normalized expression for the form factor. At infinite temperature annealed and quenched averages are the same.}. One can compute this numerically, by taking an average over many samples of the form factor which is constructed using the eigenvalues of sample matrices drawn from the ensemble. In the context of a random spectrum, the early time behaviour of the form factor has the property of being ``self-averaging'' which means that the form factor of a single sample is close to the average over many samples. However, at late times the form factor of a single sample will deviate significantly from the average and will no longer be self-averaging. In the case of classical Gaussian ensembles whose averaged form factor was studied in \cite{Cotler:2016fpe}, it was shown that the averaged late time behaviour after the initial decay took the form of a ramp followed by a plateau(e.g., see Figure \ref{ExmapleGUEIntro}). 
The existence of the ramp can be attributed to the nearest neighbor spacing statistics of the eigenvalues of the matrices drawn from the classical Gaussian ensembles. Furthermore, the spectrum of the SYK model (which is a toy model for near extremal black hole physics) had similar features to the spectrum of random matrices belonging to classical Gaussian ensembles\footnote{In Section \ref{WignerSurmiseSection} we will discuss how the SYK model form factor is differs in certain details with the form factor associated with matrices pulled from classical Gaussian ensembles.}. These findings led to the idea that large AdS black holes are described by theories that have a spectrum resembling the spectrum of random matrix theories.

In particular, the results reviewed in this subsection suggest that black holes are dual to quantum systems that have a discrete and chaotic energy spectrum\footnote{It should also be noted that there is an additional constraint that the system should saturate the Lyapunov bound discussed in \cite{Maldacena:2015waa}. We will address this point in the discussion of echoes in quantum chaotic systems at the end of the paper.}. In light of this, it is natural to ask if it is possible for a quantum chaotic system to exhibit echoes in its thermalization behaviour. If the answer is yes, then one might view such a system as potential candidate for describing black holes with microstructure near the horizon\footnote{Here we are specifically referring to microstructure that gives rise to echoes. It is possible that a black hole could have microstructure near its horizon but not exhibit echoes when perturbed. We will discuss this possibility in more detail at the end of the paper.}. In our exploration we will primarily be interested in how various aspects of the spectrum of a system manifest in the early and late time behaviour of the form factor and use this as a way to understand how perturbations to such systems thermalize. In the context of black hole echoes, we will be interested in finding a system with a quantum chaotic spectrum that gives rise to regular but decaying oscillations in the form factor which might be interpreted as echoes being generated by some microstructure near the horizon of a black hole in the bulk.

\subsection{Overview of the Paper}
In Section \ref{Section2RandomSpectrumandFF}, we will introduce a model for a random spectrum and derive closed form expressions for the averaged spectral form factor in terms of the nearest neighbor spacing statistics. In particular, Section \ref{Fixed NNS model}, will go over the details of the model we adopt to describe a random spectrum and also go over the basics of computing various averaged quantities in such a model. In Section \ref{SpectralFFNNSFormulaDerivation}, we use our model to derive a closed form expression for the averaged spectral form factor in terms of integrals which involve the nearest neighbor spacing (NNS) distribution. 

In Section \ref{FormFactsCommonNNSDist}, we consider various common NNS distributions and compute the associated averaged form factor using the formulas we derived in Section \ref{Section2RandomSpectrumandFF}. In Section \ref{DeltaFunctionFormFactor}, we analyze the form factor of a non-random evenly spaced spectrum and show the form factor is periodic with a period that is inversely proportional to the spacing between the energy levels. In Section \ref{PoissonSpacingSection}, we analyze the form factor associated with having a Poisson NNS distribution (which would serve as a model for the spectrum of a generic quantum integrable system). We show that the averaged form factor monotonically decreases and saturates to a non-zero value at late times. In Section \ref{WignerSurmiseSection}, we compute the averaged form factor of a system which has NNS statistics following the Wigner surmise (which serve as a model for the spectrum of a quantum chaotic system that follows the statistics of classical Gaussian ensembles). We find that the form factor has an initial dip followed by a ramp and plateau, which is consistent with expectations from the studies of classical Gaussian ensembles, as shown in Figure \ref{ExmapleGUEIntro}. In Section \ref{GammaDisSectionFF}, we introduce the gamma distribution as a simpler alternative to the Wigner surmise NNS distribution. We show that, in the appropriate regimes, the form factor associated with the gamma distribution will still have a ramp and plateau. We also show that there are other regimes in which the NNS distribution contains a large ``gap.'' We show the existence of the gap results in regular decaying oscillations in the averaged form factor after the initial dip. We argue that similar oscillations in the form factor will occur for large values of the Dyson index in the Wigner surmise distribution (Equation \ref{WigSurGenC}). These larger values of the Dyson index can be obtained from generalizations of classical Gaussian ensembles and are commonly refereed to as ``$\beta$-ensembles,'' which are know to arise from tri-diagonal random matrices \cite{Dumitriu_2002}.

In Section \ref{SystemOscilFFSection}, we study the spectrum of a many-body system composed of identical oscillators coupled to each other by a matrix belonging to a classical Gaussian ensemble. In the weak coupling regime, we study the role the chaotic interactions play in the splitting of degenerate energy states. We show that within each degenerate sector, the splitting of the energy levels will generally give rise to chaotic spectrum statistics within each sector. We verify this by numerically studying the form factor and spectral density in the weak and strongly coupled regime. We show that the averaged spectral form factor of such systems at weak coupling exhibit regular decaying oscillations at early times followed by a ramp and plateau at late times - which is consistent with random matrix theory models that describe black holes. We then propose such systems of weakly coupled oscillators as toy models for describing a membrane which gives rise to decaying echoes. The echoes repeat at approximately the fundamental frequency of the oscillators. We identify the parameters in the oscillator system with those of the bulk geometry+membrane system. Through this identification, we show that the question of whether the oscillator system is strongly or weakly coupled depends on how far the membrane is placed from the horizon. 

In Section \ref{ConclusionSec}, we conclude by summarizing the major findings and discuss future research directions.

\section{A Simple Model for Random Spectra}
\label{Section2RandomSpectrumandFF}
As stated in the introduction, quantum chaotic systems are generally conjectured to have spectrums that follow the statistics of random matrix theories (most notably Gaussian ensembles). One of the most important pieces of statistical data in the spectrum of a random matrix ensemble is the NNS statistics between eigenvalues. The NNS statistics encode non-trivial correlations between eigenvalues that give rise to quantum chaotic or integrable dynamics. In the following subsections, we will introduce a simple model for a random spectrum which allows us to specify the NNS statistics of the system we are trying to model. Such an approach is unorthodox from the perspective of random matrix theory. Usually, one specifies an ensemble of random matrices and then derives the spectral statistics of the ensemble. Here, we take a different approach and generate a random spectrum by specifying the NNS statistics. The choice of NNS statistics is then assumed to correspond to some choice of random matrix ensemble (for example, by using the Wigner surmise for the NNS distribution, our model should approximately describe the statistical behaviour of eigenvalues pulled from a Gaussian ensemble). In Section \ref{FormFactsCommonNNSDist} we will address the issue of how well these models reproduce various aspects of the spectral form factor of certain matrix ensembles.

\subsection{Random Spectrum with Fixed NNS (``i.i.d Model'')}
\label{Fixed NNS model}
We begin the construction of our random spectrum model by expressing its energy levels as follows:
\begin{equation}
\label{RandomSpecDef}
    E_n=E_{gs}+\sum_{k=1}^n\delta E_k,
\end{equation}
where $E_{gs}$ is the ground state energy, which we assume is fixed to a constant $E_{gs}=E_0$, and $\delta E_k$ are independent-identically-distributed (i.i.d) random variables which follow a probability distribution $\mathcal{P}$. By assuming that $\mathcal{P}$ is a distribution with non-zero support for $\delta E_k\in \mathbb{R}^+$ we get an ordered set of energy levels in the spectrum $E_0\leq E_1 \leq E_2 \leq \cdot\cdot\cdot \leq E_N$. Throughout the rest of this paper we will refer to this model as the \textbf{i.i.d model}.

Since we choose $\delta E_k$ to be i.i.d random variables we can express the joint probability density of the energy levels $\{E_k\}_{k=1}^N$ of the spectrum as\footnote{Note that under the change of variables $\delta E_k=E_{k}-E_{k-1}$ we have the equivalence of probability measures $dE_1\cdot\cdot\cdot dE_N =d\delta E_1\cdot\cdot\cdot d\delta E_N$. This is due to the Jacobian being a lower(or upper) triangular matrix filled with $1$'s.}:
\begin{equation}
    P(E_1,E_2,...,E_N)=\prod_{k=1}^N\mathcal{P}(E_k-E_{k-1}).
\end{equation}
Using this, we can define the average of some function, $f(E_1,..,E_N)$, associated with the random spectrum:
\begin{equation}
    \braket{f}=\int_{-\infty}^\infty dE_1\cdot\cdot\cdot dE_N f(E_1,..,E_N)P(E_1,..,E_N).
\end{equation}
For example, the spacing distribution between two eigenvalues $E_i$ and $E_j$ in terms of the joint probability distribution function is given by the following integral expression:
\begin{equation}
    p_{i,j}(s)=\int_{-\infty}^\infty dE_1\cdot\cdot\cdot dE_N P(E_1,..,E_N)\delta\left(s-|E_i-E_{j-1}|\right).
\end{equation}
Using the formula for the spectrum defined by Eq. (\ref{RandomSpecDef}), we can see that the spacing distribution between the adjacent energy levels $E_m$ and $E_{m+1}$ is\footnote{The result that the NNS distribution is equal to $\mathcal{P}$ relies on the assumption that $\mathcal{P}(x)=\Theta(x) \mathcal{P}(x)$ i.e. the probability distribution can only have non-zero support for non-negative $x$ values, this allows us to know the ordering of the variables $E_0\leq E_1\leq \cdot\cdot\cdot \leq E_N$.}:
\begin{equation}
\begin{split}
    &p_{m,m-1}(s)= \int_{-\infty}^\infty dE_1\cdot\cdot\cdot dE_N P(E_1,..,E_N)\delta\left(s-(E_m-E_{m-1})\right)\\
    &=\int_{-\infty}^\infty d\delta E_1\cdot\cdot\cdot d\delta E_N\prod_{k=1}^N \mathcal{P}(\delta E_k)\delta(s-\delta E_m)=\mathcal{P}(\delta E_m). \\
    \end{split}
\end{equation}
This proves that the NNS distribution between any nearest neighbor pair is given by $\mathcal{P}$ in the i.i.d model. 

We can also define the average spectral density by doing the following integral:
\begin{equation}
\begin{split}
\label{GeneralDOS}
    &\braket{\rho(E)}=\int_{-\infty}^\infty dE_1\cdot\cdot\cdot dE_N P(E_1,..,E_N)\sum_{m=0}^N\delta(E-E_m) \\
    &=\delta(E-E_0)+\sum_{m=1}^N\int_{-\infty}^\infty dE_1\cdot\cdot\cdot dE_N \prod_{k=1}^N\mathcal{P}(E_k-E_{k-1})\delta(E-E_m). \\
    \end{split}
\end{equation}

The averaged double spectral density is also another useful quantity defined by:
\begin{equation}
\begin{split}
\label{GenDoubleDOS}
    &\braket{\rho(E)\rho(E')}=\sum_{m=0}^N\sum_{p=0}^N\int_{-\infty}^\infty dE_1\cdot\cdot\cdot dE_N P(E_1,..,E_N)\delta(E-E_m)\delta(E'-E_p)\\
    &=\delta(E-E_0)\delta(E'-E_0)+ \delta(E-E_0)\sum_{p=1}^N\int_{-\infty}^\infty dE_1\cdot\cdot\cdot dE_NP(E_1,...,E_N)\delta(E'-E_p) \\
     &+\delta(E'-E_0)\sum_{m=1}^N\int_{-\infty}^\infty dE_1\cdot\cdot\cdot dE_NP(E_1,...,E_N)\delta(E-E_m)\\
    &+\sum_{m=1}^N\sum_{p=1}^N\int_{-\infty}^\infty dE_1\cdot\cdot\cdot dE_NP(E_1,...,E_N)\delta(E-E_m)\delta(E'-E_p).\\
    \end{split}
\end{equation}
Using this we can also define the connected double spectral density given by:
\begin{equation}
\begin{split}
    &\braket{\rho(E)\rho(E')}_{conn.}=\braket{\rho(E)\rho(E')}-\braket{\rho(E)}\braket{\rho(E')}\\
    &=\sum_{m,p=1}^N \int_{-\infty}^\infty dE_1\cdot\cdot\cdot dE_N P(E_1,..,E_N)\delta(E-E_m)\delta(E'-E_p)\\
    &-\left[\sum_{m=1}^N\int_{-\infty}^\infty dE_1\cdot\cdot\cdot dE_N P(E_1,..,E_N)\delta(E-E_m)\right]\left[\sum_{p=1}^N\int_{-\infty}^\infty dE_1\cdot\cdot\cdot dE_N P(E_1,..,E_N)\delta(E'-E_p)\right].\\
    \end{split}
\end{equation}
\subsection{Averaged Spectral Form Factor in the i.i.d Model}
\label{SpectralFFNNSFormulaDerivation}
In this subsection, we will derive a closed form expression for the annealed averaged normalized spectral form factor associated with the type of spectrum we discussed in subsection \ref{Fixed NNS model}. The normalized annealed averaged spectral form factor is given by the following expression:
\begin{equation}
\begin{split}
    &g(\beta,t)=\frac{\braket{\mathcal{Z}(\beta+it)\mathcal{Z}(\beta-it)}}{\braket{\mathcal{Z}(\beta)^2}}\\
    &\mathcal{Z}(\beta\pm it)=\sum_{n=0}^N e^{-(\beta\pm it) E_n}.\\
    \end{split}
\end{equation}
Following the derivation provided in Appendix \ref{FormFactorDer}, we will find the following expression for the averaged spectral form factor\footnote{Note, the expression for the form factor does not rely on the assumption of $\mathcal{P}$ begin zero for negative arguments. So in principle one could actually have $\mathcal{P}$ with support at negative values (you will just extend the lower limit of integration to $-\infty$ for the expressions describing the form factor) but then the interpretation of $\mathcal{P}$ being the NNS distribution is not as transparent. Although, we suspect that as long as $\mathcal{P}$ is an even function then it can still roughly be interpreted as the NNS distribution between eigenvalues in this model.}:
\begin{equation}
\begin{split}
\label{AveragedSpectralFF}
    &\braket{\mathcal{Z}(\beta+it)\mathcal{Z}(\beta-it)}=e^{-2\beta E_0}\left(1+\frac{b(1-b^N)}{1-b}+\frac{b^*(1-(b^*)^N)}{1-b^*}+B_1\right)\\
    &=e^{-2\beta E_0}\left(\frac{1-a^{N+1}}{1-a}+\frac{b}{1-b}\left[\frac{ a-b+a^{N+1}(b-1)+b^{N+1}(1-a)}{(1-a)(a-b)} \right]\right)\\
    &+e^{-2\beta E_0}\frac{b^*}{1-b^*}\left[\frac{ a-b^*+a^{N+1}(b^*-1)+(b^*)^{N+1}(1-a)}{(1-a)(a-b^*)} \right]\\
    &a=\braket{e^{-2\beta \delta E}}=\int_{0}^\infty \mathcal{P}(x)e^{-2\beta x}dx\\
    &b=\braket{e^{-(\beta+it)\delta E}}=\int_{0}^\infty \mathcal{P}(x)e^{-(\beta+it)x}dx\\
    &b^*=\braket{e^{-(\beta-it)\delta E}}=\int_{0}^\infty \mathcal{P}(x)e^{-(\beta-it)x}dx.\\
    \end{split}
\end{equation}
We can simplify the complicated expression under the assumption that $|a|<1$, $|b|<1$, and $N\gg 1$ (thermodynamic regime). In this case, at leading order, we ignore terms of the order $\mathcal{O}(a^N)$ and $\mathcal{O}(b^N)$ to obtain the following approximation for the normalized form factor:
\begin{equation}
\begin{split}
\label{ThermodynamicFF}
    &g(\beta,t)\approx \frac{1+\frac{b}{1-b}+\frac{b^*}{1-b^*}}{1+\frac{2b_0}{1-b_0}}\\
    &b=\int_{0}^\infty \mathcal{P}(x)e^{-(\beta+it)x}dx\\
    &b_0=\int_{0}^\infty\mathcal{P}(x)e^{-\beta x}.\\
    \end{split}
\end{equation}
Note that the approximation given in Eq. (\ref{ThermodynamicFF}) is only expected to be useful at finite temperatures for sufficiently large values of $N$. It is expected to fail when $\beta=0$. When $\beta=0$ we formally take the limit of Eq. (\ref{AveragedSpectralFF}) above as $a\rightarrow 1$ to get the following normalized average form factor at $\beta=0$:
\begin{equation}
    \begin{split}
    \label{FFbeta0}
        &g(\beta=0,t)=\frac{N+1+\left[\frac{b}{(1-b)^2}\left( b^{N+1}+N-b(N+1) \right)+\frac{b^*}{(1-b^*)^2}\left( (b^*)^{N+1}+N-b^*(N+1) \right)\right]}{(N+1)^2}\\
        &b=\int_{0}^\infty \mathcal{P}(x)e^{-itx} dx\\
        &b^*=\int_{0}^\infty \mathcal{P}(x)e^{itx} dx.\\
    \end{split}
\end{equation}
In cases where $b$ and $b^*$ go to zero as $t\rightarrow \infty$ it is easy to see that\footnote{In many of the examples we consider the assumption that $b$ and $b^*$ vanish in the large $t$ limit will be valid. However, there are examples where the assumption fails. In Sec. \ref{DeltaFunctionFormFactor} we have $\mathcal{P}(x)$ equal to a delta function, which results in an oscillating form factor that does not settle any particular value in the large $t$ limit.}:
\begin{equation}
    \lim_{t\rightarrow\infty }g(\beta=0,t)=\frac{1}{N+1}.
\end{equation}
At infinite temperature we identify the entropy as $S=\ln(N+1)$ and conclude that after a sufficiently long time, the infinite temperature averaged form factor will plateau toward a value $e^{-S}$. In a similar manner, we can approximate the expression for the large time value of the form factor (again we assume $b$ and $b^*$ vanish in the $t\to \infty$ limit) at finite temperature for large values of $N$ using the approximation given in Eq. (\ref{ThermodynamicFF}):
\begin{equation}
    \lim_{t, N\to \infty}g(\beta,t)\approx \frac{1-b_0}{1+b_0}=\frac{1-\int_{0}^\infty\mathcal{P}(x)e^{-\beta x}dx}{1+\int_{0}^\infty\mathcal{P}(x)e^{-\beta x}dx}.
\end{equation}

\section{Spectral Form Factor of Common NNS Distributions}
\label{FormFactsCommonNNSDist}
Thus far, we have proposed the i.i.d model for a random spectrum which involves specifying the NNS statistics of the system we are interested in modelling. In the following sections we will use the i.i.d model to compute the averaged spectral form factor associated with various choices of NNS statistics. In particular, we will investigate how well the i.i.d model (with a Wigner surmise NNS distribution) reproduces the features of the spectral form factor for Gaussian ensembles in Section \ref{WignerSurmiseSection}. Aside from the more common NNS statistics that are found in integrable and quantum chaotic systems, we will also study generalizations of these statistics in Section \ref{GammaDisSectionFF} and explore how the behaviour of the form factor changes.

\subsection{Delta Function (Non-Random Evenly Spaced Spectrum)}
\label{DeltaFunctionFormFactor}
The most trivial spectrum that one can generate using the i.i.d model is an evenly spaced spectrum (with no degeneracy) where the NNS distribution is given by the Dirac delta distribution below:
\begin{equation}
    \mathcal{P}(x=\delta E)=\delta(x-\Delta E),
\end{equation}
where $\Delta E\geq 0$, in this case the integrals in Eq. (\ref{AveragedSpectralFF}) are simply given by\footnote{It is clear from this that $b$ and $b^*$ do not go to zero in the large $t$ limit this is an example in which the form factor does not plateau towards a fixed value.}:
\begin{equation}
    \begin{split}
        &a=e^{-2\beta\Delta E}\\
        &b=e^{-\Delta E(\beta+it)}\\
        &b^*=e^{-\Delta E(\beta-it)}.\\
    \end{split}
\end{equation}
The normalized averaged spectral form factor equals:
\begin{equation}
\begin{split}
   &g(\beta,t)=\mathcal{N}\left[\frac{e^{2\beta\Delta E}+e^{-2N\beta\Delta E}-2e^{-\beta(N-1)\Delta E}\cos((N+1)\Delta E t)}{1+e^{2\beta\Delta E}-2e^{\beta\Delta E}\cos(\Delta E t)}\right]\\
   &\mathcal{N}=\frac{1+e^{2\beta\Delta E}-2e^{\beta\Delta E}}{e^{2\beta\Delta E}+e^{-2N\beta\Delta E}-2e^{-\beta(N-1)\Delta E}}.\\
   \end{split}
\end{equation}
In this case we can see from the expression above that the form factor is periodic with a period:
\begin{equation}
    \tau=\frac{2 \pi}{\Delta E}.
\end{equation}

\subsection{Poisson Distribution (Quantum Integrable Systems)}
\label{PoissonSpacingSection}
For sufficiently large diagonal matrices whose diagonal elements are i.i.d random variables, it is known that the NNS distribution of eigenvalues follows a Poisson distribution \cite{abulmagd2011size,livan2017introduction}. Furthermore, the Poisson distribution also appears when studying the energy spacing statistics of a wide variety of integrable systems \cite{Berry1977LevelCI,Guhr:1997ve,Atas_2013}. This gives us reason to consider the spectral form factor of a spectrum generated by the i.i.d model whose NNS distribution is given by the Poisson distribution\footnote{Usually one defines a dimensionless spacing $s=\delta E/\sigma$ and discusses NNS in terms of $s$ rather than $\delta E$, but here we will work directly with $\delta E$ and $\sigma$.}:
\begin{equation}
\label{PoissonNNSDist}
    \mathcal{P}(x=\delta E)=\Theta(x)\frac{e^{-x/ \sigma}}{\sigma}.
\end{equation}
The average spacing between adjacent eigenvalues is given by $\braket{\delta E}=\sigma$. 

\begin{figure}[H]
\centering
\includegraphics[width=120mm]{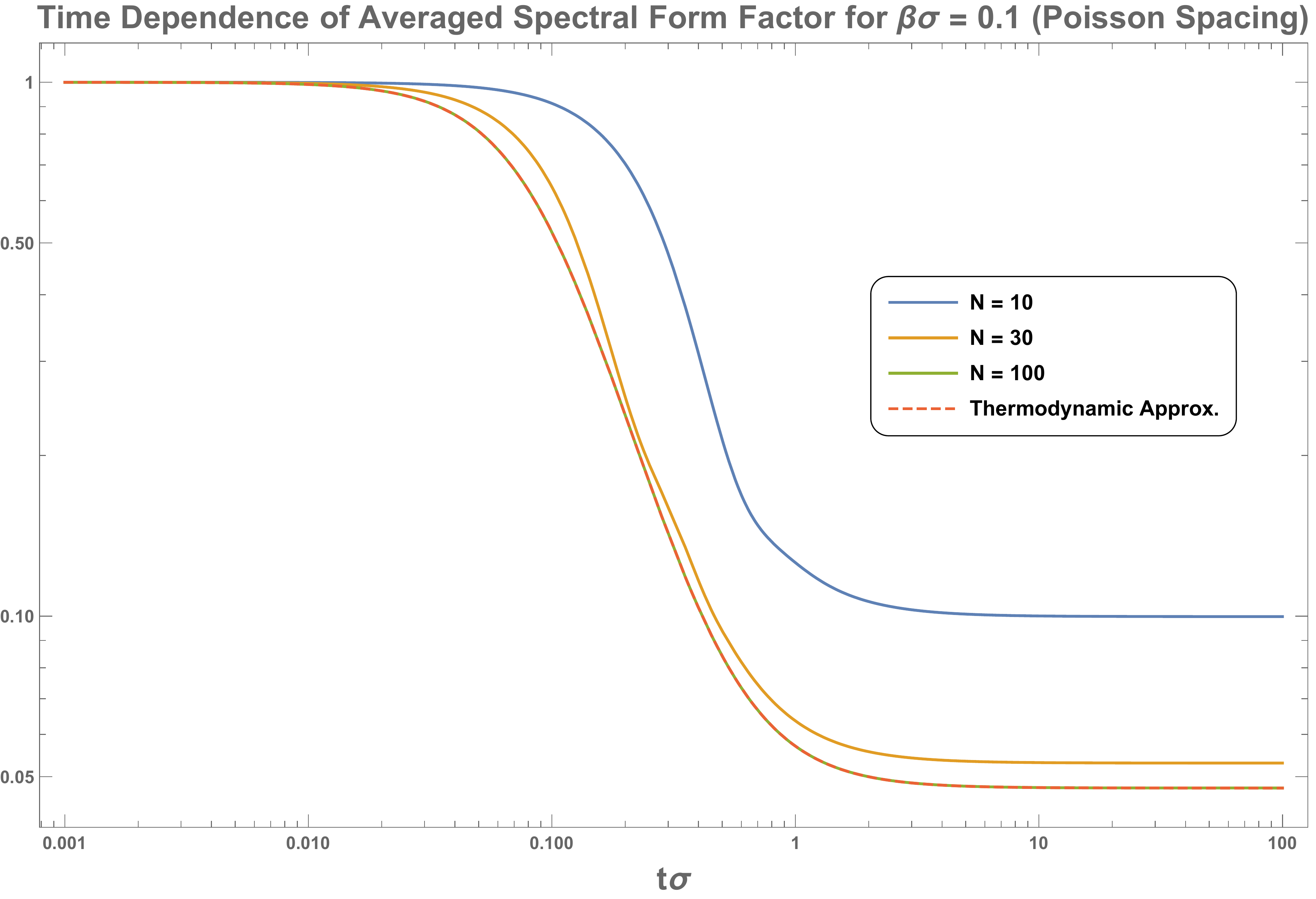}
\caption{Above is a Log-Log plot of the averaged spectral form factor of a spectrum generated by the i.i.d model with Poisson NNS distribution given by Eq. (\ref{PoissonNNSDist}). The temperature is fixed at $\beta\sigma=0.1$ and we vary $N$. The plot illustrates how the form factors for larger values of $N$ converge towards the thermodynamic approximation given in Eq. (\ref{ThermodyPoiFormFac}).\label{ThrmPoisFFPlot}}
\end{figure}

\begin{figure}[]
\centering
\includegraphics[width=120mm]{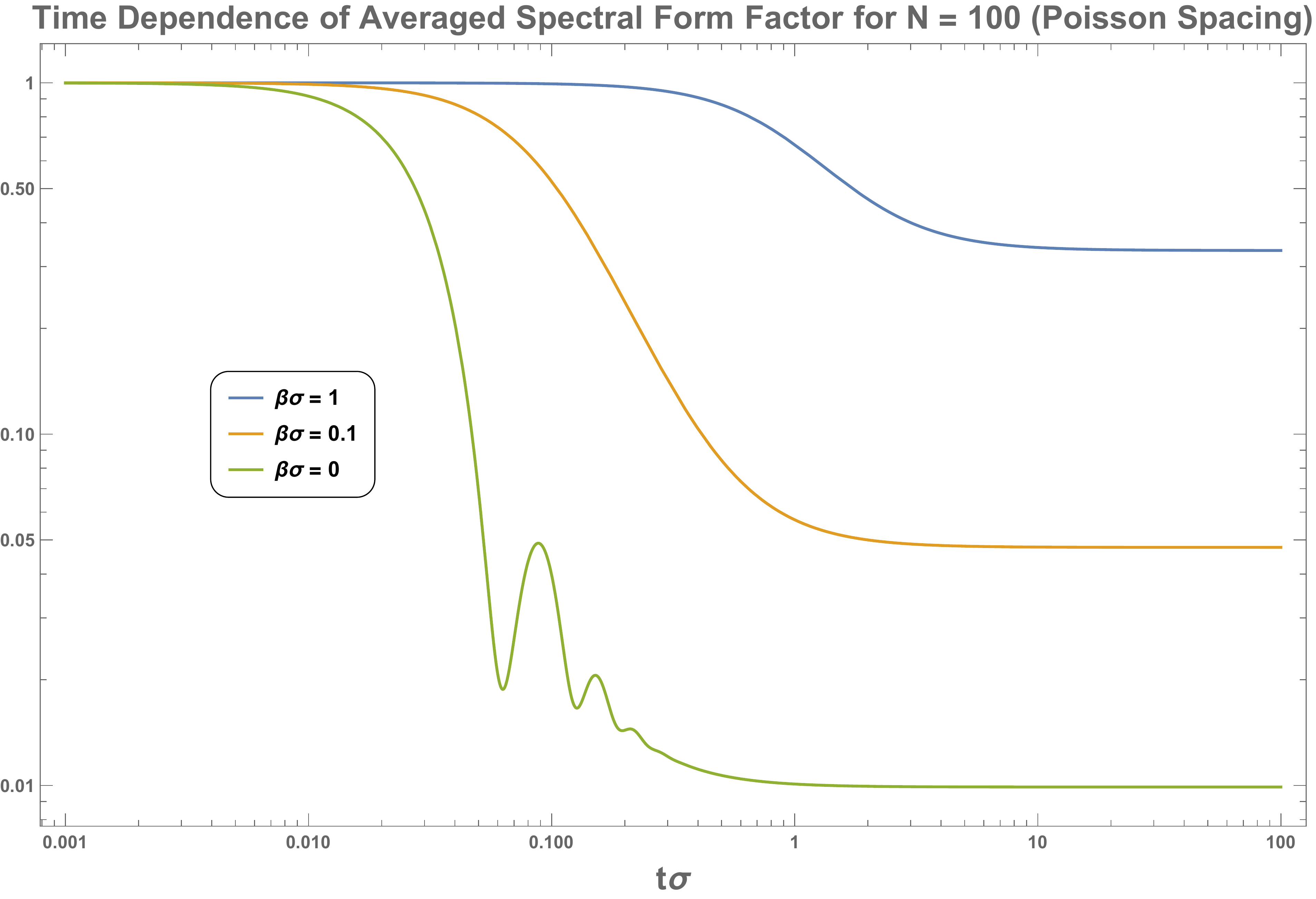}
\caption{Above is a Log-Log plot of the averaged spectral form factor of a spectrum generated by the i.i.d model with Poisson NNS distribution given by Eq. (\ref{PoissonNNSDist}). The value of $N$ is fixed at $N=100$ and the form factor is plotted for different temperatures ranging from $\beta\sigma=0$ to $\beta\sigma=1$. \label{PoissonFFDiffTempPlot}}
\end{figure}

We compute the integral expressions for $a,b,b^*$, they are given by:   
\begin{equation}
    \begin{split}
        &a=\frac{1}{1+2\beta\sigma}\\
        &b=\frac{1}{1+\sigma\left(\beta+it\right)}\\
        &b^*=\frac{1}{1+\sigma\left(\beta-it\right)}.\\
    \end{split}
\end{equation}
Notice that, in this case $b$ and $b^*$ do go to zero in the limit as $t\to \infty$, this means that the normalized average spectral form factor at infinite temperature goes to $(N+1)^{-1}$. At infinite temperature the form factor is given as:
\begin{equation}
\begin{split}
    &g(\beta=0,t)=\frac{1}{N+1}+\frac{2-\left(1+ix\right)^{-N}-\left(1-ix\right)^{-N}}{(1+N)^2x^2}\\
    &x=t\sigma.\\
    \end{split}
\end{equation}
For finite temperature, the expression for the form factor is not as simple - so we will not explicitly write it here. However, because $|b|<1$ and $|a|<1$ we can use the approximated form factor in the large $N$ regime given by Eq. (\ref{ThermodynamicFF}) to give a simple approximation for the form factor:
\begin{equation}
\label{ThermodyPoiFormFac}
    g(\beta,t)\approx \frac{1+\frac{2\beta\sigma}{\sigma^2(\beta^2+t^2)}}{1+\frac{2}{\beta\sigma}}.
\end{equation}
We compare the approximated expression for the form factor given in Eq. (\ref{ThermodyPoiFormFac}) to the full expression we have in Eq. (\ref{AveragedSpectralFF}) in Figure \ref{ThrmPoisFFPlot}.
We can see that for sufficiently large $N$ the approximated expression converges toward the exact expression for the form factor. At lower temperatures, convergence will occur at smaller values of $N$.  
We also consider the averaged spectral form factor at various temperatures for a fixed value of $N=100$ in Figure \ref{PoissonFFDiffTempPlot} \footnote{One can check that similar looking plots can be made for higher and lower values of $N$.}. 
The expression for the asymptotic value of the form factor is given by the following expression:
\begin{equation}
\begin{split}
\label{AsympValPoi}
    &\lim_{t\to\infty}g(\beta,t)=\lim_{t\to\infty}\frac{\braket{\mathcal{Z}(\beta+it) \mathcal{Z}(\beta-it)}}{\braket{\mathcal{Z}(\beta)^2}}\\
    &=\frac{\tilde{\beta}(1+\tilde{\beta})^N\left(\left(1+2\tilde{\beta}\right)^{N+1}-1\right)}{\left(1+\tilde{\beta}\right)^N\left(2+3\tilde{\beta}\right)+\left(1+2\tilde{\beta}\right)^{N+1}\left[\left(1+\tilde{\beta}\right)^N\left(2+\tilde{\beta}\right)-4\right]}\\
    &\tilde{\beta}=\beta\sigma.\\
\end{split}
\end{equation}
We can see that the form factor decays from its initial normalized value towards a non-zero value given by Eq. (\ref{AsympValPoi}). Another interesting point to make based on the plots we made is that the plateau phase occurs at a time scale $t\gtrsim \sigma^{-1}=\braket{\delta E}^{-1}$, where $\braket{\delta E}$ is the average spacing between eigenvalues.    

At infinite temperature, we see oscillations, these are similar to the oscillations in the form factor discussed in \cite{Cotler:2016fpe} they arise at high temperatures due to the spectral edges of the spectral density\footnote{One might wonder if these oscillations can be interpreted as ``echoes.'' The answer to this is no because echoes should persist at lower temperatures as well.}. 

\begin{figure}[H]
\centering
\includegraphics[width=120mm]{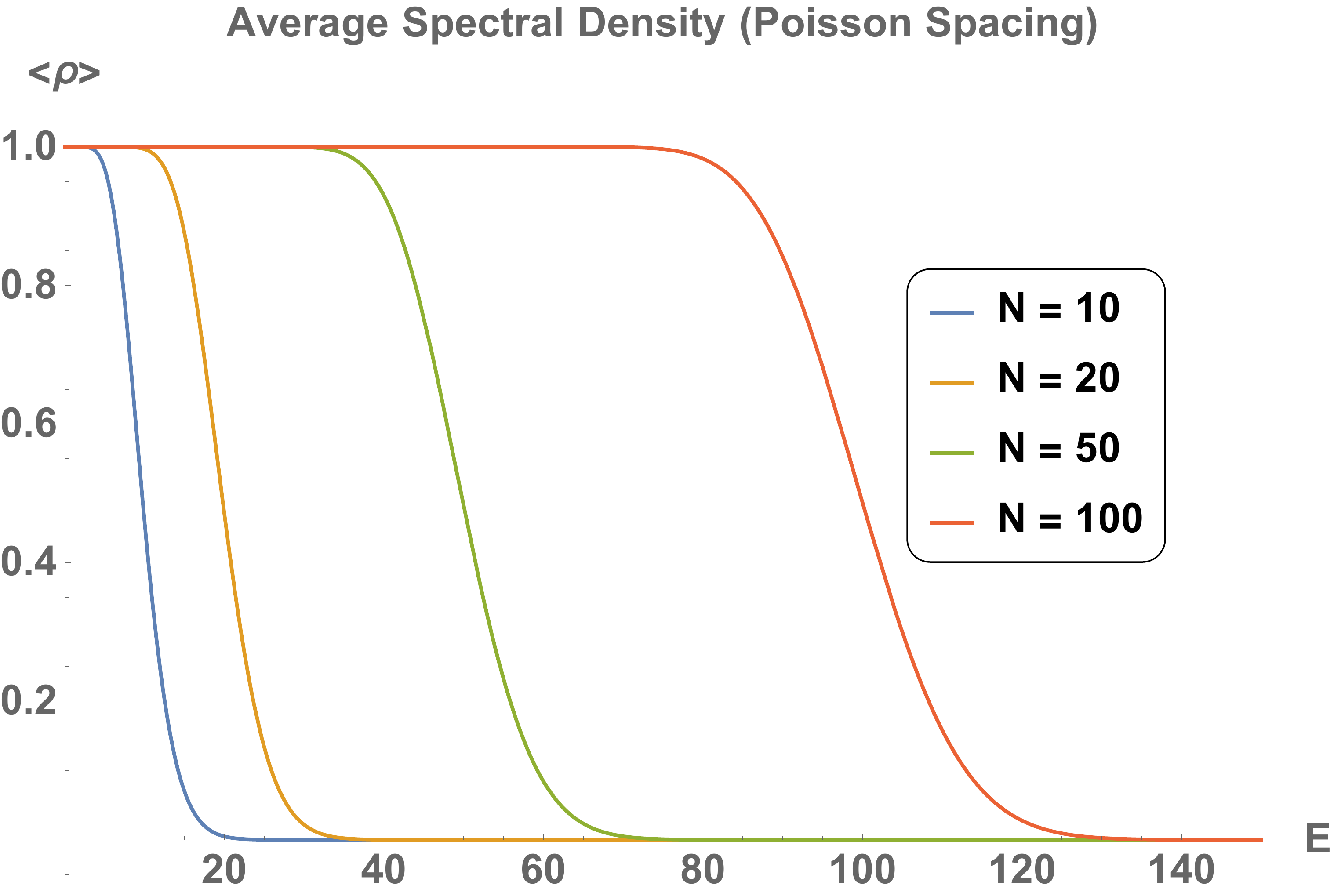}
\caption{Above we plot the average spectral density of a random spectrum generated from the i.i.d. model using the Poisson NNS distribution defined in Eq. (\ref{PoissonNNSDist}) (setting $\sigma=1$ and $E_0=0$ and changing $N$). The averaged spectral density is given by Eq. (\ref{PoissonSpecDensity}). The width roughly scales with $N$. \label{Figg4}}
\end{figure}

It is possible to explicitly calculate the averaged spectral density in the i.i.d model with Poisson NNS statistics (details are given in Appendix \ref{SpectralDensityPoissonAppendix}) it is given by the following expression: 
\begin{equation}
\label{PoissonSpecDensity}
    \braket{\rho(E)}=\delta(E-E_0)+\sum_{m=1}^N\left[\frac{(E-E_0)^{m-1}}{\sigma^{m-1}(m-1)!}\right]\Theta(E-E_0)\frac{e^{-(E-E_0)/\sigma}}{\sigma}.
\end{equation}
We can rewrite the sum in terms of the incomplete gamma function as follows:
\begin{equation}
\begin{split}
    &\braket{\rho(E)}=\delta(E-E_0)+\frac{\Theta(E-E_0)}{\sigma}\frac{\Gamma\left(N,\frac{E-E_0}{\sigma}\right)}{\Gamma\left(N,0\right)}\\
    &\Gamma(N,x)=\int_{x}^\infty t^{N-1}e^{-t}dt.\\
    \end{split}
\end{equation}
By taking the derivatives of incomplete gamma function we can show that the function is monotonically decreasing and has an inflection point at $E-E_0=(N-1)\sigma$. We expect the function to very slowly decrease up until one gets close to the inflection point. Then after the inflection point we expect the spectral density to be very small. In Figure \ref{Figg4} we plot the spectral density and verify this.
Ignoring the delta function at $E=0$, we can see the spectral density is approximately constant then quickly goes to zero near the inflection point. We can also see that as $N$ increases the nearly constant phase is extended. The nearly constant phase can be regarded as a natural consequence of the i.i.d spacing assumption in our model. We will generally expect that spectral densities generated by the i.i.d model will have an approximately constant portion far from the edges of the spectrum. We can also see that width of the density scales with $N$.

\subsection{Wigner Surmise (Quantum Chaotic Systems)}
\label{WignerSurmiseSection}
\begin{figure}[H]
\centering
\includegraphics[width=120mm]{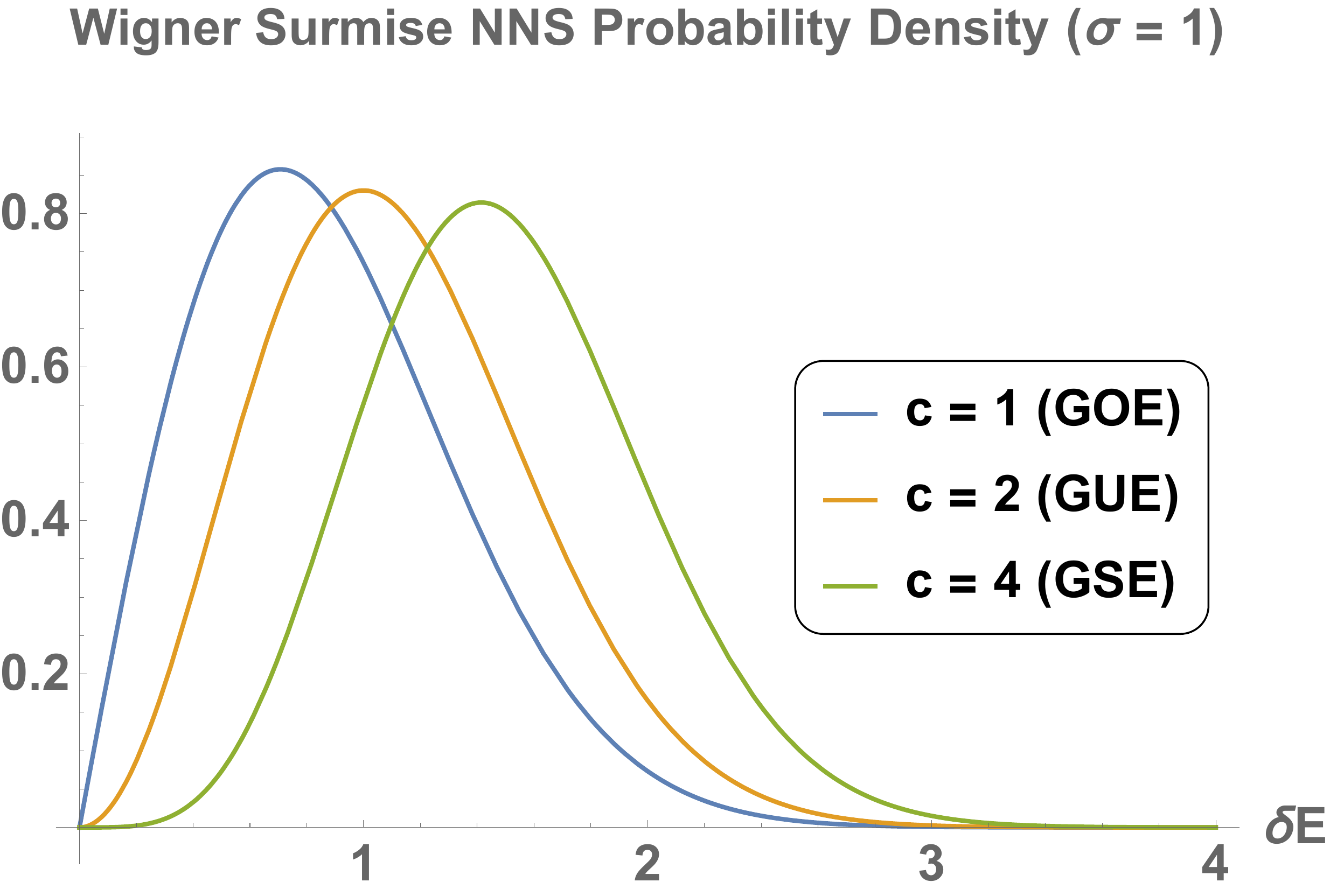}
\caption{The Wigner surmise NNS distribution for three ensembles. The Gaussian orthogonal ensemble (GOE) occurs when $c=1$. The Gaussian unitary ensemble (GUE) occurs when $c=2$. The Gaussian symplectic ensemble (GSE) occurs when $c=4$.\label{figg5}}
\end{figure}

In this section we will compute the spectral form factor in the i.i.d model when the NNS distribution is given by the Wigner surmise. The Wigner surmise is known to be a good approximation to the NNS distribution for classical Gaussian ensembles \cite{Guhr:1997ve,erdos2012universality,livan2017introduction}.   
The Wigner surmise NNS distribution is defined below\footnote{The value of $\sigma$ fixes the average spacing between eigenvalues and is usually chosen so that the spacing is unity. Just as in the Poisson spacing case we will not make assumptions on the value of $\sigma$ and simply keep $\sigma$ in our expressions.}:
\begin{equation}
\label{WigSurGenC}
    \mathcal{P}_c(x=\delta E)=\Theta(x)\frac{2x^c}{\Gamma\left(\frac{1+c}{2}\right)}\frac{e^{-x^2/\sigma^2}}{\sigma^{c+1}}.
\end{equation}
The value of $c$ (sometimes called the Dyson index) depends on the type of Gaussian ensemble one wants to consider. In particular, the three main classical Gaussian ensembles are orthogonal ($c=1$), unitary ($c=2$), and symplectic ($c=4$). In Figure \ref{figg5} we plot the distributions for the three ensembles for $\sigma=1$. 

For the sake of simplicity we will mainly focus on the Gaussian unitary ensemble, $c=2$\footnote{One can also easily do similar calculations for the other ensembles and get similar results.}. In this case we will have the following NNS distribution:
\begin{equation}
\label{GUEWigSur}
    \mathcal{P}_2(x=\delta E)=\Theta(x)\frac{4x^2}{\sigma^3\sqrt{\pi}}e^{-x^2/\sigma^2}.
\end{equation}
The average spacing between adjacent eigenvalues is given by $\braket{\delta E}=\frac{2\sigma}{\sqrt{\pi}}\sim 1.1\sigma$.

The expressions for $a,b$, and $b^*$ in this case are given by:
\begin{equation}
    \begin{split}
    \label{abbstarWigSur}
        &a=-\frac{2\beta\sigma}{\sqrt{\pi}}+e^{\beta^2\sigma^2}(1+2\beta^2\sigma^2)\left(1-{\rm erf}(\beta\sigma)\right)\\
    &b=-\frac{1}{2\sqrt{\pi}}\left[ 2\sigma(\beta+it)+\sqrt{\pi}e^{-\frac{-\sigma^2(t-i\beta)^2}{4}}\left(-2+(t-i\beta)^2\sigma^2\right)\left(1-{\rm erf}\left(\frac{1}{2}(\beta+it)\sigma\right)\right) \right]\\
    &b^*=-\frac{1}{2\sqrt{\pi}}\left[ 2\sigma(\beta-it)+\sqrt{\pi}e^{-\frac{-\sigma^2(t+i\beta)^2}{4}}\left(-2+(t+i\beta)^2\sigma^2\right)\left(1-{\rm erf}\left(\frac{1}{2}(\beta-it)\sigma\right)\right) \right].\\
    \end{split}
\end{equation}
One can easily check that for $\beta>0$ we have $|a|<1$ and $|b|<1$. This means that the thermodynamic approximation given by Eq. (\ref{ThermodynamicFF}) will be valid for sufficiently large $N$ at finite temperatures. Also, we can check that $\lim_{t\to\infty}b=0$, which tells us that the asymptotic value of the form factor at infinite temperature is $(N+1)^{-1}$. We can plot the form factor to get a sense of the general features that appear for various choices of parameters. 

\begin{figure}[]
\centering
\includegraphics[width=120mm]{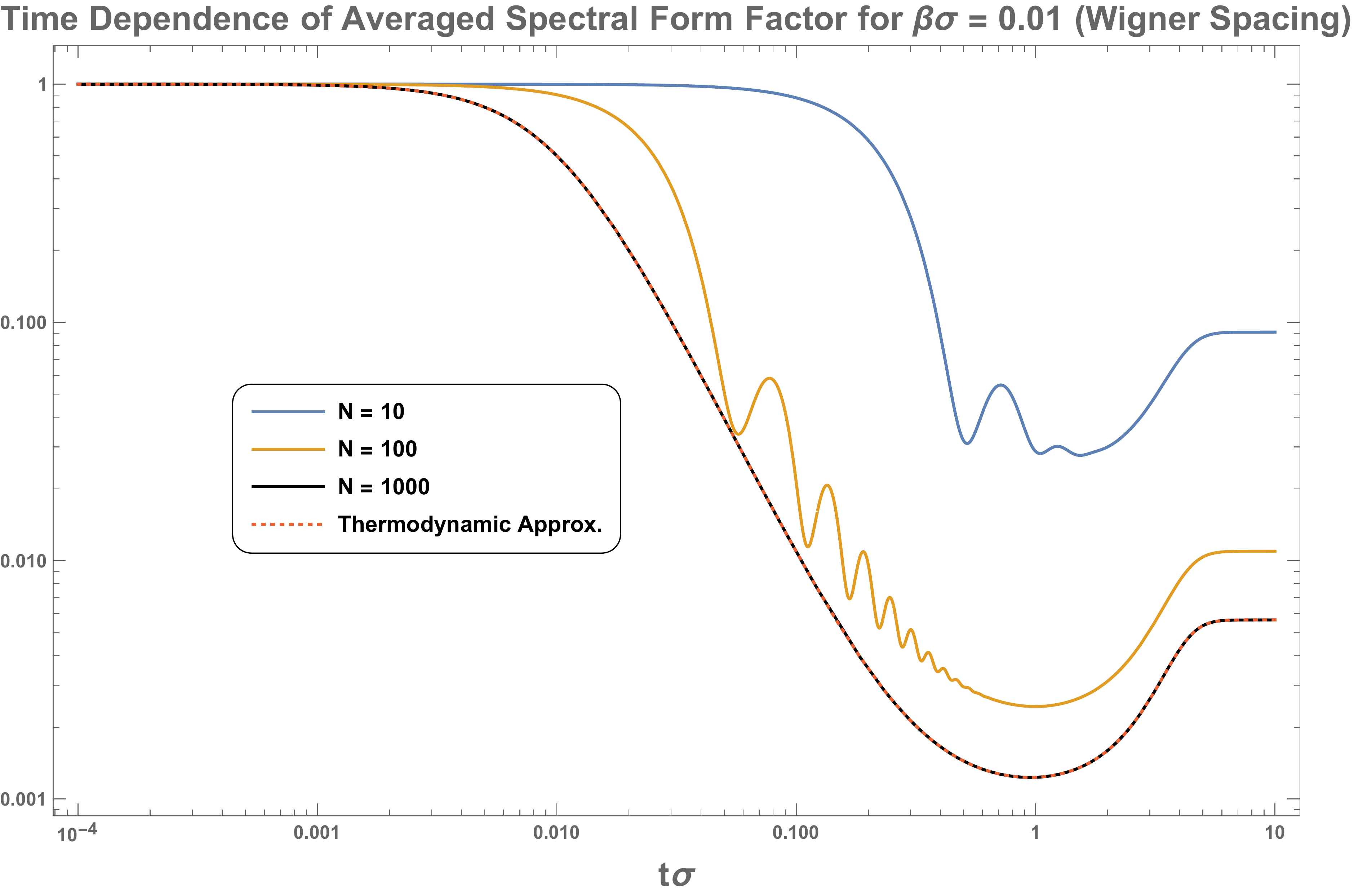}
\caption{Above is a Log-Log plot of the averaged spectral form factor of a spectrum generated by the i.i.d model with Wigner surmise NNS distribution for the GUE given by Eq. (\ref{GUEWigSur}). The temperature is fixed at $\beta\sigma=0.01$ and we vary $N$.\label{LLPlotThermWSFF}}
\end{figure}

In Figure \ref{LLPlotThermWSFF}, we compare the thermodynamic approximation given by Eq. (\ref{ThermodynamicFF}) with the exact result given by Eq. (\ref{AveragedSpectralFF}).
As expected, for sufficiently large $N$ the thermodynamic approximation converges to the exact result, and one can further verify that for lower temperatures the convergence will occur more quickly at lower values of $N$. 

\begin{figure}[]
\centering
\includegraphics[width=120mm]{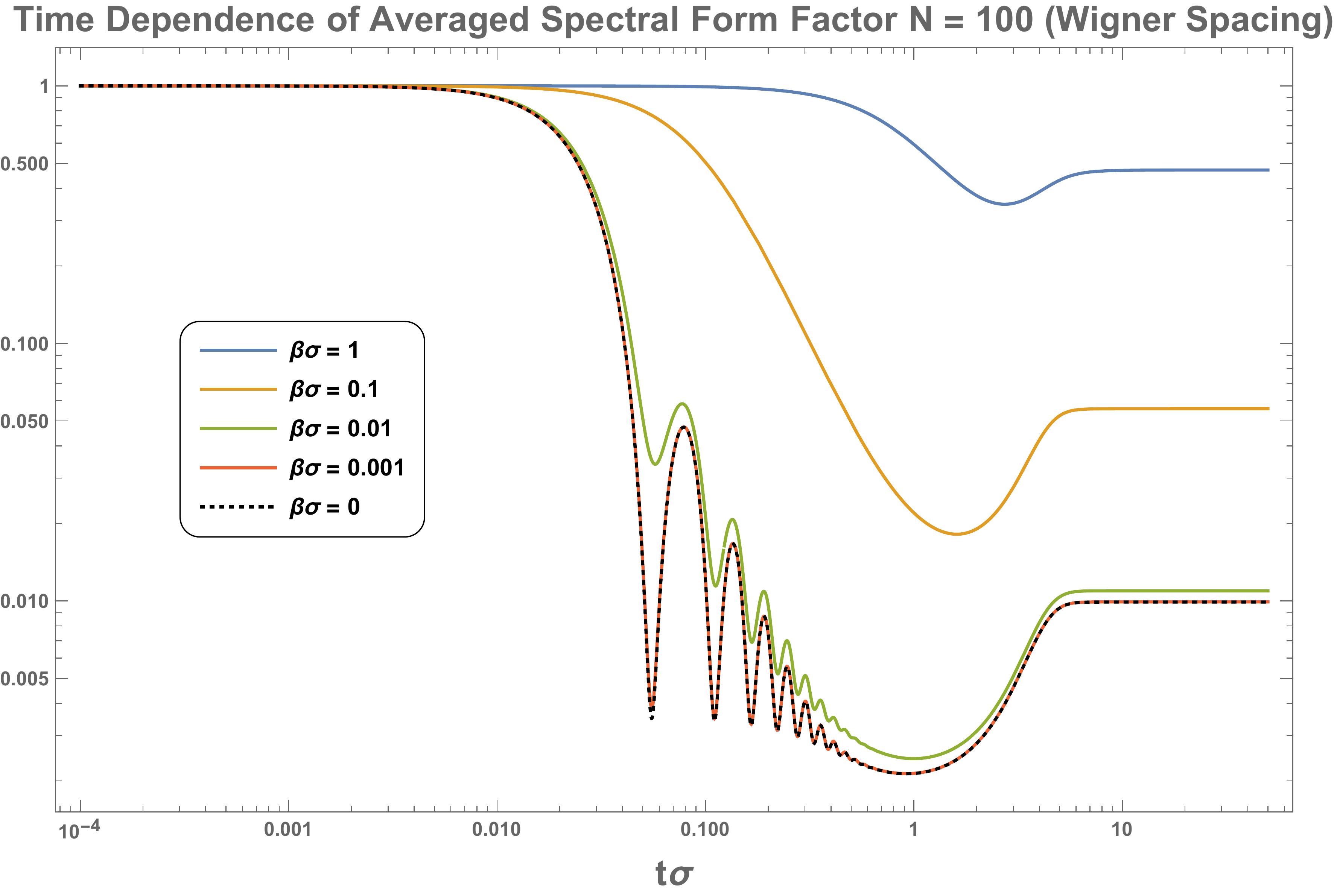}
\caption{Above is a Log-Log plot of the averaged spectral form factor of a spectrum generated by the i.i.d model with Wigner surmise NNS distribution for the GUE given by Eq. (\ref{GUEWigSur}). We fixed the value of $N=100$ and plot the averaged form factor for different temperatures.\label{WigSpVarybetaPlot}}
\end{figure}

In Figure \ref{WigSpVarybetaPlot}, we plot the exact form factor for fixed $N$ at different temperatures.
We can see that the form factor initially decreases and eventually reaches a minimum value\footnote{At high temperatures we see oscillations similar to what we saw for Poisson spacing, this is again due to the spectral edges being probed at high temperatures (not to be interpreted as ``echoes'').}. After reaching the minimum, in a time scale roughly given by $t \sim \braket{\delta E}^{-1}$, the form factor increases along a ``ramp'' and finally plateaus toward an asymptotic value that is given by taking the $t\to \infty$ limit. 

The ramp and plateau also appears in numerical calculations of the averaged form factor of the SYK model discussed in \cite{Cotler:2016fpe}. This occurs since the spectrum spacing statistics of the SYK model can be understood in terms of the spacing statistics of random matrices pulled from Gaussian ensembles. Although it is true that the form factors of the SYK and the classical Gaussian ensembles have many features in common, it is important to note that there are differences. One important distinction to make is the time scale over which the ``ramp'' begins to manifest in the form factor. In the case of the form factor generated by random matrices pulled from the GUE (and also our i.i.d. model see Appendix \ref{CompModelandGUEFF}) the beginning of the ramp occurs on time scales comparable to the Heisenberg time, which is $\mathcal{O}(\braket{\delta E}^{-1})$. This is in stark contrast to the SYK model where the ramp phase begins on time scales which are much shorter than the Heisenberg time\footnote{ \cite{Altland:2021rqn}  estimates the ramp in the SYK form factor to start at times scales of the order $\mathcal{O}(N^{1/2}\ln(N))$. This is much shorter than the Heisenberg time for the SYK model which is of the order $\mathcal{O}(e^N)$ \cite{Altland:2020ccq}. For more physical quantum chaotic theories, the time scale after which the universal ramp and plateau manifest , the so-called ``Thouless time'', is  {\it not} universal  (i.e. it changes depending on the model and the observable. We thank Julian Sonner for pointing this out).}. 

In Appendix \ref{CompModelandGUEFF}, we do a numerical study which compares the form factor generated by eigenvalues from a Hermitian random matrix to the form factor of a random spectrum generated by the i.i.d model with Wigner surmise NNS distribution. We show that if one focuses on the eigenvalues near the centre of the spectrum then our model is in reasonable agreement with actual numerical computations of the form factor associated with the truncated spectrum near the centre of the spectral density. 

\begin{figure}[]
\centering
\includegraphics[width=120mm]{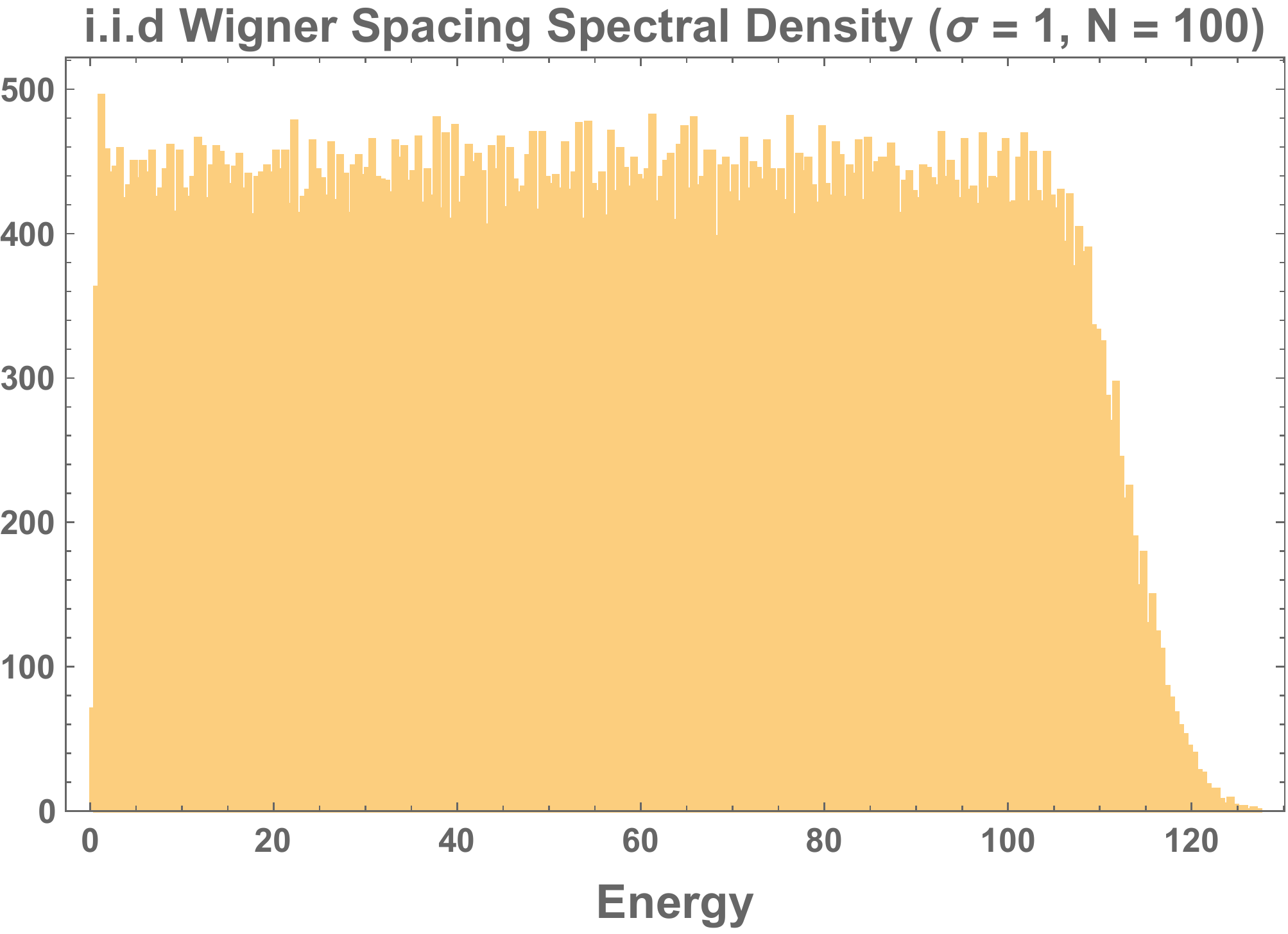}
\caption{Above is a histogram of eigenvalues generated by the i.i.d model for 100 eigenvalues (excluding the ground state at $E=0$) defined in Eq. (\ref{RandomSpecDef}). The NNS distribution is given by Eq. (\ref{GUEWigSur}) with $\sigma=1$. We expect the width of the spectral density to scale with $N$. \label{NumericWigDen}}
\end{figure}

It is not possible to get a closed form expression for the spectral density when we have Wigner surmise spacings. However, we can numerically compute the averaged spectral density by generating a large number of eigenvalues from the i.i.d model for Wigner surmise spacing. As an example, in Figure \ref{NumericWigDen}, we generate a histogram from the eigenvalues generated by the i.i.d model for $10^4$ samples with each sample having 100 eigenvalues (for simplicity we set $\sigma = 1$ and $E_0=0$). 
Unsurprisingly, we find that the density away from the edges is approximately constant and similar to the Poisson case. However, it will differ from the Poisson density near the ground state where it will be zero due to the repulsion between the ground state and the rest of the excited states (Poisson case exhibits ``attraction toward the ground state''). Another important point to make is that the averaged spectral density of the spectrum generated by the i.i.d model is not the semi-circle as it is for classical Guassian ensembles (e.g., see Figure \ref{GUEHistogramPlot}). This is not surprising since our model only captures correlations between nearest neighbors and not the longer range correlations that conspire to give the semi-circle. Nonetheless, the i.i.d model still captures the general features of the ramp and plateau in the averaged form factor.  

In the next subsection, we will introduce a new NNS distribution which retains the interesting features of the Wigner surmise (i.e. repulsion in NNS statistics) but will be easier to analytically handle in our model when computing the averaged spectral density in the i.i.d model. We will also discuss generalizations of classical Gaussian ensembles and how the late time behaviour of the form factor of such ensembles differ from the usual ramp and plateau behaviour of the form factor of classical Gaussian ensembles.

\subsection{Form Factor of Gamma NNS Distribution and $\beta$-Ensembles}
\label{GammaDisSectionFF}
In the previous subsection, we looked at the Wigner surmise as a canonical example of what the NNS distribution of eigenvalues of a chaotic system looks like. Although we are able to compute the form factor, many other quantities of interest such as the spectral density are difficult to compute in closed form. This is primarily due to the $e^{-x^2}$ in the NNS density makes the integrals in Eq. (\ref{GeneralDOS}) difficult to compute. To facilitate more explicit calculations - while still retaining the essential repulsion between eigenvalues - we will consider a slightly different NNS distribution. We define the gamma distribution for NNS:
\begin{equation}
\label{GammaDist}
    \mathcal{P}_q(x=\delta E)=\Theta(x)\frac{x^qe^{-x/\sigma}}{\Gamma(1+q)\sigma^{1+q}}.
\end{equation}
The value of $q$ (plays the role of the Dyson index in the Wigner surmise) fixes the degree of repulsion between eigenvalues. For a fixed value of $q$, $\sigma$ fixes the average spacing between nearest neighbor pairs to $\braket{\delta E}=(1+q)\sigma$.

At leading order, it is clear that such a distribution will contain the same repulsion behaviour near zero spacing as the Wigner surmise with $c$ in Eq. (\ref{WigSurGenC}) identified with $q$ in Eq. (\ref{GammaDist}). The major difference being the tail; the Wigner surmise has a Gaussian tail whereas the gamma distribution has an exponentially decaying tail. The advantage to using this is that it still contains the important repulsion of a chaotic system for any $q>0$. Furthermore, due to the exponential tail of the gamma function the spectral density can be computed exactly.  

We begin by computing the necessary integrals that define the averaged form factor in the i.i.d model:
\begin{equation}
    \begin{split}
        &a=(1+2\beta\sigma)^{-(1+q)}\\
        &b=\left(1+\sigma(\beta+it)\right)^{-(1+q)}\\
        &b^*=\left(1+\sigma(\beta-it)\right)^{-(1+q)}.\\
    \end{split}
\end{equation}
We can clearly see that $|a|,|b|<1$ at finite temperature for $q\geq 0$, therefore in the large $N$ regime we can use the thermodynamic expression of the form factor given in Eq. (\ref{ThermodynamicFF}). We find:
\begin{equation}
\begin{split}
\label{ThermoDyFFGammaDis}
    &g(t,\beta)\approx \frac{\left(1+\beta\sigma\right)^{q+1}-1}{\left(1+\beta\sigma\right)^{q+1}+1}\left[ 1+\frac{1}{\left[1+\sigma(\beta+it)\right]^{q+1}-1}+\frac{1}{(\left[1+\sigma(\beta-it)\right]^{q+1}-1} \right].\\
    \end{split}
\end{equation}
The plateau at sufficiently large $N$ and finite temperature regime has an approximate height equal to:
\begin{equation}
    \lim_{t \to \infty}g(t,\beta)\approx \frac{\left(1+\beta\sigma\right)^{q+1}-1}{\left(1+\beta\sigma\right)^{q+1}+1}.
\end{equation}

We plot the averaged form factor in the i.i.d model using Eq. (\ref{AveragedSpectralFF}) for various choices of parameters. 
\begin{figure}[H]
\centering
\includegraphics[width=120mm]{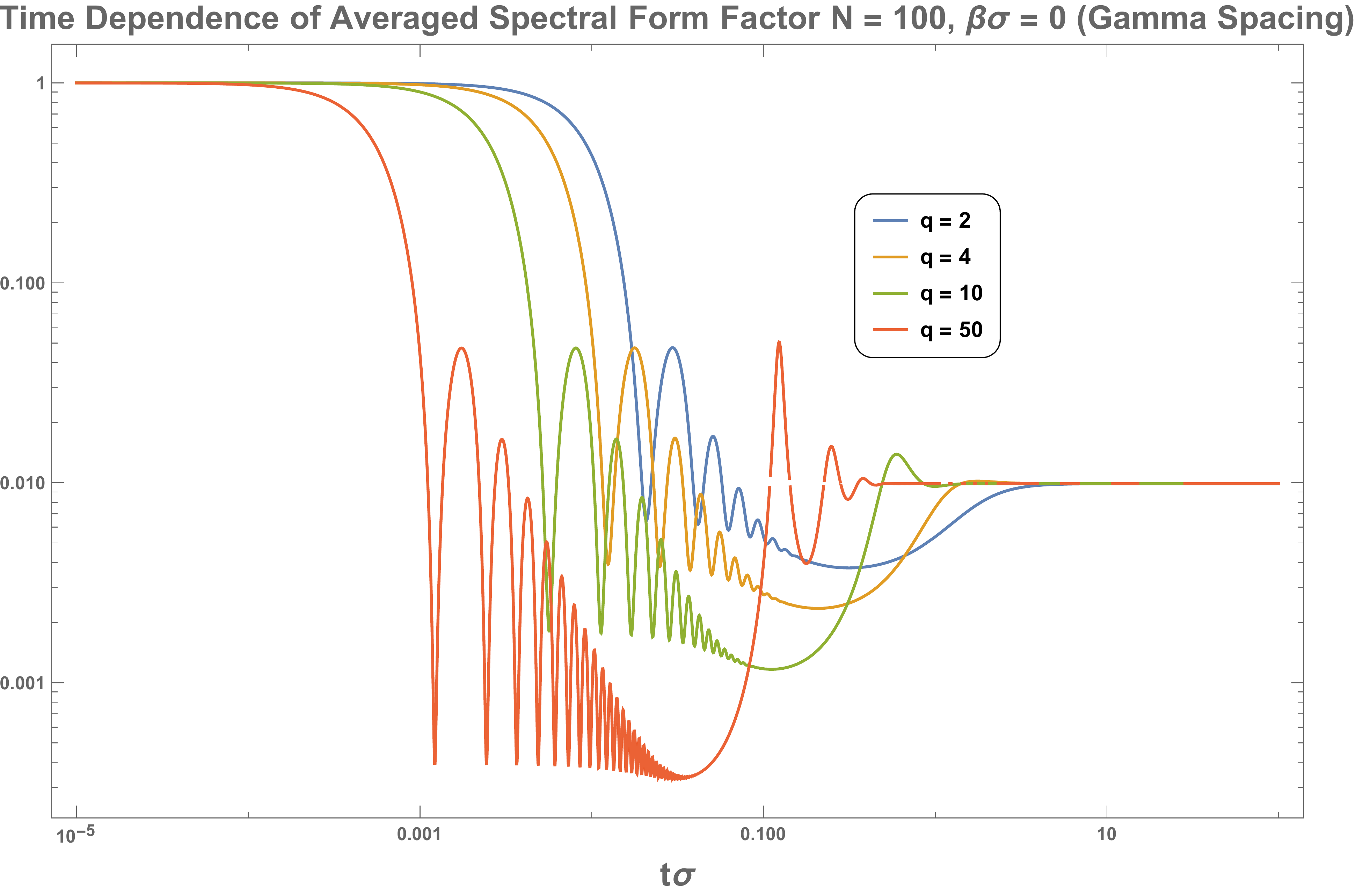}
\caption{Above is a Log-Log plot of the averaged spectral form factor at, infinite temperature, for a spectrum generated by the i.i.d model with a NNS distribution given by the gamma distribution defined in Eq. (\ref{GammaDist}). The value of $N$ is fixed and the form factor is plotted for different values of $q$. The energy scale, $\sigma$, is related to the average energy spacing between nearest neighbor eigenvalues, $\braket{\delta E}$, through the following relation $\braket{\delta E}=(q+1)\sigma$.  \label{FormFactPlotGammaDist}}
\end{figure}

\begin{figure}[]
\centering
\includegraphics[width=120mm]{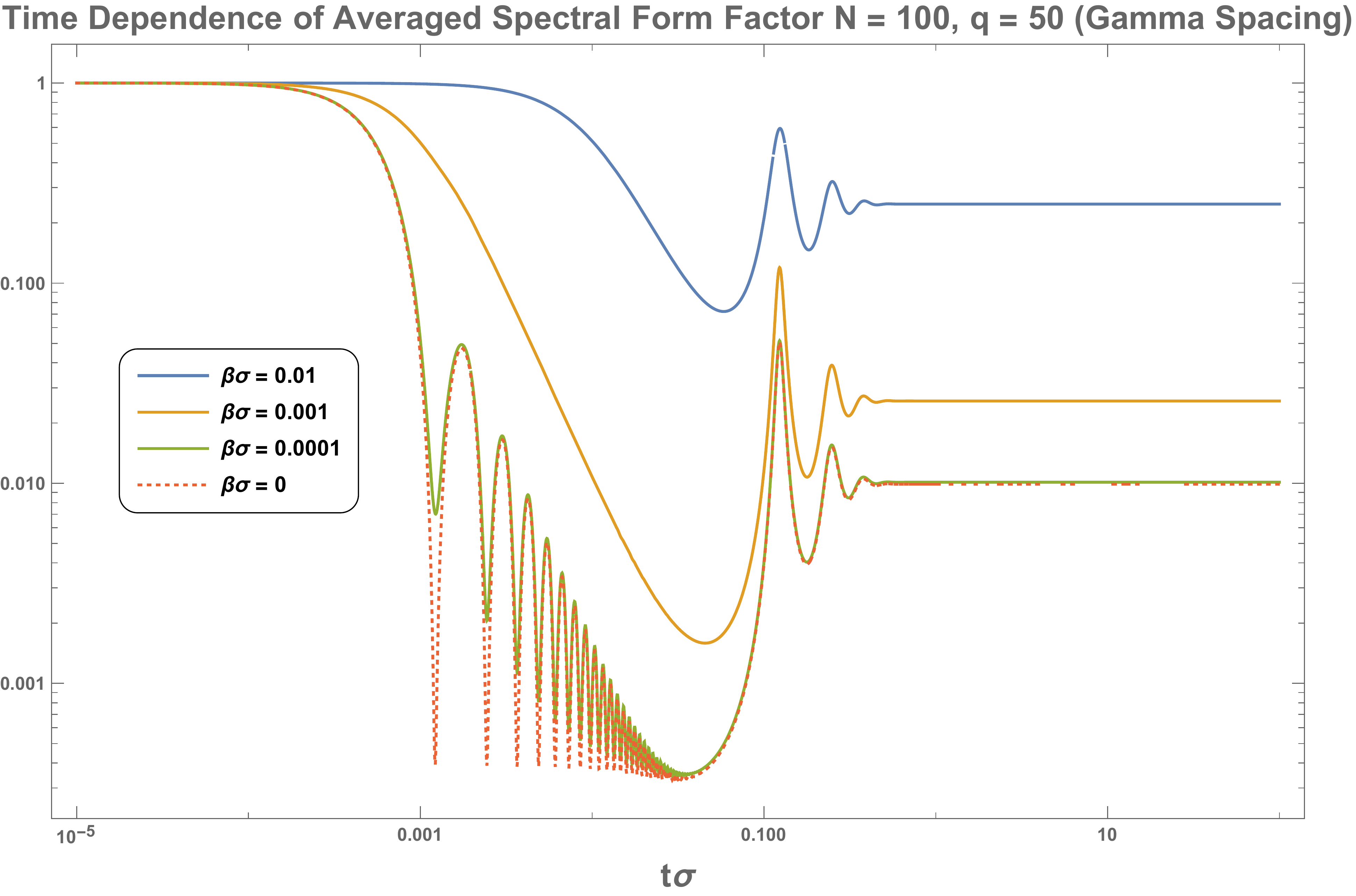}
\caption{Above is a Log-Log plot of the averaged spectral form factor at, for a spectrum generated by the i.i.d model with a NNS distribution given by the gamma distribution defined in Eq. (\ref{GammaDist}). In this plot $q=50$ and $N=100$ are fixed and the temperature is varied. The energy scale, $\sigma$, is related to the average energy spacing between nearest neighbor eigenvalues, $\braket{\delta E}$, through the following relation $\braket{\delta E}=(q+1)\sigma$.\label{FormFactPlotGammaDistq50Plot}}
\end{figure}

In Figure \ref{FormFactPlotGammaDist}, we plot the averaged form factor at infinite temperature with $N=100$ and vary $q$. At lower values of $q>1$, we still see that ramp and plateau (like the Wigner surmise), however as we increase the value of $q$ we start to see oscillations after the initial dip before saturation to the plateau. In Figure \ref{FormFactPlotGammaDistq50Plot}, we verify that these late time oscillations, at large $q$ persist at lower temperatures with the period roughly being the same as the temperature varies.

The existence of these oscillations for larger values of $q$ can be attributed to the NNS distribution localizing far from the origin and forming a substantial ``gap'' between the origin and the main distribution. One can check that the form factor of any system that has NNS density with a large gap will exhibit oscillations (although depending on the details of the NNS distribution the shape of the oscillations as well has how long they persist will vary). In particular, if one allows the Dyson index, $c$, in Eq. (\ref{WigSurGenC}) to be sufficiently large one can also check that similar oscillations will arise for the Wigner surmise. 

It is also useful to compare the averaged form factor with the form factor of a single sample in our model. 
\begin{figure}[]
\centering
\includegraphics[width=120mm]{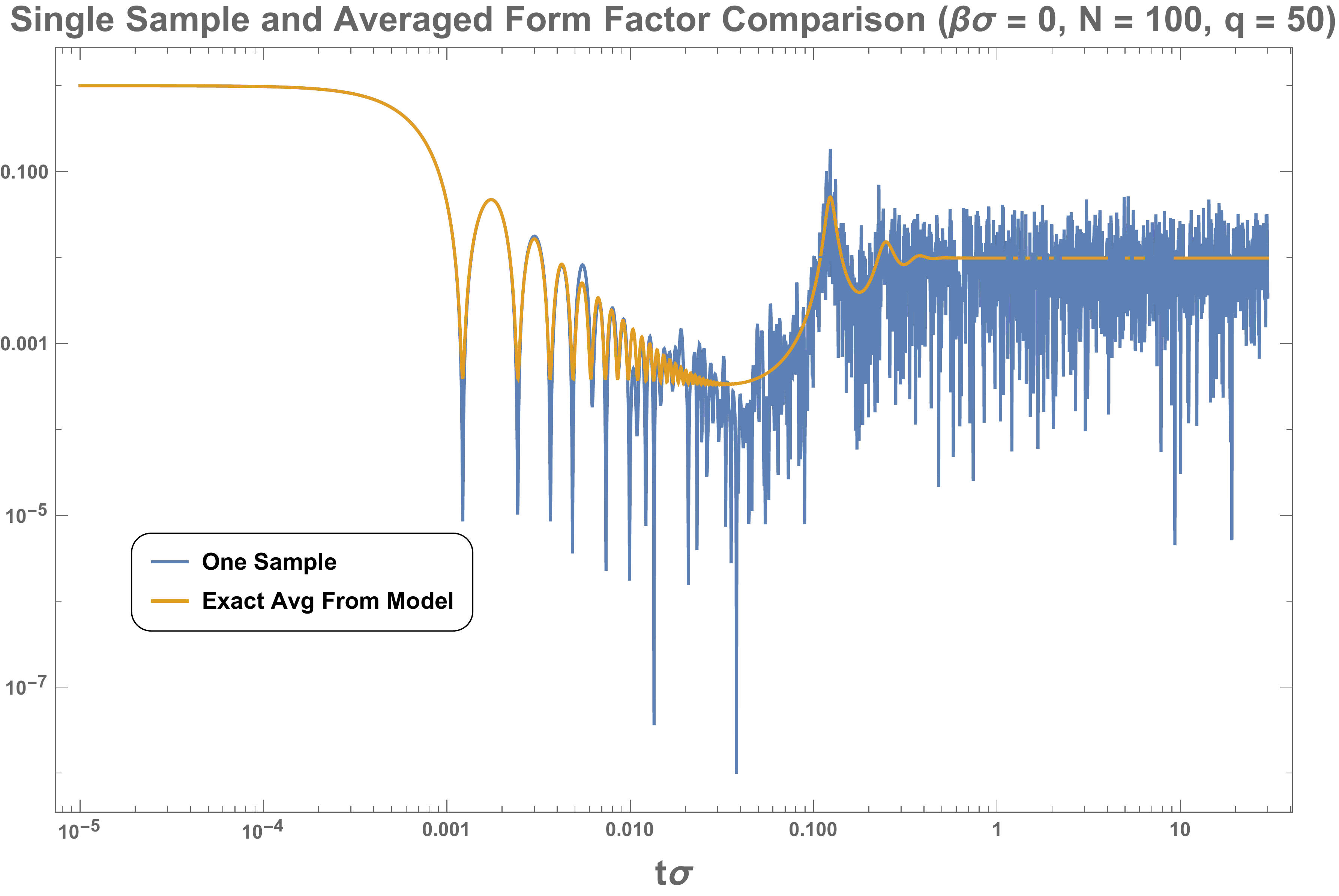}
\caption{Above is a Log-Log plot that compares the spectral form factor of a single sample (blue curve), and the averaged form factor (yellow curve) from the i.i.d model with gamma distribution spacing defined in Eq. (\ref{GammaDist}),  with $q=50$, $N = 100$, and $\beta\sigma=0$. \label{SingleSampVsAvgGama}}
\end{figure}
In Figure \ref{SingleSampVsAvgGama}, we can see that the regular late time oscillations are not self averaging. This is similar to the nature of the ramp and plateau behaviour studied in form factors of Gaussian ensembles \cite{Cotler:2016fpe}. The ramp and plateau manifest most clearly after averaging over many samples but is more difficult to ascertain when one analyzes only a single sample. 

Due to the simple expression of the form factor in the large $N$ and finite temperature regime it is possible to get a sense of where the oscillations occur by finding local extrema. By taking the first derivative of the expression in Eq. (\ref{ThermoDyFFGammaDis}) and setting it to zero we find the following condition:
\begin{equation}
    \begin{split}
    \label{GammaDisExtremaEq}
        &\frac{z^q}{\left[z^{q+1}-1\right]^2}=\frac{(z^*)^q}{\left[(z^*)^{q+1}-1\right]^2}\\
        &z=1+\sigma(\beta+it)\\
        &z^*=1+\sigma(\beta-it).\\
    \end{split}
\end{equation}
In general, the solution to this equation cannot be found exactly (in  Appendix \ref{ExtremaNNSGamma} we discuss the special cases when $q=1,2,4$ where exact solutions can be found). In the $q\gg 1$ regime it can be argued (see Appendix \ref{ExtremaNNSGamma} for the argument) that the period of the visible oscillations in the form factor will roughly be given by:
\begin{equation}
    \tau \approx \frac{2\pi(1+\beta\sigma)}{q\sigma}\approx \frac{2\pi(1+\beta\sigma)}{\braket{\delta E}}\approx\begin{cases}
\frac{2\pi}{\braket{\delta E}},\; \beta\sigma \ll 1\\
\frac{2\pi\beta\sigma}{\braket{\delta E}},\; \beta\sigma \gg 1\\
\end{cases}\\ .
\end{equation}
Above, we used that the average spacing between adjacent energy levels in our model is $\braket{\delta E}=(q+1)\sigma\approx q\sigma$.

\begin{figure}[H]
\centering
\includegraphics[width=120mm]{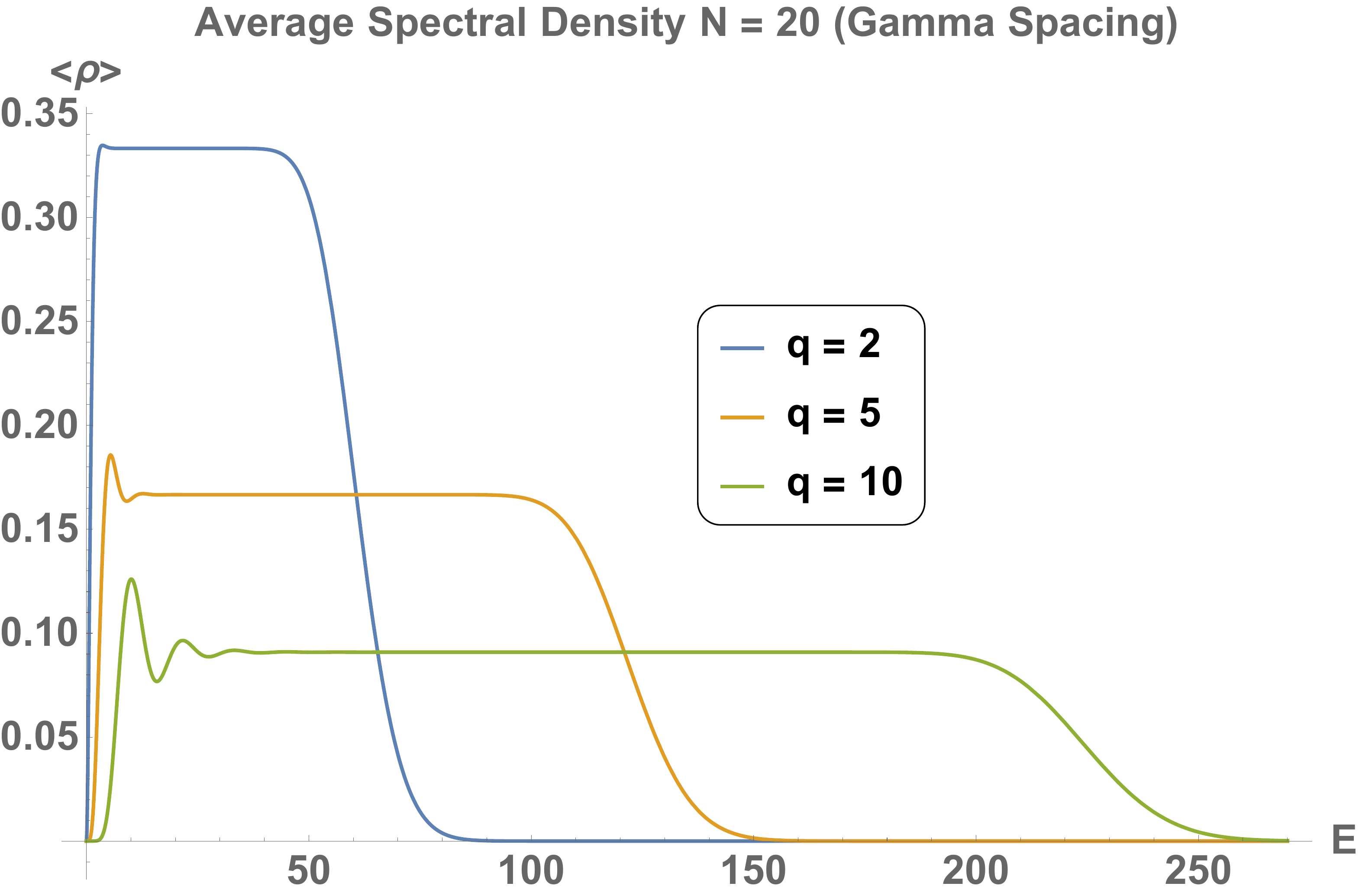}
\caption{The above plot depicts the averaged spectral density in Eq. (\ref{SpectralDensityGammaDis}). We fix $N=20$ and plot the density at various values of $q$ (we set $\sigma=1$ and $E_0=0$). The width the spectrum roughly scales as $N(q+1)$ for $N\gg 1$. \label{figg12}}
\end{figure}

Now that we have discussed the general features of the form factor we consider the averaged spectral density - which is given by (see Appendix \ref{SpectralDenAppendixGD} for details of the calculation):
    \begin{equation}
    \label{SpectralDensityGammaDis}
     \braket{\rho(E)}=\delta(E-E_0)+\frac{1}{\sigma}\sum_{m=1}^N\left[\frac{(E-E_0)^{mq+m-1}}{\Gamma\left[m(1+q)\right]\sigma^{m(1+q)-1}}\right]\Theta(E-E_0)e^{-(E-E_0)/\sigma}.
\end{equation}
Ignoring the delta function at the ground state, we can see the left most edge of the spectral density near $E=E_0$ is dominated by the first term in the sum over $m$ and vanishes at zero. To get an estimate for how wide the spectral density is, we analyze the $m=N$ term in the sum. This will give us a rough sense of what the tail of the spectral density looks like:
\begin{equation}
\label{DensityOfTail}
    \braket{\rho_{tail}(E)}\approx\frac{e^{-(E-E_0)/\sigma}}{\sigma\Gamma(N(1+q))}\left(\frac{E-E_0}{\sigma}\right)^{N(1+q)-1}.
\end{equation}
This function will give a bell shaped curve. We will be interested in the right most inflection point of the bell curve to give us an idea of where the edge of the spectral density is. This involves solving $\frac{d^2}{dE^2}\rho_{tail}(E)=0$ which yields a simple quadratic equation. The larger root gives us an estimate for where the right edge of the spectral density is:
\begin{equation}
    \frac{E_{edge}-E_0}{\sigma}=N(1+q)-1+\sqrt{N(1+q)-1}.
\end{equation}
So, for $N \gg 1$ and $q \geq 1$, the width of the spectral density roughly scales as $N(q+1)$. In Figure \ref{figg12} we can verify these findings by a simple plot of the spectral density with $\sigma=1$ and $N=20$ for various values of $q$.
We can see that for larger values of $q$ the low energy part of the spectrum exhibits visible oscillations. This is due to the fact that there is little overlap between the lower $m$ terms that make up the spectral density. A straightforward calculation of the local maximum of each term reveals that the distance between the visible local maxima is roughly $q+1$. 

It is interesting to ask what type of random matrix ensembles give rise to NNS statistics with a more general Dyson index. One answer to this question can be found in the study of ``$\beta$-ensembles'' which are discussed in \cite{Dumitriu_2002}. These ensembles have been shown to arise from $N\times N$ tri-diagonal random matrices of the form\footnote{Note: $\beta$ is not the temperature in this context. It is a parameter that plays the role of the Dyson index in the usual classical Gaussian ensembles.}:
\begin{equation}
H_\beta = \frac{1}{\sqrt{2}}\begin{bmatrix} 
    N(0,2) & \chi_{(N-1)\beta} & 0 & 0  &\cdots & 0 \\
   \chi_{(N-1)\beta} & N(0,2) & \chi_{(N-2)\beta} &0  &\ddots  &\vdots  \\
   0& \chi_{(N-2)\beta} & N(0,2) & \ddots & \ddots & 0 \\
    0 & 0 & \ddots & \ddots & \chi_{2\beta} & 0 \\
     \vdots & \ddots & \ddots &  \chi_{2\beta} &  N(0,2) & \chi_{\beta} \\
      0 & \cdots & 0 & 0 & \chi_{\beta} & N(0,2) \\
    \end{bmatrix},
\end{equation}
where $N(0,2)$ denotes random real variables pulled from a normal distribution with variance of $2$ and mean of $0$. The $\chi_m$ denotes positive real random variables pulled from the Chi-distribution with $m$ degrees of freedom\footnote{A random variable $y$ following the Chi-distribution with $m$ degrees of freedom has a probability density function given by $\chi_{m}(y)=\Theta(y)\frac{2}{\Gamma(m/2)}y^{m-1}e^{-y^2}$.}. It can be proved that the joint eigenvalue density (up to a normalization constant) of these matrices is given by:
\begin{equation}
    P_\beta(\lambda)\sim\prod_{i<j}\left|\lambda_i-\lambda_j\right|^\beta e^{-\sum_{i}\lambda_i^2/2}.
\end{equation}
The joint eigenvalue density for the classical Gaussian ensembles are reproduced when $\beta=1,2,4$ - but more general values of $\beta$ can also occur\footnote{It was shown in \cite{Dumitriu_2002} that the matrices in the classical Gaussian ensembles can be written in the tri-diagonal form. For example, a matrix pulled from the GOE ensemble can be tridiagonalized and the matrix will belong to a ``$\beta=1$-ensemble.''}. Since $\beta$ controls the degree of repulsion between eigenvalues, it seems reasonable to identify it with the Dyson index, $c$, in the Wigner surmise for NNS distribution. This conjecture was explored numerically in \cite{Le_Ca_r_2007}, where it was found that a (generalized) gamma distribution provided reasonable approximation to the NNS distribution of $\beta$-ensembles. Based on the results we obtained using the i.i.d model in this section, we suggest that with sufficiently large values of $\beta$ (here $\beta$ is not inverse temperature it is the Dyson index which labels the $\beta$-ensemble) the form factor of $\beta$-ensembles will have late time oscillations in the ensemble averaged form factor.  

In the next subsection, we briefly explore a class of simple oscillator systems that demonstrate sharp gaps in the NNS statistics. In these systems one can also see regular decaying oscillations in the late time behaviour similar to the ones we studied here. 

\subsection{Chaotic Perturbation of a Harmonic Oscillator}

One concrete example of a class of systems that have large gaps in the NNS distribution are interacting oscillator systems whose Hamiltonian may be written in the form:  
\begin{equation}
    \mathcal{H}=\mathcal{H}_0+\lambda \mathcal{H}_{chaos},
\end{equation}
where $\mathcal{H}_0$ is a Hamiltonian with a regularly spaced spectrum. $\mathcal{H}_{chaos}$ is a chaotic Hamiltonian in the sense that the NNS statistics of the eigenvalues of $\mathcal{H}_{chaos}$ obey a Wigner surmise type distribution. The parameter $\lambda$ controls the relative strength of the two terms. 

\begin{figure}[H]
\centering
\includegraphics[width=120mm]{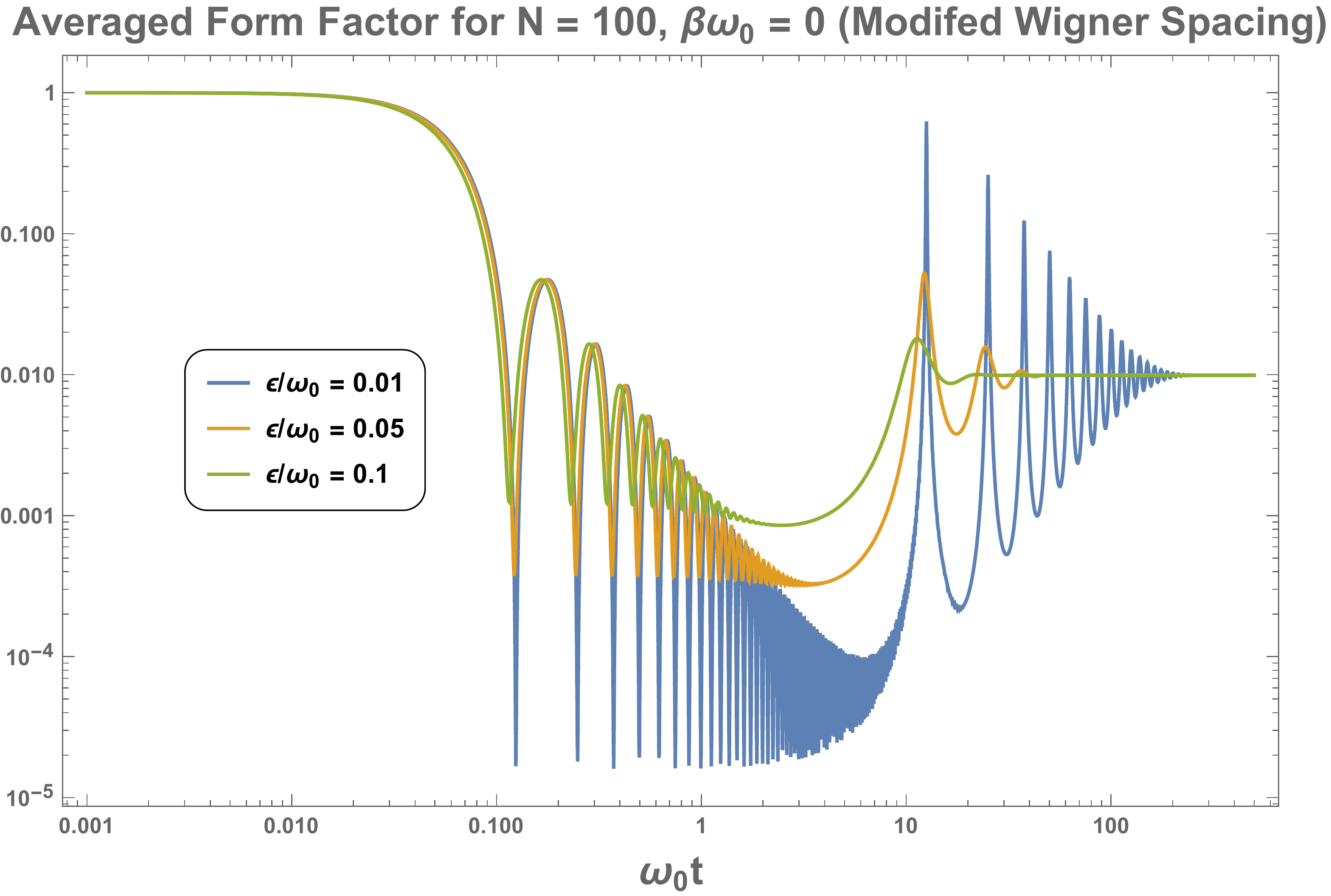}
\caption{Above is a plot of the infinite temperature average form factor of a spectrum generated by our i.i.d model ($N=100$) with the NNS distribution (modified Wigner surmise) given in Eq. (\ref{NNSDistEvenPlusRand}) for various choices of the dimensionless ratio $\epsilon/\omega_0$. \label{MWSFFPlot} }
\end{figure}

One familiar example of a system that has an evenly spaced spectrum is the one dimensional Harmonic oscillator vibrating at a frequency $\omega_0$. The introduction of the a chaotic term in such a scenario would be analogous to adding an additional potential term in the Hamiltonian which generates chaotic dynamics. Concretely, we write:
\begin{equation}
    \mathcal{H}=\omega_0\mathcal{H}_0+\epsilon\mathcal{H}_{chaos},
\end{equation}
where $\omega_0$ is some characteristic energy scale of the unperturbed oscillator and $\epsilon$ is a characteristic energy scale  for the chaotic Hamiltonian. In terms of matrices we let $\mathcal{H}_0$ be an $N\times N$ diagonal matrix with entries $\left(\mathcal{H}_0\right)_{ij}=N^{-1}(i-1) \delta_{ij}$ (we choose to normalize by $N$ so that the width of the unperturbed spectrum does not change). We model $\mathcal{H}_{chaos}$ using a random matrix. Roughly speaking, the NNS distribution between adjacent eigenvalues will be delta functions when $\epsilon=0$ once the chaotic term is turned on we will expect the delta functions to broaden and look similar to a kind of translated Wigner surmise. This results in a large gap in the NNS distribution which would lead to oscillations in the form factor. To get approximation for what the NNS distribution might look like, it is useful to compute the NNS distribution for $2\times 2$ matrices. This is a straightforward exercise which is done explicitly in Appendix \ref{NNSEvenPlusRndDerivationAppendix}, the final result is given by:
\begin{equation}
\label{NNSDistEvenPlusRand}
  \mathcal{P}(s)=\frac{s}{\sqrt{\pi}\epsilon\omega_0} \left(e^{\frac{\omega_0s}{2\epsilon^2}}-1\right) e^{-\left(\frac{2s+\omega_0}{4\epsilon}\right)^2}.
\end{equation}
Using the i.i.d model with the NNS distribution given by Eq. (\ref{NNSDistEvenPlusRand}) we can compute the form factor. Figure \ref{MWSFFPlot} plots the form factor at infinite temperature for various choices of $\epsilon/\omega_0$.
We can see that oscillations appear for smaller values of $\epsilon/\omega_0$ and begin to go away as the chaotic term becomes larger. The decay of the oscillations occur due to the random chaotic term in the Hamiltonian. In this scenario the parameter $\omega_0 /\epsilon$ characterizes the sharpness and size of the gap in the NNS distribution. Using the results we found in the previous subsection for gamma NNS statistics, we expect the period of the oscillations in the form factor in the $\epsilon/\omega_0 \ll 1$ regime to be approximated by:
\begin{equation}
    \tau \approx\frac{2\pi(1+\beta\omega_0)}{\braket{\delta E}}\approx \frac{4\pi(1+\beta\omega_0)}{\omega_0},  
\end{equation}
where we used the fact that $\braket{\delta E}\approx \omega_0/2$ when $\epsilon/\omega_0\ll 1$.

\section{Systems with Self-Averaging Oscillations in the Form Factor (Echoes)}
\label{SystemOscilFFSection}
In the previous section, we studied features of the averaged form factor and spectral densities associated with different choices of NNS distributions using the i.i.d model. For the i.i.d. model of random spectra we generally found that the spectral form factor decays until a time scale roughly given by $t\sim \braket{\delta E}^{-1}$, where $\braket{\delta E}$ is the average nearest neighbor spacing between eigenvalues. At time scales $t\ll \braket{\delta E}^{-1}$ the spectral density can be approximated as smooth and gives rise to self averaging behaviour at early times. After this time scale, the discreteness of the system manifests and the form factor is no longer self averaging. The averaged late time behaviour over many samples depends on the details of the NNS distribution for the eigenvalues of the system. From the examples we studied, we had three classes of behaviour after the initial dip:

\begin{enumerate}
   \item Continues to decrease and plateaus to a value (e.g., Poisson NNS where eigenvalues exhibit ``attraction'').  \item Increases for a duration of time  along a ``ramp'' followed by a plateau (e.g., Wigner surmise NNS where eigenvalues exhibit ``repulsion'').
   \item Increases and exhibits damped oscillations toward a plateau value (e.g., large $q$ gamma NNS where the eigenvalues exhibit ``enhanced'' repulsion resulting in large/sharp gaps between adjacent eigenvalues).
\end{enumerate}
Case 1 describes the NNS statistics of generic quantum integrable systems and has no ``ramp'' in the form factor \footnote{Systems with constant energy spacings, such as harmonic oscillator are exceptions. Such systems have form factors that oscillate forever and never saturate to a particular value.}. Case 2 describes NNS statistics of quantum chaotic systems whose spectral form factors contain a ``ramp.'' In both these cases, we can see that  there are no regular oscillations. In Case 3, there are oscillations in the averaged spectral form factor. These oscillations, however, occur in the late time behaviour of the form factor when it is no longer self-averaging (e.g., Figure \ref{SingleSampVsAvgGama}). Furthermore, the time scale over which these oscillations manifest are of the order of the Heisenberg time, $\braket{\delta E}^{-1}$. In the context of black holes one usually regards the energy spacing between black hole microstates to be of the order $e^{-S}$. This gives a Heisenberg time of the order $e^S$. So if we want to interpret the late time oscillations as echoes then we would have to wait a time scale of the order of the Heisenberg time which is extremely long for a macroscopic black hole and for practical purposes unobservable\footnote{There might be a ``caveat'' to these findings. In particular, we only discussed the late time oscillations in the context of the i.i.d. model. If we decided to use the i.i.d. model (with Wigner surmise spacing) to describe the spectrum of the SYK model we would correctly predict the existence of a ramp and plateau at late times. However, we would fail in predicting the time scale when the ramp would start, i.e. the Thouless time. In the same manner, just because the i.i.d. model predicts late time oscillations to manifest on time scales of the order of the Heisenberg time, it does not rule out the possibility of having a theory where the form factor exhibits late time oscillations on much shorter time scales which could potentially be observable in experiments.}.

In this section, we shall revisit one of the key assumption of Sections \ref{Section2RandomSpectrumandFF} and \ref{FormFactsCommonNNSDist}, i.e. that the NNS distribution are  independent identically distributed (i.i.d). As an example, the i.i.d assumption can be significantly violated if there are approximate degeneracies in the spectrum - causing many states to cluster around the same value while also being widely separated from other clusters of states. An example of this phenomenon is the hydrogen atom, where the splitting of degenerate states is controlled by the fine structure constant, $\alpha$. We can see that the assumption of ``independent-identically-distributed'' is not valid anymore, as the energy spacing is different by a factor of $\alpha$, depending on whether states belong to the same, or a different degenerate sub-sector. We shall see that this structure can also lead to self-averaging oscillations in the spectral form factor.  What one needs is to have an averaged spectral density that contains regularly spaced ``spikes'' (or near-degenerate sub-sectors).  The introduction of random interactions (regardless of how small) will generally lead to chaotic level splittings (a broadening of the ``spikes'') within each degenerate sub-sector. We will discuss one simple but explicit example in the following subsection which illustrates this. 

\subsection{Randomly Coupled Identical Fermionic Oscillators}
\label{ChaoticFermiOscillSection}
In this subsection, we will analyze a system of $N/2$ Fermionic oscillators ($N$ is even) which are coupled by a random matrix pulled from the GUE \footnote{Although we choose to work with fermionic oscillators, similar techniques can also be used to model a system of bosonic oscillators. We primarily choose fermionic degrees of freedom since they can be discussed in the context of the SYK model (a toy model describing the near horizon dynamics of extremal black holes) where the eigenstates of fermionic oscillator are the pure states studied in  \cite{Kourkoulou:2017zaj}. Furthermore, eigenstates can also be mapped to binary strings making this an interesting ``qubit'' model.}. 

The Hamiltonian describing $N/2$ uncoupled Fermionic oscillators, which we denote as $\mathcal{H}_{FHO}$, can be written as:
\begin{equation}
\label{FHOHamil}
    \mathcal{H}_{FHO}=\frac{1}{2}\sum_{k=1}^{N/2}\omega_k[b_k^\dagger,b_k],
\end{equation}
where $b_k$ and $b_k^\dagger$ are operators which satisfy the anti-commutation relations, $\{b_k,b_l^\dagger\}=\delta_{kl}$, $\{b_k,b_l\}=\{b_k^\dagger,b_l^\dagger\}=0$, and $\omega_k$ is the frequency of oscillation of the $k$-th oscillator (set $\hbar=1$). Before analyzing the effect of the GUE coupling we review the eigenstates of $\mathcal{H}_{FHO}$.

We denote the ground state of the FHO Hamiltonian as $\ket{0}$, it satisfies the following identity:
\begin{equation}
    b_k\ket{0}=0.
\end{equation}
The operators $b_k$ and $b_k^\dagger$ play the role of annihilation and creation operators respectively. We can check that the following identities are true:
\begin{equation}
\begin{split}
     &\left[\mathcal{H}_{FHO},b_k^\dagger\right]=\omega_kb_k^\dagger\\
     &\left[\mathcal{H}_{FHO},b_k\right]=-\omega_k b_k.\\
\end{split}
\end{equation}
We can create any state in the Fock space by starting with the ground state and acting with the creation and annihilation operators. There are exactly $2^{N/2}$ distinct states in the Hilbert space since we can only act once on the ground state using $b_k^\dagger$ for a particular $k$. These states (by convention) can be written as:
\begin{equation}
    \ket{\vec{n}}=\ket{n_1,n_2,...,n_{N/2}}=    \left(b^\dagger_1\right)^{n_1} \left(b^\dagger_2\right)^{n_2}  \cdot\cdot\cdot \left(b^\dagger_{N/2}\right)^{n_{N/2}}\ket{0},
\end{equation}
where $n_1,..,n_{N/2}=0$ or $1$ (we define $\left(b_k^\dagger\right)^0=\mathcal{I}$, the identity operator). When we act with a creation operator we find that:
\begin{equation}
\begin{split}
&b_k^\dagger\ket{n_1,n_2,..n_k..,n_{N/2}}=b_k^\dagger\left(b_1^\dagger\right)^{n_1}\left(b_2^\dagger\right)^{n_2}\cdot \cdot \left(b_k^\dagger \right)^{n_k} \cdot \cdot \left(b_{N/2}^\dagger\right)^{n_{N/2}}\ket{0}\\
&=(-1)^{\sigma_k}\delta_{0,n_k}\ket{n_1,n_2,..,n_k+1,..,n_{N/2}},\\
\end{split}
\end{equation}
where $\sigma_k=\sum_{i=1}^{k-1}n_i$. Similarly acting with an annihilation operator gives:
\begin{equation}
\begin{split}
&b_k\ket{n_1,n_2,..n_k..,n_{N/2}}=b_k\left(b_1^\dagger\right)^{n_1}\left(b_2^\dagger\right)^{n_2}\cdot \cdot \left(b_k^\dagger \right)^{n_k} \cdot \cdot \left(b_{N/2}^\dagger\right)^{n_{N/2}}\ket{0}\\
&=(-1)^{\sigma_k}\delta_{1,n_k}\ket{n_1,n_2,..,n_k-1,..,n_{N/2}}.\\
\end{split}
\end{equation}
We define the $k$-th number operator $\mathcal{N}_k$ as:
\begin{equation}
\begin{split}
    &\mathcal{N}_k=b_k^\dagger b_k\\
    &\mathcal{N}_k\ket{\vec{n}}=n_k\ket{\vec{n}}.\\
    \end{split}
    \end{equation}
    This means the Fock states are eigenstates of the number operator and since the Hamiltonian can also be written in terms of the number operator the Fock states diagonalize the Hamiltonian. In particular, we have:  
\begin{equation}
\begin{split}
    &\mathcal{H}_{FHO}=\frac{1}{2}\sum_{k=1}^{N/2}\omega_k\left(2\mathcal{N}_k-1\right)\\
    &\Rightarrow \mathcal{H}_{FHO}\ket{\vec{n}}=\frac{1}{2}\sum_{k=1}^{N/2}\omega_k(2n_k-1)\ket{\vec{n}}.\\
    \end{split}
\end{equation}
To simplify our considerations we will set $\omega_k=\omega_0$ for all $k$ - sometimes referred to as a mass term. In this case, we will have a regularly spaced spectrum which is easy to analyze.
We define $p=0,1,2,..,N/2$ as the number of ``occupied'' sites for a particular state (i.e. the number of 1s in the ket). Then the energy of such a state is given by:
\begin{equation}
\label{FHOENergy}
    E(p)=\omega_0\left(p-\frac{N}{4}\right).
\end{equation}
The number of microstates with the energy $E(p)$ is given by:
\begin{equation}
\label{NumberMicrostatesFHO}
    \Omega(p)=\frac{(N/2)!}{p!(N/2-p)!}.
\end{equation}
Using these results we can find the canonical partition function for the system given by:
\begin{equation}
\label{PartFuncFHO}
    \mathcal{Z}(\beta)=\sum_{p=0}^{N/2} \Omega(p)e^{-\beta E(p)}=e^{\frac{N\beta\omega_0}{4}}\left[1+e^{-\beta\omega_0}\right]^{N/2}=\left[2\cosh\left(\frac{\beta\omega_0}{2}\right)\right]^{N/2}.
\end{equation}
The spectral form factor for this system will oscillate with a period given by:
\begin{equation}
    \tau =\frac{2\pi}{\omega_0}.
\end{equation}
Now that we have discussed the details of the FHO, we address what happens when we add a random chaotic coupling. In particular, we will consider:
\begin{equation}
\label{ChaoticCoupledOscilHam}
    \mathcal{H}=\mathcal{H}_{FHO}+\epsilon\mathcal{H}_{chaos}.
\end{equation}
We know that the unperturbed energies are given by Eq. (\ref{FHOENergy}) and we know that these energies are degenerate. We will label each degenerate sector by $p$, which is equal to the number of occupied sites. We define the degenerate eigenstates in the $p$-th degenerate sector as $\{\ket{\vec{n}_p}\}_{p=1}^{\Omega(p)}$. Within the $p$-th degenerate sector we would go about computing the matrix elements:  
\begin{equation}
\braket{\vec{m}_p|\mathcal{H}_{chaos}|\vec{n}_p}.
\end{equation}
Diagonalizing a matrix with the matrix elements given above would tell us the information about how the degenerate energies within the $p$-th sector split. At leading order in degenerate perturbation theory, the splitting of the energy levels only depend on the details of the sub-block in the degenerate sector. This means that first order theory will contain no information about correlations between different degenerate energy sectors. In other words, at leading order we can treat our random matrix as block diagonal with each block having the same size as the particular degenerate sector. More specifically, we can write:
\begin{equation}
    \mathcal{H}=\bigoplus_{p=0}^{N/2}\left[E(p)\mathcal{I}(p)+\epsilon \mathcal{H}_{chaos}(p)\right],
\end{equation}
where $\mathcal{I}(p)$ is an identity matrix of dimension $\Omega(p)\times \Omega(p)$ and $\mathcal{H}_{chaos}(p)$ is a sub-block of the chaotic interaction Hamiltonian of dimension $\Omega(p)\times \Omega(p)$. 

\begin{figure}
\centering
\includegraphics[width=150mm]{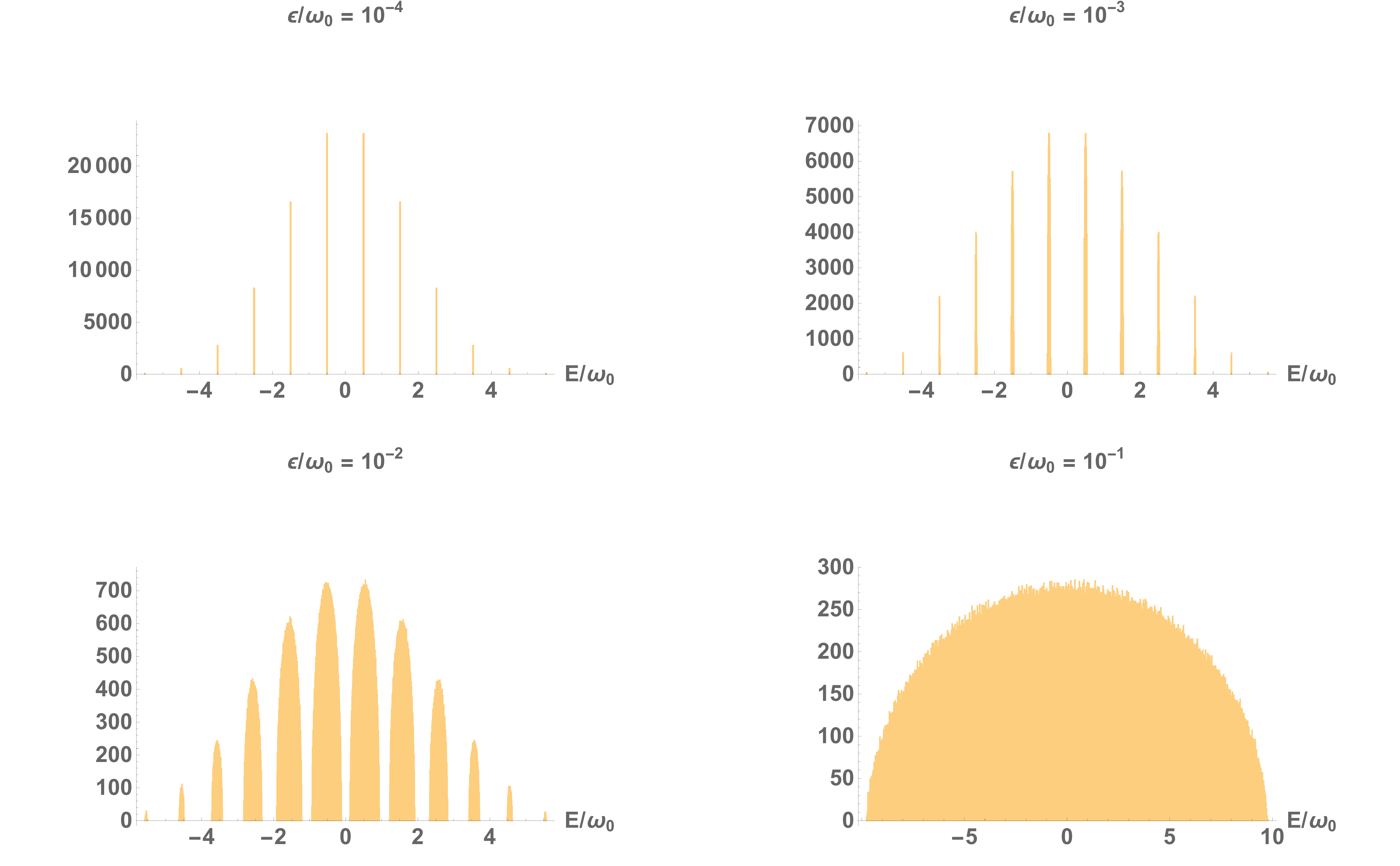}
\caption{Above is depiction of the averaged spectral density (histogram of eigenvalues) of a system of 11 identical fermionic oscillators coupled together with random interactions represented by a matrix pulled from a GUE for various values of $\epsilon/\omega_0$. The top two plots labelled by $\epsilon/\omega_0=10^{-4}, 10^{-3}$ depict the spectral density at weak coupling. The bottom two plots labelled by $\epsilon/\omega_0=10^{-2}, 10^{-1}$ depict the spectral density at intermediate coupling ($\epsilon/\omega_0=10^{-2}$) and strong coupling ($\epsilon/\omega_0=10^{-1}$).    \label{FHOPlusGUESpecDenPlot}}
\end{figure}

To simplify our considerations, we will assume that the $\mathcal{H}_{chaos}$ is given by a random matrix pulled from the GUE\footnote{The elements of the GUE are Hermitian matrices. The off-diagonal elements will be of the form $x+iy$ where $x$ and $y$ are random variables pulled from a Gaussian distribution with a mean of zero and variance of $1/2$. The diagonal elements are real random variables pulled from a Gaussian with mean of zero and variance of 1. In this case it is well known that the width of the semi-circle spectral density goes as $4\sqrt{N}$ \cite{livan2017introduction}. This width of the semi-circle can be scaled by multiplying the matrices in the ensemble by a factor c then the new width is $4c\sqrt{N}$.\label{GUEparaFootnote}}. If $\mathcal{H}_{chaos}$ is a random matrix from the GUE, then we know the splitting of degenerate states within a sector will occur so that the NNS statistics are consistent with the GUE (i.e. you will have Wigner surmise spacing). Furthermore, we expect the spectral density to spread out from a delta function at the degenerate energies to ``semi-circles'' centered around the degenerate energies. The width of the spread for the $p$-th degenerate sector will roughly go as:
\begin{equation}
\label{widthofSectorEst}
    W(p)\approx 4\epsilon\sqrt{\Omega(p)}.
\end{equation}
This allows us to define a rough criterion in which we expect perturbation theory to hold: 
\begin{equation}
\label{PertRegimeCond}
    \frac{\max_p W(p)}{\omega_0}=\frac{4\epsilon\sqrt{\max_p\Omega(p)}}{\omega_0}\ll 1.
\end{equation}
This essentially states we want to work in a regime where the spread has a width much smaller than the spacing between energy levels of the unperturbed oscillator. Figure \ref{FHOPlusGUESpecDenPlot} depicts numerically generated spectral density at $N=22$ for different choices of the parameter $\epsilon/\omega_0$.
One can check that in the specific case we plotted the condition given in Eq. (\ref{PertRegimeCond}) for being in the perturbative regime translates to $\epsilon/\omega_0\ll 10^{-2}$. We can see that the average spectral densities on the top row of Figure \ref{FHOPlusGUESpecDenPlot} satisfy the condition and the spread within the degenerate sectors is much smaller than the spacing between different degenerate sectors. In contrast, the plots on the bottom row do not satisfy the perturbative condition and we can see that at $\epsilon/\omega_0=10^{-2}$, the spread within the degenerate sectors start to become comparable to the spacing between them. The spread will increase as we increase $\epsilon/\omega_0$ and there will be substantial overlap between sectors, eventually leading to a point where the overall density simply looks like that of the GUE at $\epsilon/\omega_0=10^{-1}$. In Appendix \ref{SpreadAnalysisAppendix}, we do a numerical analysis of the exact width of the spread within the degenerate sectors and compare to what we expect from first order degenerate perturbation theory, finding reasonable agreement for $\epsilon/\omega_0\leq 10^{-2}$.

Now that we have discussed the spectral density, we can turn our attention to the spectral form factor. Within the perturbative regime, there are two distinct time scales. One is given by the oscillator frequency $\omega_0^{-1}$ and the other scale is given by $\epsilon^{-1}$ which gives the strength of the chaotic perturbation (roughly related to the average spacing of eigenvalues within a degenerate sector). At time scales $t\ll \epsilon^{-1}$, the form factor will be dominated by the coarse grained spectrum which is a sharply spiked spectrum with regular spacing. Due to this we expect to see oscillations in the form factor with a period that is well described by the unperturbed form factor, $\tau=2\pi\omega_0^{-1}$. We will also expect these oscillations to gradually decay due to the repulsion of adjacent eigenvalues within a degenerate sector. For time scales $t\gtrsim \epsilon^{-1}$, the form factor will become sensitive to fine structure of the degenerate sectors. In this regime, the form factor should contain a ramp and a plateau consistent with the late time behaviour of a form factor in the GUE, as we have seen before.     

\begin{figure}[H]
\centering
\includegraphics[width=150mm]{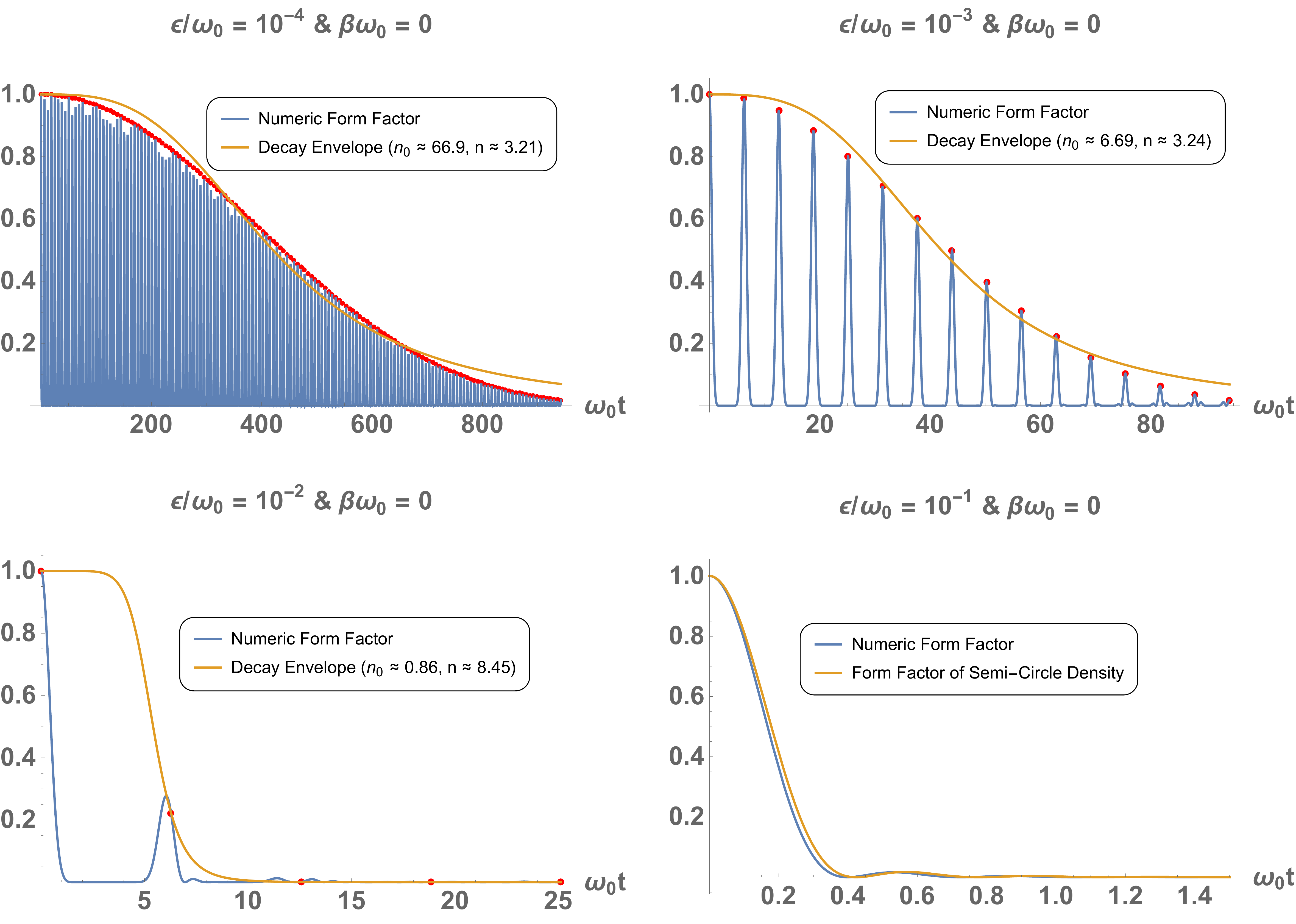}
\caption{The plots in the figure above depicts the self averaging early time behaviour in the infinite temperature form factor computed numerically for a system of 11 fermionic oscillators coupled together by a ($2^{11}\times 2^{11}$) random GUE matrix for various values of $\epsilon/\omega_0$ ranging from $10^{-4}$ to $10^{-1}$. The blue line depicts the numerical calculation of the form factor for a single sample (due to the self averaging behaviour at early times one can check the single sample will agree with the numerical average to high precision). The red dots indicate the value of the form factor at regularly spaced time intervals given by $\omega_0 t_k=2\pi k$. The yellow line is a numeric fit of the red data points to a function given by Eq. (\ref{amplitudeFit}) with fitting parameters $n_0, n$ (shown in the legend of the plots). For the bottom right plot we compare the numeric early time behaviour to the behaviour predicted by Eq. (\ref{SEMICIRCFormFact}), which is the form factor of a semi-circle spectral density.    \label{EarlyTimeOscilFig}}
\end{figure}

\begin{figure}[H]
\centering
\includegraphics[width=140mm]{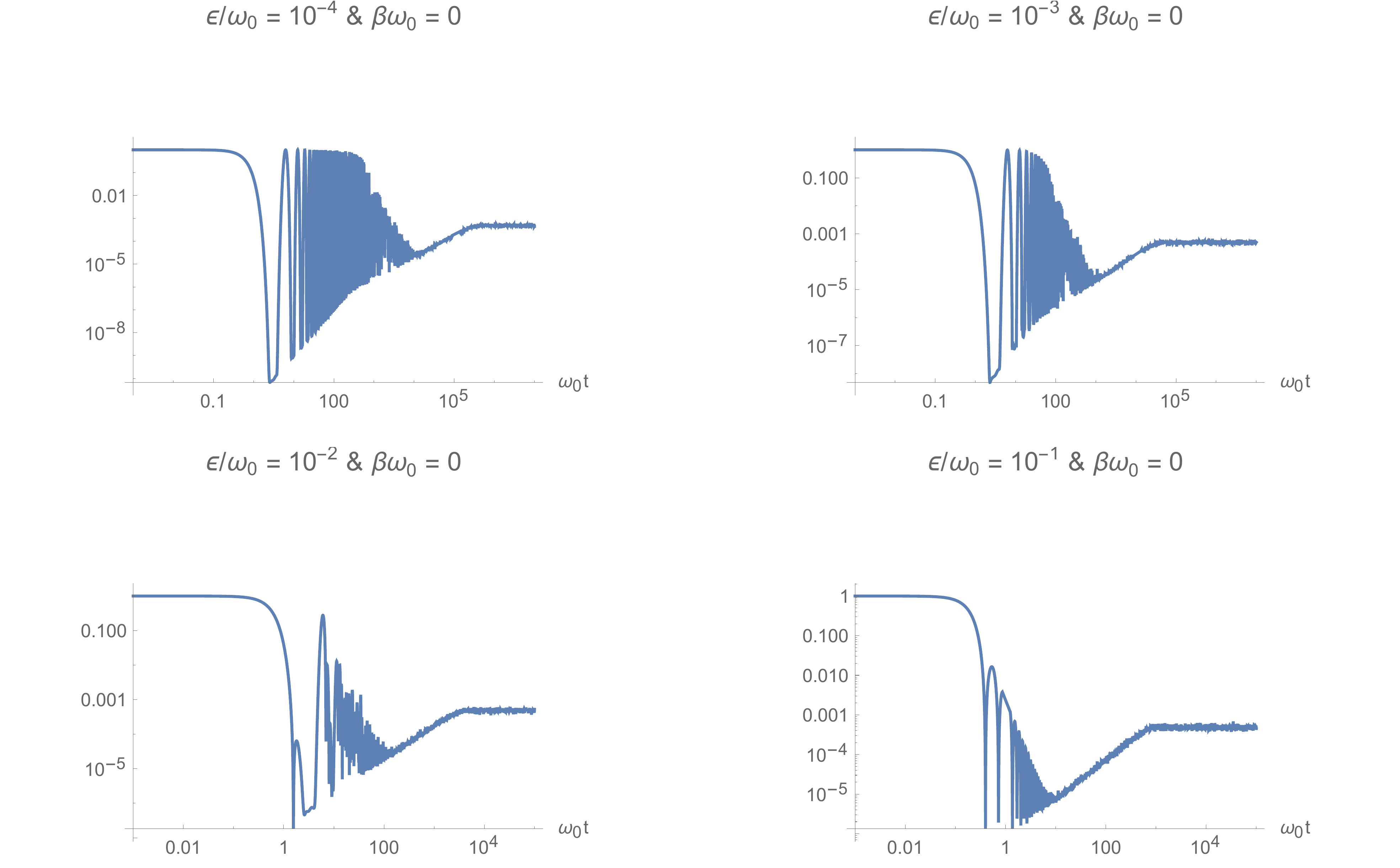}
\caption{The plots above depict the numerically averaged form factor on a Log-Log scale (averaged over 100 samples) at infinite temperature for a system of 11 fermionic oscillators coupled together by a ($2^{11}\times 2^{11}$) random GUE matrix for various values of $\epsilon/\omega_0$ ranging from $10^{-4}$ to $10^{-1}$.  \label{NumericLateTimeFHOGUEplot}}
\end{figure}

In Figures \ref{EarlyTimeOscilFig} and \ref{NumericLateTimeFHOGUEplot},  we numerically plot the form factor (averaged over 100 samples) for different values of $\epsilon/\omega_0$ at $\beta\omega_0=0$.  
The the perturbative regimes (weakly coupled) occurs when $\epsilon/\omega_0=10^{-4},10^{-3}$ and the non-perturbative regimes (strongly coupled) occur at  $\epsilon/\omega_0=10^{-2},10^{-1}$. Figure \ref{EarlyTimeOscilFig} depicts the early time behaviour of the form factor and it shows that the form factor oscillates at early times with a frequency $2\pi\omega_0^{-1}$ (the red dots indicate the value of the form factor at regular time steps given by the period $\tau=2\pi\omega_0^{-1}$). The oscillations have a decaying amplitude bounded by an envelope which can be numerically fit by:
\begin{equation}
\label{amplitudeFit}
    A(t) = \frac{1}{1+\left(\frac{\omega_0 t}{2\pi n_0}\right)^n},
\end{equation}
where $n_0$ and $n$ are fit parameters. We can see from the legend in Figure \ref{EarlyTimeOscilFig} that the amplitude of the oscillations in the weakly coupled regime decays through a power law with a power $n\sim 3$, furthermore we can see that $n_0\sim C \omega_0/\epsilon$ where $C$ is a proportionality constant that will depend on $N$\footnote{In this case it is well approximated by $C\approx \frac{1}{7 \sqrt{\max_p\Omega(p)}}\approx 0.00665$. More generally, we expect $C\sim \frac{1}{\sqrt{\max_p\Omega_p}}$ with some numerical pre-factor.}. 

In the non-perturbative regime (the bottom two plots of Figure \ref{EarlyTimeOscilFig}), we can see that the regular oscillations start to disappear and give way to the more familiar features present in the early time behaviour of the form factor for the GUE. The early time behaviour of the form factor in the case when $\epsilon/\omega_0=10^{-1}$ is well approximated by the form factor associated with a semi-circle density (SCD) given by:
\begin{equation}
\label{SEMICIRCFormFact}
    \left|\frac{\mathcal{Z}(\beta+it)}{\mathcal{Z}(\beta)}\right|^2_{SCD}=
    \left| \frac{\int_{-1}^1\sqrt{1-x^2}e^{-ixE_0(t-i\beta)}dx}{\int_{-1}^1\sqrt{1-x^2}e^{-ixE_0(-i\beta)}dx} \right|^2=\frac{\beta^2}{\beta^2+t^2}\frac{J_1\left[E_0(t-i\beta)\right]J_1\left[-E_0(t+i\beta)\right]}{\left[J_1(-i\beta E_0)\right]^2},
\end{equation}
where $E_0$ is the half the width of the semi-circle density (in our case we used $E_0/\omega_0=\sqrt{2^{13}}/10\approx 9.05$).

In Figure \ref{NumericLateTimeFHOGUEplot}, we depict the averaged form factor on a Log-Log scale at longer time scales (averaged over 100 samples). The early time behaviour contains oscillations in the weakly coupled regime which we already described in our discussion of Figure \ref{EarlyTimeOscilFig}. The averaged late time behaviour in all coupling regimes contains the ramp and plateau. This late time behaviour occurs due to the manner in which each degenerate sector of the uncoupled oscillator theory splits in the presence of the GUE matrix coupling.  

\subsection{Randomly Coupled Oscillators as a Toy Model for the Membrane}
Let us now see how our system of chaotically coupled oscillators, can act as a quantum mechanical toy model to describe gravitational waves echoes (e.g., \cite{Abedi:2020ujo}). 

To do this, we recall that gravitational wave echoes are thought to arise due to quantum gravity effects near the horizon which result in the partial reflection of perturbations. These reflections are usually modelled by cutting off the semi-classical black hole geometry close to the horizon and then placing semi-reflective boundary conditions at the cutoff. In the context of large AdS black holes, one can also place a cutoff near the horizon and echoes will arise due to the reflection of perturbations near the horizon and also at the conformal boundary \cite{Saraswat:2019npa}. In this paper, we adopt the view that the cutoff with modified boundary conditions describe some kind of dynamical membrane deep in the bulk near the horizon. This membrane has its own internal spectrum which should resemble the chaotic and discrete spectrum of a black hole.    

If one begins with a perfectly reflective membrane, then perturbations to the geometry will not decay and the geometry will have normal modes (rather than quasi-normal modes). The dynamics of the perturbations on the spacetime may be roughly modelled by some kind of oscillator system similar to the one discussed in the previous section. The characteristic frequency of the oscillators in the previous subsection would be identified with the spacing between normal modes in the cutoff spacetime. For example, for a spherical Schwarzschild AdS black hole (with $r_H/L\gg 1$), if the cutoff is a proper radial length, $l_{prop}$, from the horizon, then we expect massless scalar perturbations to make a round trip (from the cutoff to the conformal boundary and back) in a time scale roughly given by \cite{Saraswat:2019npa}:
\begin{equation}
\label{EchoTimeAdSBH}
\tau\sim \beta\ln\left(\frac{L^2}{l_{prop}^2}\right) \sim \frac{L^2}{ r_H}\ln\left(\frac{L^2}{l_{prop}^2}\right),
\end{equation}
where $r_H$ and $L$ is the horizon and AdS radii, respectively. We identify $\omega_0=2\pi\tau^{-1}$ with the characteristic oscillator frequencies\footnote{Notice that the frequencies have a linear temperature dependence, this could arise as a thermally induced mass from a renormalization procedure in QFT at finite temperature \cite{Altherr:1993tn,Laine:2016hma}.}. This will give rise to a periodic form factor with oscillations that have a period equal to $\tau$. To introduce dissipation into the system, we can further add chaotic interactions between the oscillators. In the weak coupling regime, these interactions are responsible for the decay of the regular oscillations at early times in the form factor and also will be responsible for the ramp and plateau at late times making late time thermalization similar to SYK theory. 

To interpret the energy scale $\epsilon$ in Eq. (\ref{ChaoticCoupledOscilHam}), we turn off the oscillator term by sending the cutoff to the horizon (i.e. $l_{prop}=0$ in Eq. (\ref{EchoTimeAdSBH})) so that $\omega_0=0$, and we are left with the chaotic term. This gives rise to the usual semi-circle spectral density and gives a standard SYK/JT gravity type model for a black hole - where no oscillations are present at early times. In this case, the energy scale, $\epsilon$, controls the average spacing between adjacent energy levels which is of the order $\delta E\sim \epsilon e^{-S/2}$ where $S$ is the entropy of the black hole\footnote{We obtain $\epsilon e^{-S/2}$ by recalling that the width of the semi-circle scales with the size of the matrix $\epsilon\sqrt{2^{N/2}}$. Then we take the width of the semi-circle and divide by the total number of states which is $2^{N/2}$, to get an average spacing $\delta E\sim \epsilon 2^{-N/4}$. Finally we identify $S\sim \ln\left(2^{N/2}\right)$  to obtain $\delta E\sim \epsilon e^{-S/2}$.}. At the same time, if we consider the first law of black hole thermodynamics it suggests that the average spacing between adjacent levels is of the order $\delta E_{\rm min}= \delta S_{\rm min}/\beta$. We will be agnostic about what $\delta S_{\rm min}$ should be. We know the smallest possible value of $\delta S_{\rm min}$ is of the order $e^{-S}$ but there might also be other choices (for example, $\delta S_{\rm min}\sim 1$ might also be sensible when discussing entropy changes by emission of a single Hawking quanta).

Identifying the $\delta E$ expression from the random matrix model to the $\delta E_{\rm min}$ from the first law gives $\epsilon\sim \beta^{-1}\delta S_{\rm min}e^{S/2}$. When we do this and consider the dimensionless ratio between $\epsilon$ and $\omega_0$ we have:
\begin{equation}
\frac{\epsilon}{\omega_0}\sim e^{S/2}\ln\left(\frac{L^2}{l_{prop}^2}\right)\delta S_{\rm min}.
\end{equation}
If $\delta S_{\rm min}\sim e^{-S}$ (the smallest possible change in entropy) and $l_{prop}\sim \ell_{p}$, where $\ell_{p}$ is the Planck length, then our oscillator system is in the weakly coupled regime and will exhibit echoes that manifest after a scrambling time scale. Such echoes would be observable within a reasonable window of time for astrophysical black holes, making this model interesting in the discussion of gravitational wave echoes.   

\section{Discussion and Conclusion}
\label{ConclusionSec}
The AdS/CFT correspondence formulates quantum gravitational systems in AdS in terms of CFT systems on the boundary.  In such a scenario, the statement that a black hole is a quantum chaotic system is understood in terms of the high energy eignestates of the CFT. In the context of spacing statistics of the CFT spectrum, one could imagine diagonalizing the Hamiltonian $\mathcal{H}$ to find a complete orthogonal basis of eigenstates $\ket{n}$ and eigenvalues $E_n$. Any state can then be written in terms of the eigenstates of the Hamiltonian including the black hole. 
As one goes to ``very high'' energies, one should expect the spectrum of the CFT to exhibit chaotic spacing statistics (i.e. eigenvalue repulsion). If one assumes that the a black hole is a thermal ensemble of such microstates, then how the black hole relaxes after being perturbed should be described by thermal correlation functions. The question of how quantum aspects of a black hole manifest in the ringdown and whether they are potentially observable could be addressed using this picture. 

In this paper, we explored the concept of echoes due the  black hole microstructure from the perspective of the thermalization behaviour of quantum chaotic systems. The form factor served as a convenient proxy to study this behaviour for systems with varying spectral statistics. We employed a simple model for a random spectrum that involved summing a set of independent-identically-distributed (i.i.d) random variables which represent the spacing between adjacent eigenvalues in the spectrum.  By defining a random spectrum in this manner, we were able to specify the resulting spectrum's Nearest Neighbor Spacing (NNS) statistics. Despite the simplicity of the i.i.d model, we demonstrated that it could reproduce various important late time features seen in the spectral form factor of strongly coupled chaotic systems. Most notably, by using Wigner surmise NNS statistics we were able to obtain the ramp and plateau behaviour. We also used the i.i.d model to study more general NNS distributions where the ``Dyson index'' is allowed to be any real positive number, rather than the canonical values of $1,2,$ and $4$ (corresponding to the GOE, GUE, and GSE ensembles respectively). 

We briefly discussed how more general choices of the Dyson index could correspond to more general random matrix ensembles called ``$\beta$-ensembles,'' which are constructed from certain types of tri-diagonal matrices. We also argued that for a large Dyson index, the averaged form factor of a $\beta$-ensemble would have damped oscillations at late times before approaching a plateau. The origin of these oscillations are the large regularly spaced ``gaps'' in NNS distribution which are expected to arise when the Dyson index becomes large. We further showed that such gaps could be realized in systems with an evenly spaced spectrum - like a harmonic oscillator with an additional chaotic potential. Although such oscillations might be thought of as ``echoes'' in the black hole context, such echoes would manifest on extremely long  time scales of the order $e^S$, disqualifying them as being observable in the context of astrophysical black hole  microstructure. Although, the late time oscillations found in Section \ref{GammaDisSectionFF} may not be interesting in the astrophysical context, they may be connected to the discussion of form factors in more exotic theories of gravity. In particular, in the work \cite{Collier:2021rsn}, the form factor of Narain's family of free boson CFTs (which is dual to ``U(1) gravity'' in AdS$_3$) has been shown to have a form factor that contains oscillations at late time similar to the ones we found in Section \ref{GammaDisSectionFF}. Our results suggest that such oscillations occur due to enhanced ``repulsion'' between eigenvalues in the spectrum. It would be interesting to investigate if the late time oscillations showing up in the form factor of Narain ensemble CFTs can also be traced back to enhanced repulsion statistics in the spectrum. 

Studies using the i.i.d. model in Sections \ref{Section2RandomSpectrumandFF} and  \ref{FormFactsCommonNNSDist} suggested that the quantum aspects of the thermalization behaviour of black holes usually manifest at very late times which are unobservable in the astrophysical context. A common aspect of the spectra generated by the i.i.d. model was that there was a single large cluster of microstates with very little coarse-grained structure. To get non-trivial behaviour on observable time scales, we thus proposed that there must be additional structure in the coarse grained density of states. In particular, for echoes to manifest, we showed that the coarse-grained density had to have large clusters of states widely separated from other clusters. In Section \ref{SystemOscilFFSection}, we considered a many body system of identical fermionic oscillators coupled by an interaction term modelled by a random matrix in the GUE. In the weakly coupled regime, we used degenerate perturbation theory to argue that degenerate sectors of $\mathcal{H}_{FHO}$ would split in a way consistent to the random matrix statistics of the GUE. This naturally led to regularly spaced decaying oscillations at early times followed by a ramp and plateau in the averaged form factor. We then discussed the possibility of using our model of coupled oscillators as a toy model that describe near horizon modifications that give rise to gravitational wave echoes.  

Although we were able to show that one can construct quantum chaotic systems with echoes, there are still a number of issues that need to be addressed before we can view them as reasonable models for describing black holes with microstructure near the horizon. Aside from the condition that a black hole has a quantum chaotic energy spectrum, it must also satisfy additional constraints which are usually discussed in the context of out-of-time-order correlators (OTOCs)  \cite{Maldacena:2015waa,Hashimoto:2017oit,Cotler:2017jue}. In particular, quantum chaotic systems describing black holes are expected to also saturate the Lyapunov bound discussed in \cite{Maldacena:2015waa}. We did not demonstrate that the coupled oscillator systems introduced in Section \ref{SystemOscilFFSection} satisfy this constraint. A good place to start exploring the issue of OTOCs (and also other types of correlators) is with the following concrete oscillator model\footnote{It is interesting to point out that a coupled oscillator system is similar to the kinds of models studied in \cite{Garcia-Garcia:2017bkg}, although the quadratic term is random in those models and of a more general form.}:
\begin{equation}
\label{OscPlusSYK}
    \mathcal{H}=-i\omega_0\sum_{k=1}^{N/2}\chi^{2k-1}\chi^{2k}+\epsilon \sum_{1\leq \alpha<\beta<\mu<\nu\leq N}j_{\alpha\beta\mu\nu}\chi^\alpha\chi^\beta\chi^\mu\chi^\nu,
\end{equation}
where $\chi$'s are Majorana fermions: $\{\chi^\alpha,\chi^\beta\}=\delta^{\alpha\beta}$, and $j_{\alpha\beta\mu\nu}$ is a completely anti-symmetric coupling tensor where each component is an independent random Gaussian variable with a mean of zero. The quadratic term in Eq. (\ref{OscPlusSYK}) describes $N/2$ free identical fermionic oscillators expressed in the Majorana basis\footnote{The transformation between the creation and annihilation operators defined in Eq. (\ref{FHOHamil}) and the Majorana fermions $\chi$ is $b_k=\left(\chi^{2k-1}-i\chi^{2k}\right)/\sqrt{2}$ and $b_k^\dagger=\left(\chi^{2k-1}+i\chi^{2k}\right)/\sqrt{2}$.} and the quartic interaction is the standard SYK model Hamiltonian. As before, the quantity $\epsilon/\omega_0$, would control the strength of the interactions between the oscillators. One would then have a family of theories labeled by $\epsilon/\omega_0$. In the limit when $\epsilon/\omega_0 \to\infty$ we should recover the SYK model with no echoes which saturates the Lyapunov bound (i.e. the Lyapunov exponent $\lambda_L=\frac{2\pi}{\beta}$). In the opposite limit, when $\epsilon/\omega_0\to 0$, we would have free fermionic oscillators (non-decaying periodic behaviour in form factor) and the Lyapunov exponent will be zero. In the intermediate regime, we should have regular decaying oscillations; similar to the ones discussed in Section \ref{ChaoticFermiOscillSection}. We expect the Lyapunov exponent in intermediate regimes to interpolate between the values of 0 and $2\pi/\beta$ (perhaps even in a discontinuous manner). The central question is what the Lyapunov exponent is when echoes manifest in the form factor. If we are able to show that  in the regime where echoes exist the Lyapunov bound is nearly saturated\footnote{The reason we say the Lyapunov bound is nearly saturated is due to results of a work that computes the Lyapunov exponent in a fuzzball geometry by analyzing geodesic motion in the vicinity of the fuzzball \cite{Bianchi:2020des}. The work suggests that fuzzballs tend to have a slightly smaller Lyapunov exponent when compared to their black hole counter parts. This may also be true more generally for highly compact objects that resemble black holes far away but deviate near the horizon.} then it might be plausible that the quantum chaotic system described by a Hamiltonian given in Eq. (\ref{OscPlusSYK}) describes a black hole with microstructure near the horizon which exhibits echoes when perturbed. If we find that echoes only exist in coupled oscillator models when the Lyapunov exponent is nearly zero then the modified black hole interpretation is not plausible, we will likely have to consider more complicated models where we can tune the value of the Lyapunov exponent. We leave this exploration to future work.

An important point to make in the discussion of modifications to the near horizon description of black holes and their imprints on the ringdown is that they may not manifest as dramatically as described by the models discussed in Section \ref{SystemOscilFFSection}. In particular, the model of identical oscillators coupled by a random interaction can give sharp echoes like the ones shown in Figure \ref{EarlyTimeOscilFig} at weak coupling. These models have evenly spaced clusters of states. If the clusters were not evenly spaced one might get less dramatic echoes or no discernible echoes at all. Presumably, a completely random placement of these clusters would completely ``wash out'' a clear echo signal in the form factor. It would be interesting to explore where on the spectrum between evenly spaced clusters and randomly spaced clusters echoes persist. This might give a sense of how ``resilient'' the ``echo'' phenomenon may be to deviations from the ideal evenly-spaced clusters scenario. 

Although we expect the high energy sector of a holographic CFT to exhibit chaos, the exact structure of the spectrum is unknown. The results in this paper suggest a wide variety of deviations can occur at early times before the universal ergodic behaviour manifests (i.e. ramp and plateau). The origin of possible deviations at intermediate and early times may be additional coarse-grained structure in the spectrum of states that give rise to clustering effects on energy separations  $ \gg \beta^{-1}e^{-S}$. In our randomly coupled oscillator models, degenerate states were responsible for this clustering. This is interesting because degeneracies arise due to ``symmetries'' in a system. The symmetries thus act as ``seeds'' for additional structure at intermediate energy spacings. If there are symmetries in a holographic CFT that have the same effect, then this may also give rise to additional structure in the spectrum which would imprint itself in the thermalization behaviour of a black hole and potentially be detected as echoes (or other deviations from GR predictions) that could be searched for in future gravitational wave experiments. The i.i.d. model might serve as a useful phenomenological tool to model the wave-forms for unitary black hole ringdowns.

To conclude our discussion, it is worth clarifying a few points. In this paper, we explored the possibility of echoes from black hole microstructure in the context of unitary quantum chaotic systems. Our primary conclusion is that echoes may appear, either due to enhanced repulsion between individual microstates, or the occurrence of regular spacing between clusters of microstates. Appearance of echoes in the unitary quantum description on time scales comparable to the scrambling time suggests the existence of Planck-scale microstructure near the horizon. The reason we asserted this was because the echo time scale found in the classical models with a cutoff depends on how close the cutoff is placed from the horizon. In particular, cutoffs placed a proper-radial Planck length from the horizon (i.e. $r_{\text{cutoff}}\sim r_h+\ell_p^2/\beta $) give echoes around a scrambling time scale $\beta\ln(S)$,\footnote{Note that the $S\sim L/\ell_p\sim N$ for large AdS black holes.}\cite{Saraswat:2019npa}. The randomly coupled oscillator example in Section \ref{SystemOscilFFSection} served as a simple toy model to illustrate how self-averaging echoes can arise due to non-trivial clustering of microstates. The oscillations found in Section \ref{GammaDisSectionFF}, however, are of a different type. They are not self-averaging and arose from enhanced repulsion statistics between individual adjacent microstates. Furthermore, in the context of the i.i.d. model, these non-self averaging oscillations occur on time scales of the order of the Heisenberg time ${\cal O}( e^S)$, which is typically much longer than the scrambling time\footnote{Assuming the average spacing between adjacent microstates was of the order $e^{-S}$.}. The physical interpretation of non-self averaging echoes that occur on the Heisenberg time scale is more subtle and might require a more careful understanding in terms various saddles (or even off-shell effects) that occur in the Euclidean path integral. While the time scale of $e^S$ is what one may expect from quantum tunneling from structure behind the horizon, this is more likely an upper limit, given that black hole evaporation happens on polynomial times. We hope to explore these ideas in more detail in future work.

\acknowledgments

We would like to thank Dominik Neuenfeld, Beni Yoshida, Julian Sonner, and Alexander Maloney for helpful discussions and comments. This work was supported by the University of Waterloo, Natural Sciences and Engineering Research Council of Canada (NSERC), and the Perimeter Institute for Theoretical Physics.  Research at Perimeter Institute is supported in part by the Government of Canada through the Department of Innovation, Science and Economic Development Canada and by the Province of Ontario through the Ministry of Colleges and Universities.

\appendix

\section{Deriving the Spectral Form Factor of the i.i.d Model}
\label{FormFactorDer}
In this section we derive Eq. (\ref{AveragedSpectralFF}). We begin with the expression for the partition function for a spectrum given by Eq. (\ref{RandomSpecDef}). It is given by (in this derivation we will treat $\beta$ as a general complex parameter):
\begin{equation}
    \begin{split}
       &\mathcal{Z}(\beta)=\sum_{n=0}^Ne^{-\beta\left( E_0+\sum_{k=1}^n\delta E_k \right)}=e^{-\beta E_0}\left(1+\sum_{n=1}^N\prod_{k=1}^n e^{-\beta \delta E_k}\right)\\ 
       &=e^{-\beta E_0}\left[1+e^{-\beta \delta E_1}+\left(e^{-\beta \delta E_1}e^{-\beta \delta E_2}\right)+\cdot\cdot\cdot+\left(e^{-\beta \delta E_1}e^{-\beta \delta E_2}\cdot\cdot\cdot e^{-\beta \delta E_N}\right) \right].\\
    \end{split}
\end{equation}
Now we define the following recursive indexed quantity:
\begin{equation}
\begin{split}
\label{RecursionRelation}
    &Z_m(\beta)=e^{-\beta\delta E_m}\left(1+Z_{m+1}(\beta)\right)\\
    &Z_{N+1}(\beta)=0\Rightarrow Z_N(\beta)=e^{-\beta \delta E_N},\\
\end{split}
\end{equation}
where $m=1,2,...,N-1,N$. Using these indexed quantities, we may express the partition function and its complex conjugate as follows:
\begin{equation}
    \begin{split}
        &\mathcal{Z}(\beta)=e^{-\beta E_0}\left(1+Z_1(\beta)\right)\\
        &\mathcal{Z}(\beta^*)=e^{-\beta^* E_0}\left(1+Z_1(\beta^*)\right),\\
    \end{split}
\end{equation}
where $\beta^*$ is the complex conjugate of $\beta$. We can then write:
\begin{equation}
\begin{split}
\label{Z1FormulaForZSq}
    &\braket{\mathcal{Z}(\beta)\mathcal{Z}(\beta^*)}=e^{-E_0(\beta+\beta^*)}\braket{1+Z_1(\beta)+Z_1(\beta^*)+Z_1(\beta)Z(\beta^*)}\\
    &=e^{-E_0(\beta+\beta^*)}\left[1+\braket{Z_1(\beta)}+\braket{Z_1(\beta^*)}+\braket{Z_1(\beta)Z_1(\beta^*)}\right].\\
    \end{split}
\end{equation}
To compute $\braket{Z_1(\beta)}$, we will find a general formula for $\braket{Z_m(\beta)}$. We use the recursion relation given by Eq. (\ref{RecursionRelation}) to find the following result for $\braket{Z_N(\beta)}$:
\begin{equation}
    \begin{split}
        &\braket{Z_N(\beta)}=\braket{e^{-\beta\delta E_N}}=\int_{-\infty}^{\infty}dE_1\cdot\cdot\cdot dE_N P(E_1,..,E_N)e^{-\beta\delta E_N}\\
        &=\int_{-\infty}^{\infty}d\delta E_1\cdot\cdot\cdot d \delta E_N \prod_{k=1}^N\mathcal{P}(\delta E_k) e^{-\beta\delta E_N}=b\\
    &b=\int_{-\infty}^\infty dx \mathcal{P}(x)e^{-\beta x}.\\
    \end{split}
\end{equation}
For $\braket{Z_{N-1}(\beta)}$, we have:
\begin{equation}
    \begin{split}
        &\braket{Z_{N-1}(\beta)}=\braket{e^{-\beta\delta E_{N-1}}\left(1+Z_N(\beta)\right)}=\braket{e^{-\beta \delta E_{N-1}}+e^{-\beta \delta E_{N-1}}e^{-\beta \delta E_N}}\\
        &=\braket{e^{-\beta \delta E_{N-1}}}+\braket{e^{-\beta \delta E_{N-1}}e^{-\beta \delta E_N}}=b+b^2.\\
    \end{split}
\end{equation}
Continuing inductively we can show that:
\begin{equation}
    \braket{Z_{N-k}(\beta)}=\braket{e^{-\beta\delta E_{N-k}}+\sum_{\ell=1}^k\prod_{j=0}^\ell e^{-\beta\delta E_{N-k+j}}}=b\sum_{p=0}^k b^p=\frac{b\left(1-b^{k+1}\right)}{1-b}.
\end{equation}
Using these results we can conclude that:
\begin{equation}
\begin{split}
    &\braket{Z_m(\beta)}=\frac{b(1-b^{N-m+1})}{1-b}\\
    &\braket{Z_m(\beta^*)}=\frac{b^*(1-(b^*)^{N-m+1})}{1-b^*},\\
    \end{split}
\end{equation}
where $b^*$ is the complex conjugate of $b$. We can immediately conclude that:
\begin{equation}
\begin{split}
\label{AveragedPartitionFunc}
    &\braket{\mathcal{Z}(\beta)}=e^{-\beta E_0}\left[1+\frac{b(1-b^N)}{1-b}\right]\\
    &\braket{\mathcal{Z}(\beta^*)}=e^{-\beta^* E_0}\left[1+\frac{b^*(1-(b^*)^N)}{1-b^*}\right].\\
    \end{split}
\end{equation}
In a similar manner, we consider the quantity $B_m=\braket{Z_m(\beta)Z_m(\beta^*)}$:
\begin{equation}
\begin{split}
    &B_m=\braket{Z_m(\beta)Z_m(\beta^*)}=\braket{e^{-\delta E_m(\beta+\beta^*)}\left(1+Z_{m+1}(\beta)\right)\left(1+Z_{m+1}(\beta^*)\right)}\\
    &=\braket{e^{-\delta E_m(\beta+\beta^*)}}+\braket{e^{-\delta E_m(\beta+\beta^*)}Z_{m+1}(\beta)} + \braket{e^{-\delta E_m(\beta+\beta^*)}Z_{m+1}(\beta^*)}\\
    &+\braket{e^{-\delta E_m(\beta+\beta^*)}Z_{m+1}(\beta)Z_{m+1}(\beta^*) }.\\
    \end{split}
\end{equation}
Using the fact that $Z_{N+1}(\beta)=0$, we obtain the following result for $B_N$:
\begin{equation}
\begin{split}
    &B_N=\braket{e^{-\delta E_N(\beta+\beta^*)}}=\int_{-\infty}^\infty dx\mathcal{P}(x)e^{-x(\beta+\beta^*)}=a.\\
    \end{split}
\end{equation}
Using the fact that $Z_{m+1}$ is only a function of $\{\delta E_{i}\}_{i=N}^{m+1}$, we have the following recursion relation:
\begin{equation}
\begin{split}
\label{ZsqRecurRelation}
    &B_m=a\left(1+\braket{Z_{m+1}(\beta)}+\braket{Z_{m+1}(\beta^*)}+B_{m+1}\right)\\
    &=a\left( 1+\frac{b(1-b^{N-m})}{1-b}+\frac{b^*(1-(b^*)^{N-m})}{1-b^*}+B_{m+1} \right)\\
    &a=\braket{e^{-\delta E_m(\beta+\beta^*)}}=\int_{-\infty}^{\infty}dx \mathcal{P}(x)e^{-x(\beta+\beta^*)}\\
    &B_{N+1}=0\Rightarrow B_{N}=a.\\
    \end{split}
\end{equation}
Using the inductive relation above it is straightforward to show that:
\begin{equation}
\begin{split}
    &B_{N-k}=a\left[a^k+\sum_{p=0}^{k-1}a^p f_{N-k+p}\right]\\
    &f_m=1+\frac{b(1-b^{N-m})}{1-b}+\frac{b^*(1-(b^*)^{N-m})}{1-b^*}.\\
    \end{split}
\end{equation}
One can then explicitly evaluate the geometric sums involved and arrive at the following expression for $B_m$:
\begin{equation}
\begin{split}
    &B_m=\frac{a-a^{N-m+2}}{1-a}+\frac{ab}{1-b}\left[\frac{a(1-a^{N-m}-b^{N-m+1})-b(1-b^{N-m}-a^{N-m+1})}{(1-a)(a-b)}\right]\\
    &+\frac{ab^*}{1-b^*}\left[\frac{a(1-a^{N-m}-(b^*)^{N-m+1})-b^*(1-(b^*)^{N-m}-a^{N-m+1})}{(1-a)(a-b^*)}\right].\\
    \end{split}
\end{equation}
We can check that this expression satisfies the recursion relation given in Eq. (\ref{ZsqRecurRelation}). By setting $m=1$ we can obtain $B_1$:
\begin{equation}
    B_1=\frac{a}{1-a}\left[1-a^N+\frac{b\left( a-b+a^N(b-1)+b^N(1-a) \right)}{(1-b)(a-b)}+\frac{b^*\left( a-b^*+a^N(b^*-1)+(b^*)^N(1-a) \right)}{(1-b^*)(a-b^*)}\right].
\end{equation}
Now that we have calculated all the necessary averages we can plug them into the Eq. (\ref{Z1FormulaForZSq}) to find the following result:
\begin{equation}
\begin{split}
    &\braket{\mathcal{Z}(\beta)\mathcal{Z}(\beta^*)}=e^{-(\beta+\beta^*) E_0}\left(1+\frac{b(1-b^N)}{1-b}+\frac{b^*(1-(b^*)^N)}{1-b^*}+B_1\right)\\
   &=e^{-(\beta+\beta^*) E_0}\left(\frac{1-a^{N+1}}{1-a}+\frac{b}{1-b}\left[\frac{ a-b+a^{N+1}(b-1)+b^{N+1}(1-a)}{(1-a)(a-b)} \right]\right)\\
    &+e^{-(\beta+\beta^*) E_0}\frac{b^*}{1-b^*}\left[\frac{ a-b^*+a^{N+1}(b^*-1)+(b^*)^{N+1}(1-a)}{(1-a)(a-b^*)} \right]\\
    &a=\braket{e^{-(\beta+\beta^*) \delta E}}=\int_{-\infty}^\infty \mathcal{P}(x)e^{-(\beta+\beta^*) x}dx\\
    &b=\braket{e^{-\beta\delta E}}=\int_{-\infty}^\infty \mathcal{P}(x)e^{-\beta x}dx\\
    &b^*=\braket{e^{-\beta^* \delta E}}=\int_{-\infty}^\infty \mathcal{P}(x)e^{-\beta^* x}dx.\\
    \end{split}
\end{equation}
Making the replacements $\beta\rightarrow \beta+it$ and $\beta^*\rightarrow \beta-it$ gives us the expression for spectral form factor given in Eq. (\ref{AveragedSpectralFF}).
\section{Average Spectral Density For Poisson Spacing}
\label{SpectralDensityPoissonAppendix}
In this section we go over the details of the computing the integrals involved in computing expression for the average spectral density for a Poisson spacing distribution. To begin we can write the JPDF for the energy levels given as:
\begin{equation}
\begin{split}
    &P(E_1,..,E_N)=\prod_{k=1}^N\left[\Theta(E_k-E_{k-1})\frac{e^{-(E_k-E_{k-1})/\sigma}}{\sigma}\right]\\
    &=\frac{e^{-(E_N-E_0)/\sigma}}{\sigma^N}\prod_{k=1}^N\Theta(E_k-E_{k-1}).\\
    \end{split}
\end{equation}
Using this we can write the average spectral density as:
\begin{equation}
\begin{split}
    &\braket{\rho(E)}=\delta(E-E_0)+\int_{-\infty}^{\infty}dE_1\cdot\cdot\cdot dE_NP(E_1,..,E_N)\sum_{m=1}^N\delta(E-E_m)\\
    &=\delta(E-E_0)+\sum_{m=1}^N\int_{-\infty}^\infty dE_1\cdot\cdot\cdot dE_N \frac{e^{-(E_N-E_0)/\sigma}}{\sigma^N}\delta(E-E_m)\prod_{k=1}^N\Theta(E_k-E_{k-1}).\\
    \end{split}
\end{equation}
With some work we can derive the following identities:
\begin{equation}
\begin{split}
    &\int_{-\infty}^\infty dE_1\cdot\cdot\cdot dE_{m-3}\prod_{k=1}^{m-2}\Theta(E_k-E_{k-1})=\Theta(E_{m-2}-E_0)\frac{(E_{m-2}-E_0)^{m-3}}{(m-3)!}\\
    &\int_{-\infty}^\infty dE_{m+2}\cdot\cdot\cdot dE_N e^{-(E_N-E_0)/\sigma}\prod_{k=m+2}^N\Theta(E_k-E_{k-1})=\sigma^{N-m-1}e^{-(E_{m+1}-E_0)/\sigma}.\\
    \end{split}
\end{equation}
Using these integral identities we can show that:
\begin{equation}
    \begin{split}
       &\int_{-\infty}^\infty dE_1\cdot\cdot\cdot dE_N \frac{e^{-(E_N-E_0)/\sigma}}{\sigma^N}\delta(E-E_m)\prod_{k=1}^N\Theta(E_k-E_{k-1})\\
       &=\Theta(E-E_0)\frac{(E-E_0)^{m-1}}{\sigma^m(m-1)!}e^{-(E-E_0)/\sigma}.\\
    \end{split}
\end{equation}
We therefore conclude that the average spectral density is given by:
\begin{equation}
    \braket{\rho(E)}=\delta(E-E_0)+\sum_{m=1}^N\left[\frac{(E-E_0)^{m-1}}{\sigma^m(m-1)!}\right]\Theta(E-E_0)e^{-(E-E_0)/\sigma}.
\end{equation}
This gives the averaged spectral density given in Eq. (\ref{PoissonSpecDensity}). By integrating $\braket{\rho(E)}$ over $E$ we correctly get $N+1$ which is the total number of states.
\section{Form Factor of GUE vs i.i.d Model with Wigner Spacing}
\label{CompModelandGUEFF}
In this section we will do a brief numerical study of how well the Wigner Surmise distribution given by Eq. (\ref{GUEWigSur}) fits the NNS distribution of eigenvalues of a $100\times 100$ random Hermitian matrix (i.e. matrices in GUE). We will numerically compute the averaged form factor of the Gaussian unitary ensemble and compare to the form factor expression given by Eq. (\ref{AveragedSpectralFF}) with $a,b,b^*$ given by Eq. (\ref{abbstarWigSur}). By doing this, we will get a sense of how well our naive model can capture certain aspects of the true form factor associated with matrices in GUE.

\begin{figure}[]
\centering
\includegraphics[width=120mm]{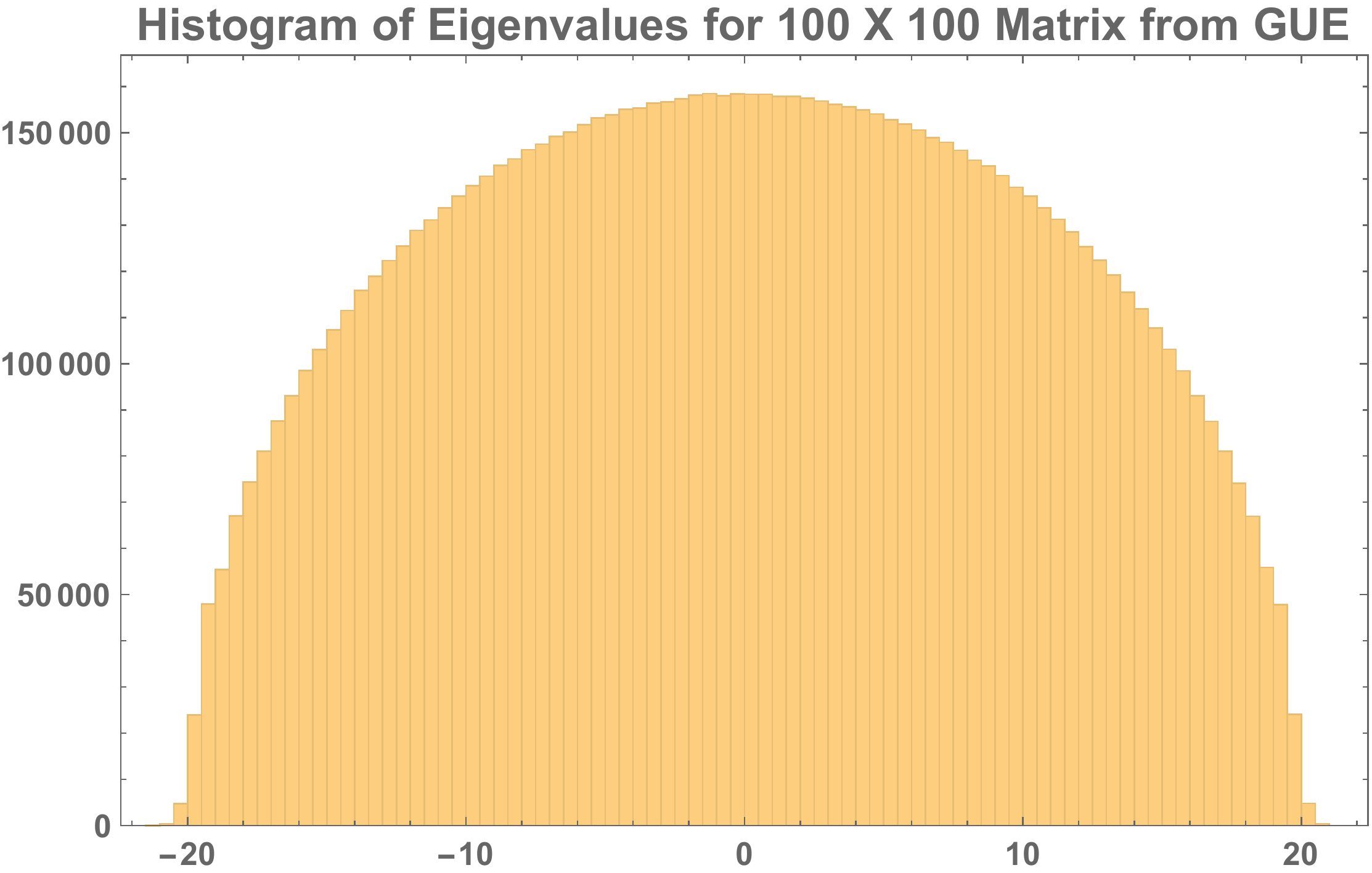}
\caption{Histogram for eigenvalues of a $100\times 100$ random matrix pulled from the GUE. The histogram is generated from the eigenvalues of $10^5$ random samples.\label{GUEHistogramPlot}}
\end{figure}

\begin{figure}[]
\centering
\includegraphics[width=120mm]{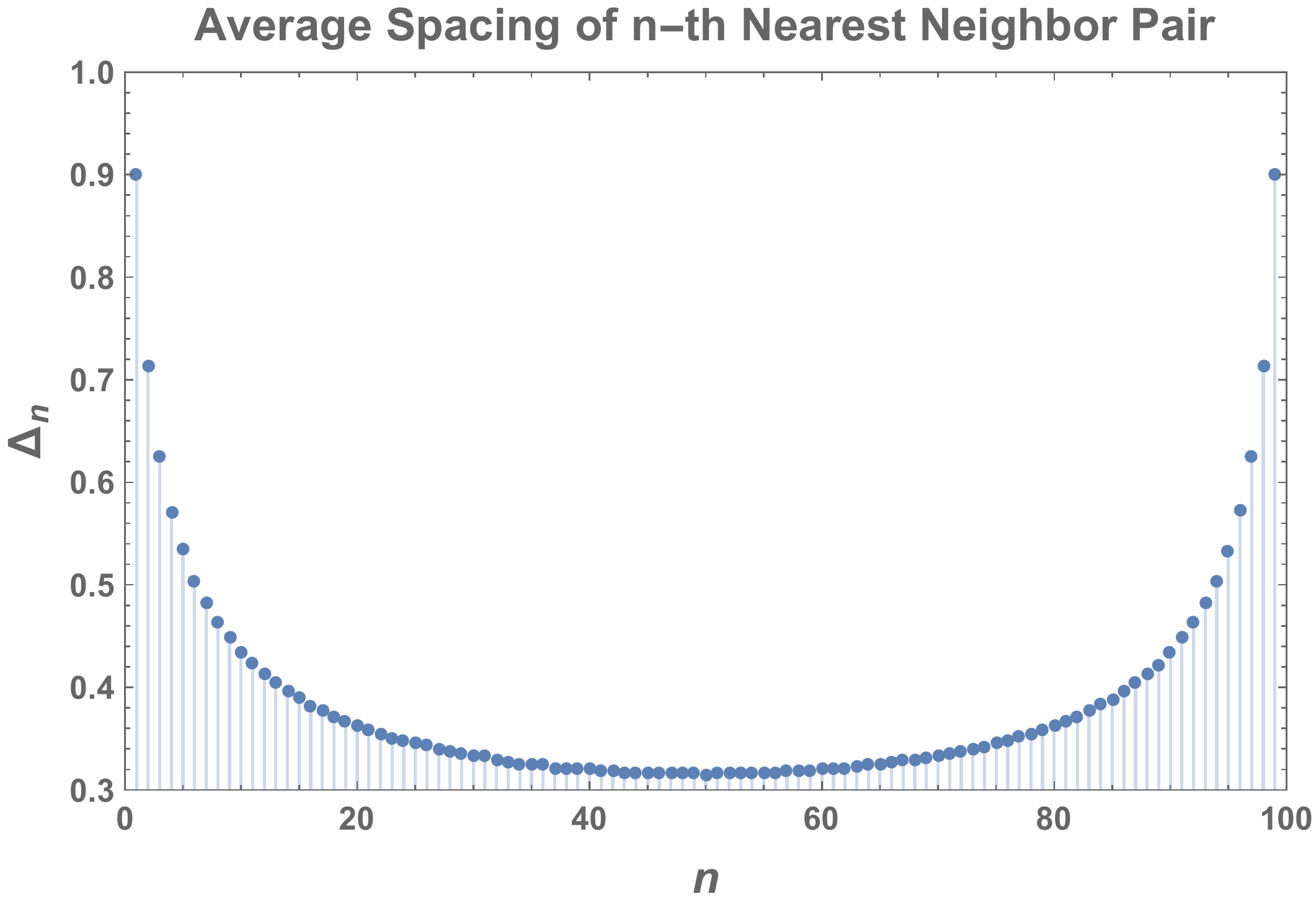}
\caption{We plot the average (over $10^5$ samples) spacing, $\Delta_n$, between the $n$-th nearest neighbor pairs of eigenvalues in the GUE ($100\times 100$ matrices). We can see that near the edge of the GUE spectrum, the average spacing between nearest neighbor pairs changes quickly.  Near the centre of the spectral density of the GUE the average spacing changes slowly.    \label{AVGSpacingPairplot}}
\end{figure}

The numerical calculation is done by defining $10^5$, $100\times 100$ random Hermitian matrices. The diagonal entries $M_{ii}$ are real and are pulled from the following Gaussian distribution:
\begin{equation}
    P(M_{ii})=\frac{1}{\sqrt{2\pi}}e^{-\frac{1}{2}M_{ii}^2}.
\end{equation}
The off diagonal entries are complex entries of the form $x+iy$ and the real and imaginary parts of the entries are pulled from the following Gaussian distribution:
\begin{equation}
\begin{split}
    &P(x)=\frac{1}{\sqrt{\pi}}e^{-x^2}\\
    &P(y)=\frac{1}{\sqrt{\pi}}e^{-y^2}.\\
\end{split}
\end{equation}
We can diagonalize all $10^5$ matrices to get $10^7$ eigenvalues and we will obtain the histogram (i.e. averaged spectral density) for the eigenvalues depicted in Figure \ref{GUEHistogramPlot}.
From our sample of eigenvalues we can determine the NNS statistics. An interesting quantity to analyze is the sample averaged spacing of nearest neighbors, and how this changes between which nearest neighbor pairs we choose. We order the eigenvalues so that $E_1\leq E_2\leq \cdot\cdot\cdot \leq E_{100}$, then we define the average spacing between the $i$-th pair as:
\begin{equation}
    \Delta_i=\frac{1}{10^5}\sum_{n=1}^{10^5}\left(E_{i+1}^{(n)}-E_i^{(n)}\right),
\end{equation}
where $i=1,2,3,..,99$. This represents the average (over $10^5$ samples) of the spacing between nearest neighbor pairs throughout the spectrum. We can do a discrete plot of this as a function of $i$ to obtain Figure \ref{AVGSpacingPairplot}.
We can see that for eigenvalue pairs located near the center, the average spacing between nearest neighbor pairs is roughly constant but as we approach the edges the spacing changes more quickly between one pair and the next. The range of values which the sample averaged spacings can take roughly lie in the interval  $\Delta_i \in[0.32,0.90]$. Based on this, we can see that the spacing statistics of Gaussian ensembles is not i.i.d (except approximately, near the centre of the spectrum where the average spacing changes slowly). Therefore, we expect the model introduced in Section \ref{Fixed NNS model} to give an accurate description of the spectrum of Gaussian ensembles near the centre of the spectrum and far from the edges\footnote{More generally we expect our models with the i.i.d assumption to describe the statistics of eigenvalues located near local the extrema of the spectral density of a random matrix theory.}. Although, we do not expect our model to describe the edges of Gaussian ensembles to a high degree of accuracy it is still interesting to explore how the form factor of our model spectrum compares to the form factor of an actual Gaussian ensemble as we include eigenvalues near the edge of the spectrum.

To do this we define the following numerically calculated quantity:
\begin{equation}
\label{GUENumMicrocanonFF}
    \frac{\braket{\mathcal{Z}_{[a,b]}(\beta+it) \mathcal{Z}_{[a,b]}(\beta-it)}}{\braket{\mathcal{Z}_{[a,b]}(0)^2}}=\frac{\sum_{l=1}^S\sum_{n,m=a}^{b}e^{-(\beta+it)E_n^{(l)}}e^{-(\beta-it)E^{(l)}_m}}{\sum_{l=1}^S\sum_{n,m=a}^{b}e^{-\beta E_n^{(l)}}e^{-\beta E^{(l)}_m}}.
\end{equation}
This is computing the form factor in the GUE which contains the $a$-th through $b$-th eigenvalues anneal averaged over $S$ samples. This is essentially describing a ``microcanonical'' anneal averaged form factor which focuses on eigenvalues within a window $[a,b]$. We want to compare this numerical result to our model which involves the i.i.d assumption on the NNS distribution.   

\begin{figure}[H]
\centering
\includegraphics[width=150mm]{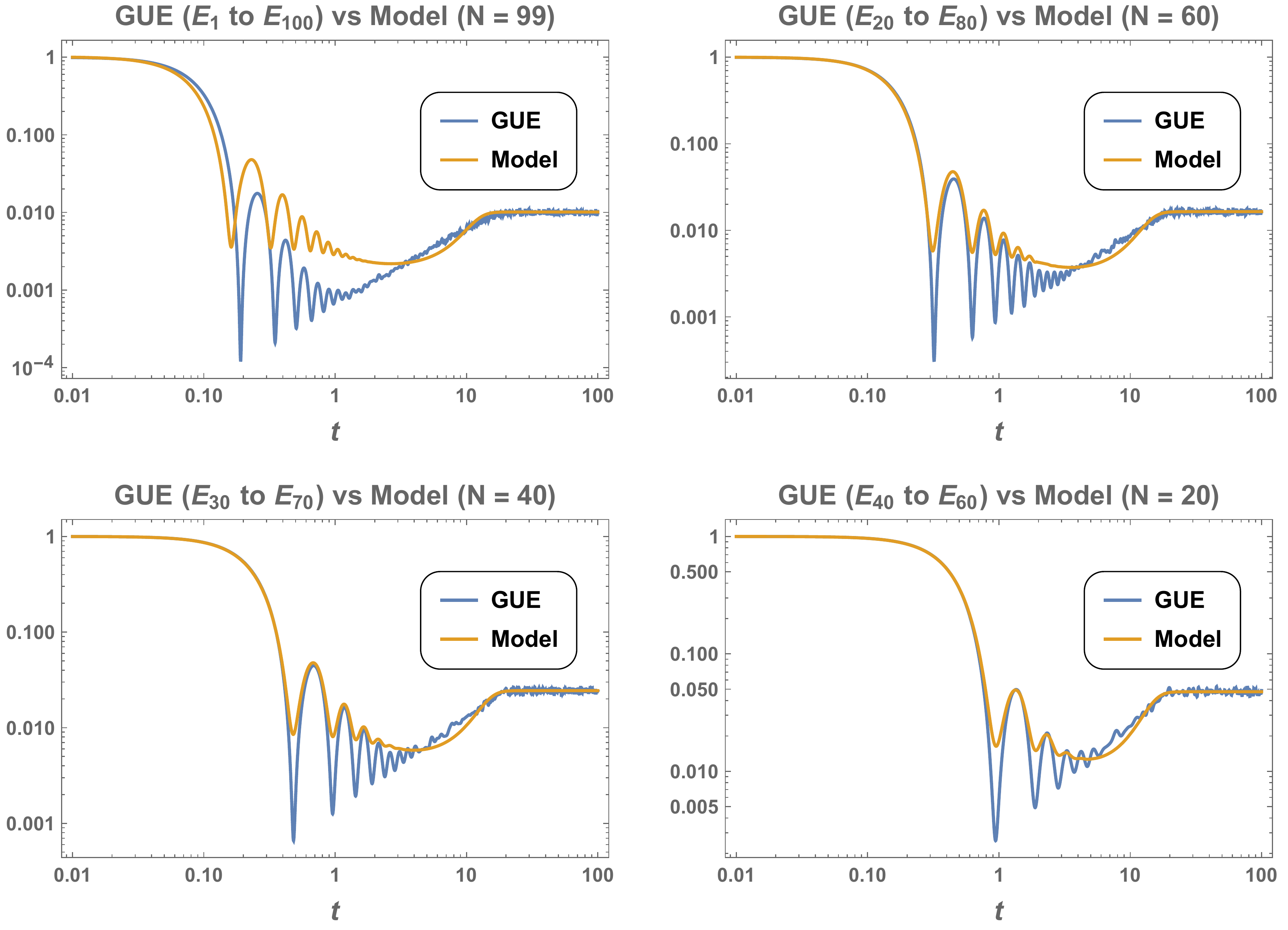}
\caption{Plot depicts a comparison of the infinite temperature form factor of the GUE computed numerically using Eq. (\ref{GUENumMicrocanonFF}) and our i.i.d model with NNS distribution given by Eq. (\ref{microNNSGUE}). The top left plot depicts the a comparison with all eigenvalues in the GUE spectrum. Top right plot depicts a comparison with GUE form factor for roughly 60\% of the eigenvalues within the centre of the spectrum. Bottom left depicts a comparison with GUE form factor for roughly 40\% of the eigenvalues within the centre of the spectrum. Bottom right depicts a comparison with GUE form factor for roughly 20\% of the eigenvalues within the centre of the spectrum. We see that as the window focuses on the centre eigenvalues the deviations between the form factor for the i.i.d model and GUE become smaller at all time scales.   \label{TrunNumFFWSCompPlot}}
\end{figure}

We must specify the following two parameters $c$ and $\sigma$ for the NNS distribution given in Eq. (\ref{GUEWigSur}). Since we are working with the GUE we should set $c=2$ which leaves us one free parameter $\sigma$. Recall that $\sigma$ controls the average spacing between eigenvalues in our model. In particular, for our model the with $c=2$ the relation between the average spacing, $\Delta$, and the parameter $\sigma$ is simply:
\begin{equation}
    \sigma=\frac{\sqrt{\pi}}{2}\Delta.
\end{equation}
We will fix $\sigma$ by analyzing the $\Delta_i$'s in the numerical simulation of the GUE. As we already showed in Figure \ref{AVGSpacingPairplot} the average spacing between eigenvalue pairs is roughly constant near the centre of the spectrum. Motivated by this observation we define following quantity which involves data of the eigenvalues $E_a,E_{a+1},...,E_{b-1},E_b$:
\begin{equation}
    \bar{\Delta}_{[a,b]}=\frac{1}{b-a}\sum_{i=a}^{b-1}\Delta_i.
\end{equation}
In the above equation, $a$ and $b$ label the index of the $a$-th and $b$-th eigenvalues (which are ordered $E_a\leq E_b$ if $a<b$). The quantity $\Delta_{[a,b]}$ defines the sample averaged spacing, averaged over pairs within an interval that contains $E_a,E_{a+1},...,E_{b-1},E_b$ eigenvalues. We define the averaged $\sigma$ over the range of the eigenvalues as, $\bar{\sigma}_{[a,b]}=\frac{\sqrt{\pi}}{2}\Delta_{[a,b]}$. Then for a particular set of eigenvalues in the GUE with indices in the closed interval $[a,b]$ we identify the $\sigma=\bar{\sigma}_{[a,b]}$. This gives the following NNS distribution for our model spectrum:
\begin{equation}
\label{microNNSGUE}
    \mathcal{P}_{[a,b]}(x)=\Theta(x)\frac{4x^2}{\sqrt{\pi}\sigma_{[a,b]}^3}e^{-x^2/\sigma^2_{[a,b]}}.
\end{equation}
We use this NNS distribution along with $N=b-a$ and plug all this into Eq. (\ref{AveragedSpectralFF}) and plot it along-side the ``microcanonical'' GUE form factor defined by Eq. (\ref{GUENumMicrocanonFF}). At infinite temperature we get the plots given in Figure \ref{TrunNumFFWSCompPlot}. 

The top left plot in Figure \ref{TrunNumFFWSCompPlot} depicts the numerically averaged form factor of the GUE (blue line) for the entire spectrum (i.e. it includes $E_1$ to $E_{100}$). We compare with our i.i.d model form factor (the yellow line) with $N=99$ and NNS given by Eq. (\ref{microNNSGUE}) with $\sigma_{[1,100]}\approx 0.345$. We can see that the plateau of the model matches the plateau of the model but at earlier times it deviates away from the GUE.

The top right plot depicts the GUE form factor with the edges of the spectrum cut off - only retaining the eigenvalues $E_{20}$ to $E_{80}$. We compare with the i.i.d model form factor with parameters $N=60$ and $\sigma_{[20,80]}\approx 0.292$. We see that the plateaus still match but now the early time behaviour is matching more closely than before.

The bottom left plot depicts the GUE form factor with the edges of the spectrum cut off even further than, only retaining the eigenvalues $E_{30}$ to $E_{70}$. We compare with the i.i.d model form factor with parameters $N=40$ and $\sigma_{[30,70]}\approx 0.285$. 

Finally, the bottom right plot depicts the GUE, retaining the eigenvalues $E_{40}$ to $E_{60}$. We compare with the i.i.d model form factor with parameters $N=20$ and $\sigma_{[40,60]}\approx 0.281$.

Together the four plots in Figure \ref{TrunNumFFWSCompPlot} plots of the form factor illustrates our claim that our model captures the behaviour spectrum of Gaussian ensembles near the centre of the semi-circle far from the edge. We can see that as we shrink our window of eigenvalues closer to the centre of the spectrum the form factor of the truncated GUE (blue lines) matches more closely with the form factor of our model (yellow line).

\section{Local Extrema of Form Factor for Gamma Distribution NNS}
\label{ExtremaNNSGamma}
In this section, we will analyze Eq. (\ref{GammaDisExtremaEq}) which is given by:
\begin{equation}
    \begin{split}
        &\frac{z^q}{\left[z^{q+1}-1\right]^2}=\frac{(z^*)^q}{\left[(z^*)^{q+1}-1\right]^2}\\
        &z=1+\sigma(\beta+it)\\
        &z^*=1+\sigma(\beta-it).\\
    \end{split}
\end{equation}
By inspection, it is clear that $t=0$ is a valid solution and corresponds to an extremal point. Furthermore, in the limit that $t\to\infty$, the equation is also satisfied, this represents the saturation of the form factor to a constant value (the plateau phase). Aside from these trivial solutions we should expect at least one other solution in the case when $q>0$. This is because for $q>0$ there is repulsion and therefore a ramp after the initial dip downward. This will give a local minimum in the form factor. Furthermore for very large values of $q\gg 1$ we also expect oscillations to exist which will give further extremal points in the form factor. 

To understand where these non-trivial extremal points occur it is useful to write the complex variable $z$ in the polar coordinate representation:
\begin{equation}
\begin{split}
    &z=Ae^{i\theta}\\
    &A=\left(1+\beta\sigma\right)\sqrt{1+\left(\frac{t\sigma}{1+\beta\sigma}\right)^2}\\
    &\theta=\arctan\left(\frac{t\sigma}{1+\beta\sigma}\right).\\
    \end{split}
\end{equation}
In this representation, Eq. (\ref{GammaDisExtremaEq}) can be written in the following form:
\begin{equation}
\label{ImPartEq0}
    -A^{2(q+1)}\sin\left[(q+2)\theta\right]+2A^{q+1}\sin(\theta)+\sin(q\theta)=0.
\end{equation}
This is difficult to solve in general so we will consider the special cases when $q=1,2,4$ where we can solve the equation exactly. We will also make some estimates in the $q\gg 1$ regime.\\

\textbf{$q=1$ Case:}\\
In this case one will find five distinct solutions. One solution is the trivial solution when $t=0$, the other four are non-trivial but only one is real and positive for $\beta\sigma>0$, it is given by: 
\begin{equation}
   t\sigma=\sqrt{\beta\sigma+\left(1+\beta\sigma\right)\left[\beta\sigma+2\sqrt{\beta\sigma(2+\beta\sigma)}\right]}\simeq \begin{cases}
\sqrt{3}\beta\sigma+\mathcal{O}(1),\; \beta\sigma\gg 1\\
\left(8\beta\sigma\right)^{1/4}+\mathcal{O}((\beta\sigma)^{3/4}),\; \beta\sigma \ll 1\\
\end{cases}\\ .
\end{equation}
\\

\textbf{$q=2$ Case:}\\
In this case one will find seven distinct solutions. One solution is the trivial solution when $t=0$, the other six are non-trivial but only one is real and positive for $\beta\sigma>0$, it is given by: 
\begin{equation}
   t\sigma=\left(1+\beta\sigma\right)\left[1-\frac{1}{\left(1+\beta\sigma\right)^3}\right]^{1/4}\simeq \begin{cases}
\beta\sigma+\mathcal{O}(1),\; \beta\sigma\gg 1\\
\left(3\beta\sigma\right)^{1/4}+\mathcal{O}((\beta\sigma)^{3/4}),\; \beta\sigma \ll 1\\
\end{cases}\\.
\end{equation}
In both cases we can see that up to a pre-factor the local minimum before the ramp at low temperatures scales with $\beta$ this is contrast to the high temperature case where it scales as $\beta^{1/4}$.\\

\textbf{$q = 4$ Case:}\\
In this case we will have two non-trivial real solutions which have closed form expressions. The exact expressions are complicated so we will not write them here, instead we will write the solutions as series expansion for small and large $\beta\sigma$. We start with the smaller solution which represents the initial dip in the form factor before the ramp given by:
\begin{equation}
   t\sigma\simeq \begin{cases}
\beta\sigma/\sqrt{3}+\mathcal{O}(1),\; \beta\sigma\gg 1\\
\left(\beta\sigma\right)^{1/4}+\mathcal{O}((\beta\sigma)^{3/4}),\; \beta\sigma \ll 1\\
\end{cases}\\.
\end{equation}
The next real solution corresponds to the small ``kink'' in the form factor that connects the ramp to the plateau which is also seen in GSE. The expression for the time when the kink occurs is well approximated in all temperature regimes by the following expression:
\begin{equation}
    t\sigma\approx \sqrt{3}(1+\beta\sigma).
\end{equation}
Overall, we see that for $q=1,2,4$ the inital dip before the plateau has the same temperature dependence up to an order one pre-factor. Furthermore, we are able to show that for the $q=4$ case the form factor has an additional local maximum which represents the  ``kink'' which connects the plateau to the ramp seen in Figure \ref{FormFactPlotGammaDist}. \\

\textbf{$q\gg 1$ Case:}\\
We know that for very large values of $q$, the form factor exhibits an initial dip followed by oscillations before saturation to a plateau. In this case, we should expect to find more than one non-trivial extremal point which describe the peaks and troughs of the oscillations. In the case when $q\gg1$, Eq. (\ref{ImPartEq0}) at leading order will read:
\begin{equation}
  \left[A^{q}-A^{-q}\right]\sin(q\theta)=2\sin(\theta).
\end{equation}
If we define $x=t\sigma/(1+\beta\sigma)$ we can explicitly write:
\begin{equation}
\label{explicitlargeqExtEq}
    \left[(1+\beta\sigma)^q(1+x^2)^{q/2}-(1+\beta\sigma)^{-q}(1+x^2)^{-q/2}\right]\sin\left[q\arctan(x)\right]=\frac{2x}{\sqrt{1+x^2}}.
\end{equation}
We can clearly see that the left hand side of Eq. (\ref{explicitlargeqExtEq}) will diverge like $x^q\sin(q\pi/2)$ as $x\to \infty$. The left hand side saturates to a value of 2 in the $x\to \infty$ limit. This means that there are a finite number of solutions for a given $q\gg 1$ - which is in agreement with the plots we have made. Clearly $x=0$ is a solution, but we expect there to be more. 

To understand how many solutions there are, we start by analyzing the number of roots (excluding $x=0$) to the expression on the left hand side of Eq. (\ref{explicitlargeqExtEq}). This is simply given by the roots of $\sin[q\arctan(x)]$ which occur when $q\arctan(x)=m\pi$ where $m\in \mathbb{N}\cup \{0\}$. Since $0\leq \arctan(x)\leq \pi/2$, the number of zeros is controlled by $q$. In particular, we denote $N_{root}$ as the number of zeros (excluding the zero at $x=0$) we have:
\begin{equation}
    N_{root}=\begin{cases}
\frac{q}{2}-1,\; q/2\in\mathbb{N}\\
\lfloor\frac{q}{2}\rfloor,\; \text{otherwise}\\
\end{cases}\\.
\end{equation}
Now consider the right hand side of Eq. (\ref{explicitlargeqExtEq}). Since the right-hand side is a monotonically increasing function which takes on values in the interval $(0,2)$. It is not difficult to see that for a sufficiently large $q$ there will be at least $N_{root}-1$ non-trivial solutions to Eq. (\ref{explicitlargeqExtEq})\footnote{By sufficiently large $q$, we mean that for a given $\beta\sigma$ we must have $q$ such that the left hand side of Eq. (\ref{explicitlargeqExtEq}) evaluated at $x=tan(\pi/(2q))$ be larger than 2. This roughly gives a condition that $q>\frac{\ln(2)}{\beta\sigma}$}. Whether there is an additional solution depends on the limits as $x\to\infty$. In particular, one can argue that the number of solutions (excluding the $x=0$ solution) is given by:
\begin{equation}
    N_{sol.}=\begin{cases}
N_{root},\; \frac{N_{root}}{2}\in\mathbb{N}\\
N_{root}-1,\; \frac{N_{root}}{2}\in (2\mathbb{N}+1)\\
\end{cases}\\.
\end{equation}
Now that we have quantified the number of solutions we estimate that the solutions for sufficiently large $q$ occur roughly at:
\begin{equation}
    x_n=\frac{t_n\sigma}{1+\beta\sigma}=\frac{n \pi}{q},\; n=0,1,2,3,..,N_{sol.}
\end{equation}
\begin{figure}[]
\centering
\includegraphics[width=120mm]{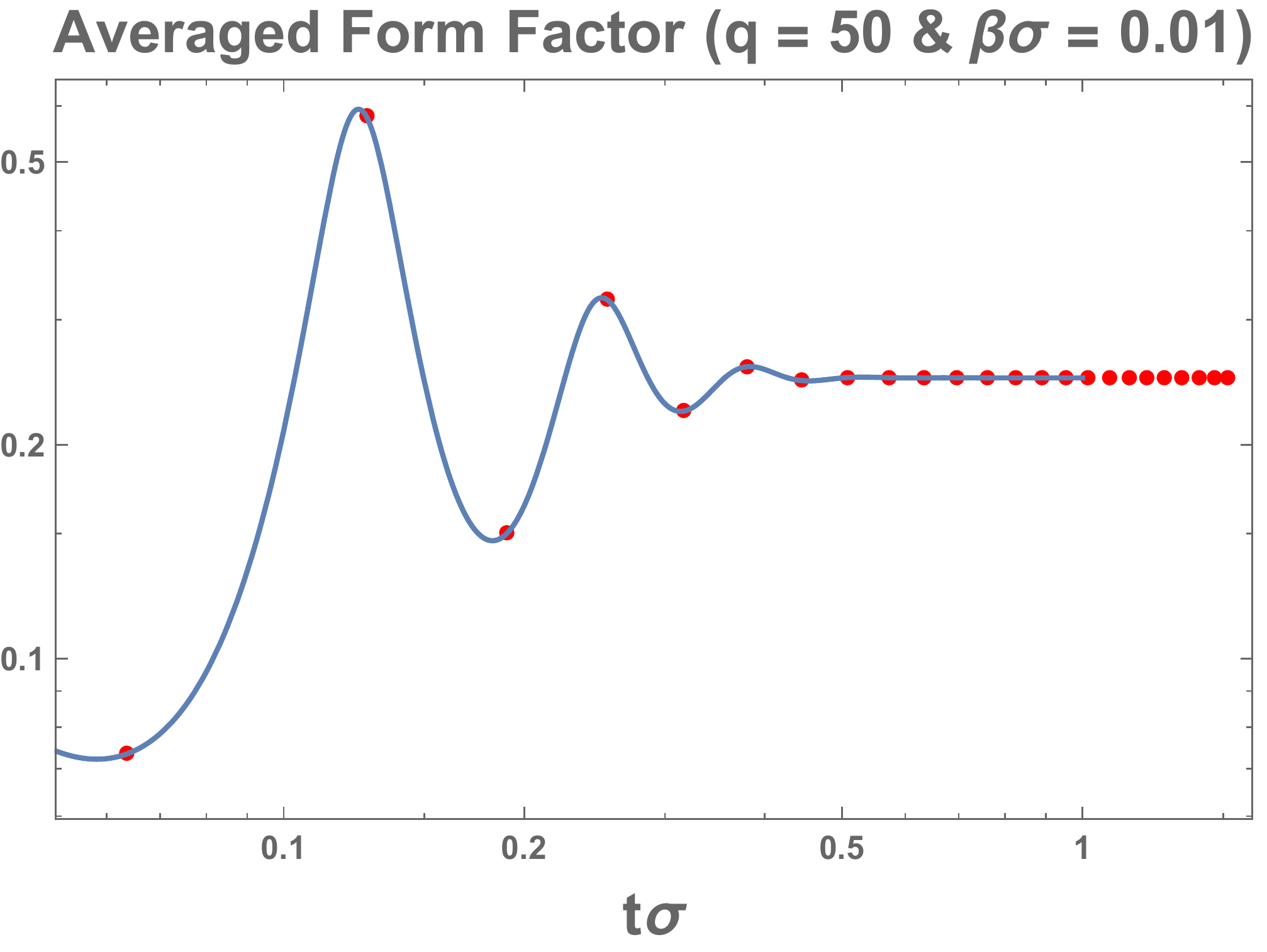}
\caption{We plot the expression of the form factor in Eq. (\ref{ThermoDyFFGammaDis}) for $q = 50$, $\beta\sigma=0.01$. The red dots indicate points on the form factor at $t_n\sigma=1.01\pi n/50$. We can see that the points roughly occur local maxima and minima of the form factor.  \label{ExtremaOnq50FFPlot}}
\end{figure}
To illustrate how accurately this approximation captures the location of the extremal points of the form factor described by Eq. (\ref{ThermoDyFFGammaDis}). We make Figure \ref{ExtremaOnq50FFPlot}, which depicts the averaged form factor with $q=50$ and $\beta=0.01$ with the red dots representing the value of the form factor at $t_n\sigma=(1+\beta\sigma)n\pi/q=(1.01)\pi n/50$. We can see that our approximation for the location of local extrema representing the peaks and troughs of the regular decaying oscillation is not perfect but it does give a reasonable estimate. Similar plots can also be made at other temperatures as well.

Based on these results we conclude that the period of the oscillations in the form factor are roughly given by:
\begin{equation}
    T\approx \frac{2\pi(1+\beta\sigma)}{q\sigma}.
\end{equation}
\section{Average Spectral Density For Gamma Distribution Spacing}
\label{SpectralDenAppendixGD}
In this section we compute the average spectral density for the Gamma Distribution spacing distribution given in Eq. (\ref{GammaDist}). To begin, we can write the JPDF for the energy levels given as:
\begin{equation}
\begin{split}
    &P(E_1,..,E_N)=\prod_{k=1}^N\left[\Theta(E_k-E_{k-1})\frac{(E_k-E_{k-1})^qe^{-(E_k-E_{k-1})/\sigma}}{\Gamma(1+q) \sigma^{1+q}}\right]\\
    &=\frac{e^{-(E_N-E_0)/\sigma}}{\left[\Gamma(1+q)\sigma^{1+q}\right]^N}\prod_{k=1}^N\Theta(E_k-E_{k-1})\left(E_k-E_{k-1}\right)^q.\\
    \end{split}
\end{equation}
Using this we can write the average spectral density as:
\begin{equation}
\begin{split}
    &\braket{\rho(E)}=\int_{-\infty}^{\infty}dE_1\cdot\cdot\cdot dE_NP(E_1,..,E_N)\sum_{m=0}^N\delta(E-E_m)\\
    &\braket{\rho(E)}=\delta(E-E_0)+\sum_{m=1}^N\int_{-\infty}^\infty dE_1\cdot\cdot\cdot dE_N \frac{e^{-(E_N-E_0)/\sigma}}{\left[\Gamma(1+q)\sigma^{1+q}\right]^N}\delta(E-E_m)\prod_{k=1}^N\Theta(E_k-E_{k-1})(E_k-E_{k-1})^q.\\
    \end{split}
\end{equation}
With some work we can derive the following identities:
\begin{equation}
    \begin{split}
        &\int_{-\infty}^\infty dE_1\cdot\cdot\cdot dE_{m-3}\prod_{k=1}^{m-2}\Theta(E_k-E_{k-1})(E_k-E_{k-1})^q\\
        &=\frac{\sqrt{\pi}\Gamma(1+q)^{m-3}}{2^{2q+1}\Gamma(3/2+q)}\frac{\Gamma\left[2(1+q)\right]}{\Gamma\left[(m-2)(1+q)\right]}\Theta(E_{m-2}-E_0)(E_{m-2}-E_0)^{(m-2)q+(m-3)}\\
        &=\frac{\Gamma(1+q)^{m-2}}{\Gamma\left[(m-2)(1+q)\right]} \Theta(E_{m-2}-E_0)(E_{m-2}-E_0)^{(m-2)q+(m-3)}\\
    \end{split}
\end{equation}
\begin{equation}
    \begin{split}
        &\int_{-\infty}^\infty dE_{m+2}\cdot\cdot\cdot dE_N e^{-(E_N-E_0)/\sigma}\prod_{k=m+2}^N\Theta(E_k-E_{k-1})(E_k-E_{k-1})^q\\
        &=\left[\Gamma(1+q)\sigma^{1+q}\right]^{N-m-1}e^{-(E_{m+1}-E_0)/\sigma}.\\
    \end{split}
\end{equation}
Using the identities above with some additional work one will find:
\begin{equation}
    \begin{split}
        &\braket{\rho(E)}=\delta(E-E_0)+\sum_{m=1}^N\left[\frac{(E-E_0)^{mq+m-1}}{\Gamma\left[m(1+q)\right]\sigma^{m(1+q)}}\right]\Theta(E-E_0)e^{-(E-E_0)/\sigma}.\\
    \end{split}
\end{equation}
This gives us the average spectral density given by Eq. (\ref{SpectralDensityGammaDis}). Furthermore, setting $q=0$ we will recover the spectral density for the Poisson NNS model discussed in Section \ref{PoissonSpacingSection}. 

\section{NNS Distribution for Oscillator with Chaotic Interactions $2\times 2$ Case}
\label{NNSEvenPlusRndDerivationAppendix}
We go over the calculation of the NNS distribution for the following matrix:
\begin{equation}
    \mathcal{H}=\frac{\omega_0}{2}\begin{bmatrix}
0 & 0 \\
0 & 1 \\
\end{bmatrix}+\epsilon\begin{bmatrix}
x_1 & x_3+ix_4 \\
x_3-ix_4 & x_2 \\
\end{bmatrix},
\end{equation}
where the real entries are drawn from the following distributions:
Where $x_1,x_2,x_3,$ and $x_4$ are real random variables which follow the following distributions:
\begin{equation}
\begin{split}
       &\mathcal{P}(x_1)=\frac{1}{\sqrt{2\pi}} e^{-x_1^2/2}\\
       &\mathcal{P}(x_2)= \frac{1}{\sqrt{2\pi}} e^{-x_2^2/2} \\
       &\mathcal{P}(x_3)=\sqrt{\frac{1}{\pi}} e^{-x_3^2} \\
       &\mathcal{P}(x_4)=\sqrt{\frac{1}{\pi}} e^{-x_4^2} \\.
\end{split}
\end{equation}
We can compute the spacing distribution between the eigenvalues by evaluating the following integral:
\begin{equation}
    \mathcal{P}(s)=\int_{-\infty}^\infty dx_1\cdot\cdot\cdot dx_4\mathcal{P}(x_1)\mathcal{P}(x_2)\mathcal{P}(x_3)\mathcal{P}(x_4)\delta\left(s-\frac{\omega_0}{2}\sqrt{\left(1+\frac{2\epsilon(x_2-x_1)}{\omega_0}\right)^2+\frac{16\epsilon^2}{\omega_0^2}(x_3^2+x_4^2)}\right).
\end{equation}
We define the following change of variables:
\begin{equation}
\begin{split}
    &x_1=\frac{z+r\cos \theta}{2\epsilon}+\frac{\omega}{2\epsilon}\\
    &x_2=\frac{z-r\cos \theta}{2\epsilon}\\
    &x_3=\frac{r\sin \theta\sin\phi}{2\epsilon}\\
    &x_4=\frac{r\sin \theta\cos \phi}{2\epsilon}.\\
    \end{split}
\end{equation}
With these variables we have $dx_1dx_2dx_3dx_4=\frac{r^2\sin \theta}{8\epsilon^4}drdz d\theta d\phi$ and the integral becomes:
\begin{equation}
\begin{split}
    &\mathcal{P}(s)=\frac{1}{16\pi^2\epsilon^4}\int_{0}^\infty dr\int_{-\infty}^\infty dz \int_{0}^\pi d\theta
\int_{0}^{2\pi} d\phi r^2\sin\theta e^{-\frac{2(r^2+z^2)+2z\omega+\omega^2+2r\omega\cos(\theta)}{8\epsilon^2}}\delta(s-r)\\
&=\frac{s}{\sqrt{\pi}\epsilon\omega_0} \left(e^{\frac{\omega_0s}{2\epsilon^2}}-1\right) e^{-\frac{(2s+\omega_0)^2}{16\epsilon^2}}.\\
\end{split}
\end{equation}
This gives the result in Eq. (\ref{NNSDistEvenPlusRand}).
\section{Numerical Analysis in the Spread of Degenerate States}
\label{SpreadAnalysisAppendix}
In this appendix, we go over numerical results which discuss the statistics of how the degenerate eigenvalues of the fermionic harmonic oscillator given in Eq. $(\ref{FHOENergy})$ split due to a perturbation by a random matrix in the GUE. The Hamiltonian in the basis of the unperturbed oscillator can be written as:
\begin{equation}
     \frac{\mathcal{H}}{\omega_0}=\bigoplus_{p=0}^{N/2}\left[\left(p-\frac{N}{4}\right)\mathcal{I}(p)\right]+\frac{\epsilon}{\omega_0} \mathcal{H}_{GUE},
\end{equation}
where $I(p)$ is an identity matrix of size $\Omega(p)\times \Omega(p)$ (recall that $\Omega(p)$ is given by Eq. (\ref{NumberMicrostatesFHO})) and $\mathcal{H}_{GUE}$ is a $2^{N/2}\times 2^{N/2}$ random matrix pulled from the Gaussian unitary ensemble (see footnote \ref{GUEparaFootnote} for details of the variance of the Gaussian used for the matrix elements). We fix $\epsilon/\omega_0$ then we numerically compute the eigenvalues over $r$ independent samples and collect all these eigenvalues (there are $S=2^{N/2}\times r$ number of them) and put them in an ordered set $\{E_i\}_{i=1}^{S}$ where $E_1\leq E_2\leq \cdot\cdot\cdot \leq E_{S}$. We define the following subsets labeled by $p$:
\begin{equation}
\begin{split}
   & A_p=\left\{ E_k:k_{\min}(p)\leq k\leq k_{\max}(p) \right\}\\
   &k_{\min}(p)=1+r\sum_{l=0}^{p-1}\Omega(l)\\
   &k_{\max}(p)=r\sum_{l=0}^{p}\Omega(l),\\
    \end{split}
\end{equation}
where $p=0,1,2,..,N/2$. The set $A_p$ simply partitions the original ordered set into subsets labeled by $p$ which are of the size $|A_p|=r\Omega(p)$. The reason this is done can be illustrated by considering what happens when $\epsilon=0$. In this case, there is no perturbation and the Hamiltonian is diagonal. If we took $r$ samples of the exact same diagonal matrix and ordered all the eigenvalues in a list we would find there are exactly $|A_p|=r\Omega(p)$ eigenvalues with the value $E(p)$. Adding a small perturbation would lead to a small spreading in the eigenvalues within each degenerate sector so the set $A_p$ is the set of eigenvalues that have ``split'' in the $p-th$ degenerate sector. We define the numerical width of the $p-th$ sector as:
\begin{equation}
\begin{split}
    &W_{num}(p)=\max(A_p)-\min(A_p)\\
    &\max(A_p)=E_{k_{\max}(p)}\\
    &\min(A_p)=E_{k_{\min}(p)}.\\
\end{split}
\end{equation}
The numerical average energy within the $p-th$ sector is given by:
\begin{equation}
    \braket{E(p)}_{num}=\frac{1}{r}\sum_{i=k_{\min(p)}}^{k_{\max}(p)}E_i.
\end{equation}
Applying this numerical procedure we obtain the following plots which compare various numerical and estimated quantities when $r=100$ and $N=22$.  
\begin{figure}[H]
\centering
\includegraphics[width=130mm]{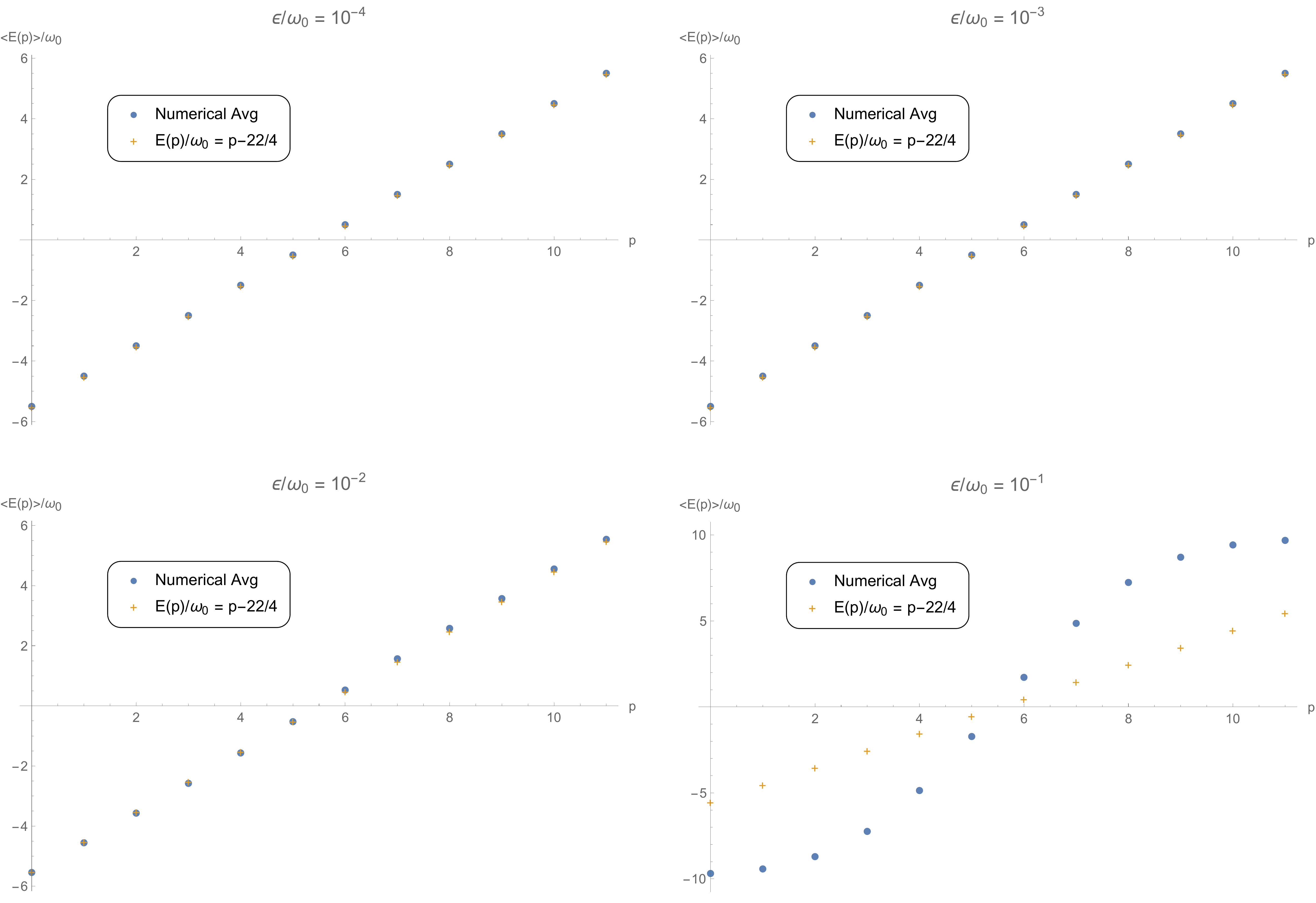}
\caption{Above are plots of the average energy within degenerate sectors labelled by $p$ at various coupling regimes ranging from the weakly coupled regime with $\epsilon/\omega_0=10^{-4} $ to the strongly coupled regime with $\epsilon/\omega_0=10^{-1}$. The solid circles are numerical computations of the average and the ``$+$'' is the energy of the degeneracy sector in the free oscillator case.  \label{AverageDegSecPlot}}
\end{figure}
We see that in the weakly coupled regime which is given by the top two plots of Figure \ref{AverageDegSecPlot} the numerical average of the energies within the sector $p-th$ sector is close to the degenerate energy of the free oscillators which is consistent with predictions of first order degenerate perturbation theory. When $\epsilon/\omega_0=10^{-2}$ we can start to see small deviations in the average energy from the unperturbed energy. By the time we get to the strongly coupled regime where degenerate sectors strongly mix there are large deviations. 
\begin{figure}[H]
\centering
\includegraphics[width=130mm]{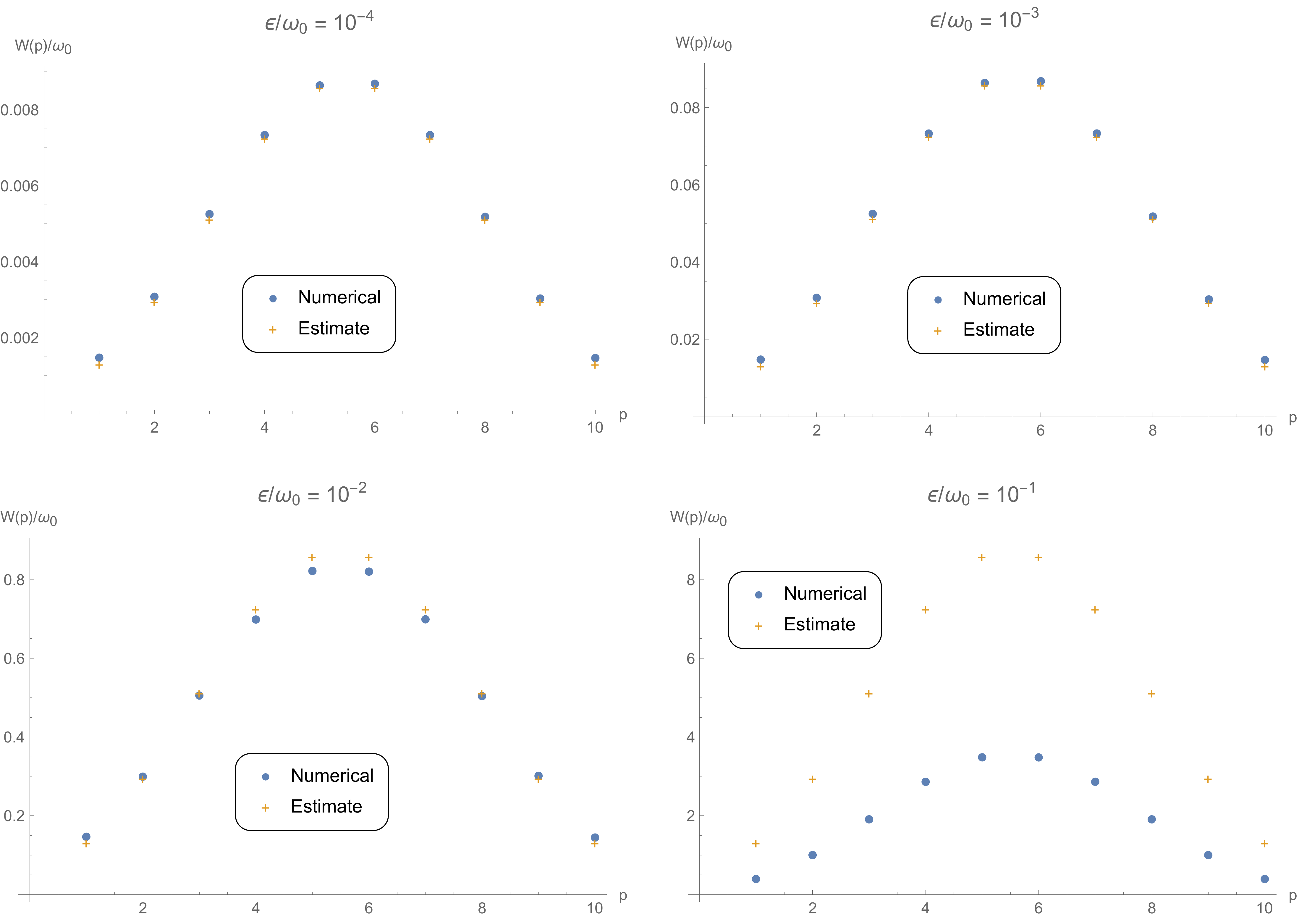}
\caption{Above are plots of the width in the spread of degenerate energy states within the $p$-th degenerate sector at various coupling regimes ranging from the weakly coupled regime with $\epsilon/\omega_0=10^{-4} $ to the strongly coupled regime with $\epsilon/\omega_0=10^{-1}$. The solid circles are numerical computations of the width and the ``$+$'' is the expectation of the width (Given by Eq. (\ref{widthofSectorEst})) from first order degenerate perturbation theory.  \label{DegSecWidthPlots} }
\end{figure}
In the weakly coupled regime which is given by the top two plots of Figure \ref{DegSecWidthPlots}, the numerical width of the spread in the energies within the $p-th$ sector is close to the estimate given by Eq. (\ref{widthofSectorEst}) which is obtained by applying first order degenerate perturbation theory. When $\epsilon/\omega_0=10^{-2}$, we can start to see small deviations from the estimate. By the time we get to the strongly coupled regime where degenerate sectors strongly mix there are large deviations.
\bibliography{Ref.bib}
\bibliographystyle{JHEP}



\end{document}